\documentclass[a4paper,11pt]{article}
\pdfoutput=1 

\usepackage{jheppub} 
\usepackage[utf8x]{inputenc}
\usepackage[T1]{fontenc} 
\newcommand{\be}{\begin{equation}}
	\newcommand{\ee}{\end{equation}}
\newcommand{\bea}{\begin{array}}
	\newcommand{\ea}{\end{array}}
\newcommand{\beqa}{\begin{eqnarray}}
	\newcommand{\eeqa}{\end{eqnarray}}
\newcommand{\nn}{\nonumber}
\usepackage{physics}
\newcommand{\del}{\partial}
\usepackage{bm}
\usepackage{bbm}
\usepackage{multirow}
\usepackage[table,xcdraw]{xcolor}
\usepackage{mathtools}
\usepackage{dsfont}
\usepackage{float}
\usepackage{caption}
\usepackage{subcaption}

\title{\boldmath Chaos from Massive Deformations of Yang-Mills Matrix Models}

\author{K. Ba\c{s}kan,}
\author{S. K\"{u}rk\c{c}\"{u}o\v{g}lu,}
\author{O. Oktay,}
\author{C. Ta\c{s}cı}

\affiliation{Middle East Technical University, Department of Physics,\\Dumlupinar Boulevard, 06800, Ankara, Turkey}

\emailAdd{kagan.baskan@metu.edu.tr}
\emailAdd{kseckin@metu.edu.tr}
\emailAdd{onur.oktay@metu.edu.tr}
\emailAdd{cankut@metu.edu.tr}

\abstract{We focus on an $SU(N)$ Yang-Mills gauge theory in $0+1$-dimensions with the same matrix content as the bosonic part of the BFSS matrix model, but with mass deformation terms breaking the global $SO(9)$ symmetry of the latter to $SO(5) \times SO(3) \times {\mathbb Z}_2$. Introducing an ansatz configuration involving fuzzy four and two spheres with collective time dependence, we examine the chaotic dynamics in a family of effective Lagrangians obtained by tracing over the aforementioned ansatz configurations at the matrix levels $N = \frac{1}{6}(n+1)(n+2)(n+3)$, for $n=1,2,\cdots\,,7$. Through numerical work, we determine the Lyapunov spectrum and analyze how the largest Lyapunov exponents(LLE) change as a function of the energy, and discuss how our results can be used to model the temperature dependence of the LLEs and put upper bounds on the temperature above which LLE values comply with the Maldacena-Shenker-Stanford (MSS) bound $2 \pi T $ , and below which it will eventually be violated.}  
	

\begin{document} 
	\maketitle
	\flushbottom

\section{Introduction}

Recently, there has been intense interest in exploring, modeling and understanding the emergence of chaos from matrix models of Yang-Mills(YM) theories \cite{Sekino:2008he, Asplund:2011qj, Shenker:2013pqa, Gur-Ari:2015rcq, Berenstein:2016zgj, Maldacena:2015waa, Aoki:2015uha, Asano:2015eha, Berkowitz:2016jlq, Buividovich:2017kfk, Buividovich:2018scl, Coskun:2018wmz}. Attention has been focused on examining the many aspects and features of chaos in the matrix models of primary interest within the context of M-theory and string theories, namely in the BFSS and BMN models \cite{Banks:1996vh, deWit:1988wri, Itzhaki:1998dd,  Berenstein:2002jq, Dasgupta:2002hx,Ydri:2017ncg,Ydri:2016dmy}. These models are supersymmetric $SU(N)$ gauge theories in $0+1$-dimensions, whose bosonic part consist of nine $N \times N$ matrices and commonly referred to as matrix quantum mechanics in the literature. BFSS model is associated to the type II-A string theory and appears as the DLCQ of the M-theory in the flat background, while the BMN model, being a deformation of the BFSS involving quadratic and qubic terms preserving the maximal amount of supersymmetry, is the discrete light cone quantization(DLCQ) of M-theory on the pp-wave background. They describe the dynamics of $N$-coincident $D0$-branes in flat and spherical backgrounds, respectively; the latter being due to the fact that fuzzy 2-spheres are vacuum configurations in the BMN model. In the gravity dual, i.e. at large $N$ and strong coupling/low temperature limit, it is known that these $D0$-branes form a black-brane phase, i.e. a string theoretic black hole \cite{Kiritsis:2007zza,Ydri:2017ncg,Ydri:2016dmy}. 

From these considerations it is apparent that a strong motivation in studying possible chaotic dynamics possessed by these matrix models comes from possibilities in acquiring an in-depth perspective and perhaps even discovering novel physical phenomena relating to the dual black-brane phases, such as their thermalization and evaporation processes. In some of the recent works, large $N$ and high temperature limit of the BFSS and BMN models are probed using numerical methods, and results indicating fast termalization are encountered \cite{Asplund:2011qj} and interpreted to indicate fast scrambling \cite{Sekino:2008he}. The latter is a conjecture on black holes, proposed by Sekino and Susskind, which may be concisely stated as the fact that chaotic dynamics sets in faster in black holes than in any other physical system and the rate at which this happens is proportional to logarithm of the number of degrees of freedom, i.e. to the logarithm of the Bekenstein-Hawking entropy of the black hole. It may be remarked that although this large $N$ and high temperature limit is distinguished from the limit in which the gravity duals are obtained, as noted in \cite{Gur-Ari:2015rcq}, in numerical studies conducted so far \cite{Anagnostopoulos:2007fw, Catterall:2008yz}, no indication of an occurrence of a phase transition between low and high temperature limits of these matrix models is reported, which makes it plausible to expect that features like fast scrambling of black holes in the gravity dual could survive at the high temperature limit too. In fact, the latter is the natural limit in which classical dynamics provides a good approximation to the matrix quantum mechanics,\footnote{It may be emphasized that, this picture is only valid for quantum mechanics but not for quantum field theory, since in the latter high temperature theory does not have a good classical limit due to the UV catastrophe, as already noted in \cite{Gur-Ari:2015rcq}.} which is free from fermions as the latter could alter or impact the dynamics of the bosonic matrices only at low temperatures. Analysis performed in  \cite{Gur-Ari:2015rcq} for the BFSS model by calculating the Lyapunov spectrum demonstrated a classical version of the fast scrambling holds with characteristic time being proportional to $\log N^2$. Another related important result obtained in \cite{Gur-Ari:2015rcq}, within this framework is the profile of the LLE w.r.t. the temperature, which is determined to be $\lambda_L = 0.2924(3)(\lambda_{'t Hooft} T)^{1/4}$ and it  is parametrically smaller than the Maldacena-Shenker-Stanford (MSS) bound, $\lambda_L \leq 2 \pi T$ \cite{Maldacena:2015waa}, on quantum chaos for sufficiently large $T$. The latter is a recent conjecture, which states that physical systems, which are holograpically dual to black holes are expected to be maximally chaotic, that is, saturating the MSS bound. This explicitly  demonstrated for the Sachdev-Ye-Kitaev (SYK) model \cite{Maldacena:2016hyu} and expected to be so for the BFSS model too. Classical description provided in \cite{Gur-Ari:2015rcq} violates this bound only at small temperatures, at $(\frac{0.292}{2 \pi})^\frac{4}{3} = 0.015$ to be precise,as may expected since the classical theory is good only in describing the dynamics in the high temperature regime. 

Authors of \cite{Asano:2015eha} considered different ansatz configurations in the BMN model at small matrix sizes of $N=2,3$ to probe chaos. In particular, they have focused on configurations of matrices which are essentially fuzzy two spheres at angular momentum $1/2$ and $1$, but with a collective dependence to distinct real functions of time such that the Gauss law constraint is readily satisfied. The effective models obtained in this manner exhibit chaos as demonstrated in \cite{Asano:2015eha}. From a general perspective, appearance of chaos in BFSS and BMN models is not all surprising, and in fact, consistent with the early investigations on chaotic motion in Yang-Mills theories in 80's \cite{Matinyan:1981dj,Savvidy:1982wx,Savvidy:1982jk} and the initial work performed on the BFSS model in \cite{Arefeva:1997oyf}. Let us also note that, BFSS as well as BMN models and their smaller subsectors at small values of $N$ are highly non-trivial many-body system whose complete solution evades us to date. Studies made so far provide some qualitative implications on the black hole phases in such models; for instance, in an $SU(2)$ YM matrix model with only two $2 \times 2$ matrices inspected in \cite{Berenstein:2016zgj}, the edge of the chaotic region is argued to correspond to the end of the black hole phase. In a recent paper \cite{Coskun:2018wmz} coauthored by one of the present authors, a Yang Mills five matrix models with a mass term is examined in detail and by considering equivariant fluctuations around its fuzzy four sphere \cite{Castelino:1997rv,Kimura:2002nq,Ydri:2016dmy} solutions an effective action is obtained. The chaotic dynamics in this model is revealed through detailed numerical analysis. Since non-trivial $D0$-brane dynamics leading to chaos require non-commutativity \cite{Aoki:2015uha}, fuzzy spheres which are described by matrices with certain non-vanishing commutation properties could serve as good candidates for background geometries to probe chaotic behaviour in YM matrix models. Dynamics of $D0-D2$- and $D0-D4$-brane systems with spherical $D2$- and $D4$-branes, i.e. on the fuzzy two- and four-spheres, are studied in the literature from various perspectives \cite{Kabat:1997im,Myers:1999ps,Brodie:2000yz, Castelino:1997rv,Fabinger:2002bk} in the past, while studies oriented toward exploring chaos in this context appears to be rather limited. 

We therefore think that it will be quite useful to further investigate the role played by fuzzy two- and four- sphere configurations in the development of chaotic dynamics in Yang Mills matrix models, in a setting which does not confine us to small matrix sizes but rather allows us to work at the large $N$ and small coupling regime, namely in the 't Hooft limit, with $\lambda_{'t Hooft} =g^2N$ held fixed, and aim at obtaining comprehensive data through which we may be able to comment further on the dynamics of $D0$-branes in the background of fuzzy spheres. In order to serve these purposes, we think that it may be useful to consider a matrix model in which mass deformation terms allow for both fuzzy two and four spheres. In particular, we take a Yang Mills matrix model with the same content as that of the BFSS, but with mass deformation terms breaking the global $SO(9)$ symmetry of the latter to $SO(5) \times SO(3) \times {\mathbb Z}_2$. Introducing a configuration involving fuzzy four and two spheres with collective time dependence as an ansatz, we study the chaotic dynamics in a family of effective Lagrangians which we obtain by tracing over these configurations at the matrix levels $N = \frac{1}{6}(n+1)(n+2)(n+3)$, for $n=1,2,\cdots\,,7$, corresponding to the sizes of fuzzy four spheres. Subsequently, we identify the fixed points of the phase space for the model at each matrix level and perform a linear stability analysis of these points. Identifying the unstable fixed points enables us to predict the energies about which the chaotic motion could start to develop. This expectation is corroborated through the numerical work we perform, in which we obtain the Lyapunov spectrum and analyze how the largest Lyapunov exponents change as a function of the energy. Using virial and equipartition theorems we derive inequalities that relate energy and temperature and show, for mass parameters $\mu_1^2=-16$ and $\mu_2^2 = -2$, the MSS bound is violated at low temperatures and at zero temperature, the latter being due to presence of negative unstable fixed point energies as we explain in detail in section 3. Examining best fitting curves to the LLE w.r.t. the $E/N^2 $ data with a particular power law function, we predict upper bounds on the temperature below which the MSS bound is violated and also obtain an upper bound on the LLE values of the form $\beta_n T^{1/4}$ in the high temperature regime, which are parametrically smaller than the MSS bound. Similar results are also determined for the models with mass-squared values $\mu_1^2=-8$ and $\mu_2^2 = 1$. For this case, there is only a single unstable fixed point in the phase space and it has zero energy at each level $n$, due to which there is no longer a violation of the MSS bound at zero temperature. We also reach the result that the critical temperatures in this case are closer to the value obtained in \cite{Gur-Ari:2015rcq} for the BFSS model. 

A natural question that arises is how to go beyond the classical descriptions given in \cite{Gur-Ari:2015rcq} for the pure BFSS model and also its massive deformations such as those studied in the present paper. Indeed, there are quite interesting very recent advances in accessing the quantum dynamics of the BFSS model in a real-time formulation \cite{Buividovich:2017kfk, Buividovich:2018scl}. In the conclusions and outlook section, we provide a detailed description of the Gaussian state approximation method introduced in these references, whose main point is to introduce a  general time-dependent Gaussian function, characterized by the wave packet centers and  dispersions, to truncate, via the use of Wick's theorem, the otherwise infinite set of equations that arise for the expectation values of polynomial operators with increasing degree in the canonical variables, starting from the expectation values of the Heisenberg equations of motion for the latter. Application of this approach to the bosonic BFSS model shows a decrease in the LLE values, and predicts vanishing LLE at a finite temperature, indicating that the LLE remains below the MSS bound at all temperatures. In section 5,  we also  sketch our ideas in some detail how the Gaussian state approximation method can be applied to the models treated in this paper to obtain the LLEs incorporating  the effects of quantum dynamics, which are expected to be lower in value compared to their classical counterparts. Last but not least, another point that deserves attention is the relation between these developments and the recent as well as somewhat earlier progress in the quantum mechanical treatment of the BFSS, BMN and related matrix models via both analytical perturbative and non-perturbative Monte Carlo methods in the Euclidean time formalism \cite{Kawahara:2006hs, Kawahara:2007fn, DelgadilloBlando:2007vx, DelgadilloBlando:2008vi, Asano:2018nol, Berkowitz:2018qhn}. The ties between Euclidean time formalism and the Gaussian state method in terms of their predictions for the observables are briefly discussed in sections 5 in the light of the results presented in \cite{Buividovich:2018scl} and \cite{Berkowitz:2018qhn}. We also suggest how future work can be oriented within this context to discover links between the temperature dependence of LLEs in real-time and the transitions between different phases of matrix models determined in the Euclidean time approach.

\section{Yang-Mills Matrix Models with Double Mass Deformation}

In this section, we introduce a Yang-Mills (YM) matrix models with two distinct mass deformation terms, which may be contemplated as a double mass deformation of the bosonic part of the BFSS model. The latter can be described by the action
\be
S_{BFSS} = \frac{1}{g^2} \int dt \, Tr \left ( \frac{1}{2}(D_t X_I)^2 + \frac{1}{4} [X_I, X_J ]^2 \right ) \,,
\label{BFSS_actn}
\ee 
where $X_I$ $(I,J=1,..,9)$ are $N \times N$ Hermitian matrices transforming under the adjoint representation of $U(N)$ as
\be
X_I \rightarrow U^{\dagger} X_I U \,, \quad U \in U(N) \,.
\ee
The covariant derivatives are given by
\be
D_t {X}_I=\del_t X_I - i \lbrack A_t, X_I \rbrack \,,
\ee
where $A_t$ is the $U(N)$ gauge field transforming as
\be
A_t \rightarrow U^{\dagger} A_t U + i U^{\dagger} \del_t U. 
\ee
Action (\ref{BFSS_actn}) is invariant under the $U(N)$ gauge transformations as well as the rigid global $SO(9)$ rotations
\be
X_I \rightarrow X_I^\prime = R_{IJ} X_J \,, \quad R \in SO(9) \,,
\ee
of the matrices $X_I$ among themselves. 
Fixing the gauge to $A_t=0$ yields the equations of motion for the $X_I$'s as
\be  
\ddot{X_I} + \lbrack X_J , \lbrack X_J , X_I \rbrack \rbrack = 0 \,,
\label{BFSSeom}
\ee
which are supplemented by the Gauss law constraint
\be
\lbrack X_I, \del_t X_I \rbrack = 0 \,.
\label{GaussLaw1}
\ee
(\ref{GaussLaw1}) is an immediate consequence of the algebraic equations of motion for the gauge field $A_t$ and amounts to having physical states as $SU(N)$-singlets in the quantized theory.

Let us note that we are going to be essentially working in the 't Hooft limit in which the 't Hooft coupling $\lambda_{' t Hooft} = g^2 N$ is kept fixed, while $N$ is taken large and $g^2$ is taken small. $\lambda_{' t Hooft}$ has units of $\mbox{(Length)}^{-3}$ and without loss of generality it can be set to unity by scaling all the dimensionful quantities in the action in units of $\lambda^{1/3}$. 

A gauge invariant double mass deformation of (\ref{BFSS_actn}) may be specified as 
\be
S_{YMM} = N \int dt \, L_{YMM} := N \int dt \, Tr \left ( \frac{1}{2}(D_t X_I)^2 + \frac{1}{4} [X_I, X_J]^2 - \frac{1}{2} \mu_1^2 X_a^2 - \frac{1}{2} \mu_2^2 X_i^2 \right ) \,, 
\label{YMDM}
\ee
where the indices $a$ and $i$ take on the values $a = 1,..,5$ and $i = 6,7,8$, respectively. In (\ref{YMDM}) the terms proportional to $\mu_1^2$ and $\mu_2^2$ are the quadratic deformations, which respect the $U(N)$ gauge symmetry, but altogether break the $SO(9)$ down to $SO(5) \times SO(3) \times {\mathbb Z_2}$. The discrete ${\mathbb Z}_2$ factor is due to the $X_9 \rightarrow - X_9$ symmetry and would be still present even if we had a separate mass term for this field too. Since, in what follows, we are going to consider the sector in which $X_9$ is set equal to the zero matrix, or equally well, taken as the zero matrix as a part of the ansatz configuration which will be introduced shortly, this ${\mathbb Z}_2$ factor will be immaterial. Equation (\ref{YMDM}) written already in the 't Hooft limit with  $\lambda_{' t Hooft}$ set to unity. We can restore $\lambda_{' t Hooft}$ by scaling  $X_i \rightarrow \lambda^{-1/3} X_i$, $A_t\rightarrow \lambda^{-1/3} A_t$, $t \rightarrow \lambda^{1/3} t$, $\mu_i^2 \rightarrow \lambda^{-2/3} \mu_i^2$, if needed.

In the $A_t=0$ gauge, the equations of motion for $X_I$ take the form
\begin{subequations}
\begin{align}
\ddot{X_a} + \lbrack X_I , \lbrack X_I , X_a \rbrack \rbrack  + \mu_1^2 X_a &= 0 \,, \label{YMDMeomA}\\ 
\ddot{X_i} + \lbrack X_I , \lbrack X_I , X_i \rbrack \rbrack +  \mu_2^2 X_i &= 0 \,, \label{YMDMeomB} \\
\ddot{X_9} + \lbrack X_I , \lbrack X_I , X_9 \rbrack \rbrack &= 0 \label{YMDMeomC}\,,
\end{align}
\label{YMDMeom} 
\end{subequations}
while the Gauss law constraint remains unchanged in the form as given in (\ref{GaussLaw1}).

The massive deformation of the BFSS model, which preserves maximal amount of supersymmetry is already known to be the BMN matrix model \cite{Berenstein:2002jq}, which possesses fuzzy two-spheres and their direct sums as possible vacuum configurations. However, in the present work, our focus is directed toward exploring the emerging chaotic dynamics from the Yang-Mills matrix models which could allow for not only fuzzy two-sphere configurations but also higher dimensional fuzzy spheres, and in particular a fuzzy four-sphere. It is possible to conceive deformations of $S_{BFSS}$ including two separate mass terms, which break the $SO(9)$ symmetry down to several different product subgroups. The underlying motivation for introducing the specific massive deformation in (\ref{YMDM}) comes from the fact that in two distinct limiting cases, the equations of motion can be solved either with fuzzy two-sphere or fuzzy four-sphere configurations. To be more precise, we have for $X_i = 0 = X_9$, (\ref{YMDMeomB}) and (\ref{YMDMeomC}) are satisfied identically, while (\ref{YMDMeomA}) takes the form 
\be
\ddot{X_a} + \lbrack X_b , \lbrack X_b , X_a \rbrack \rbrack  + \mu_1^2 X_a = 0 \,,
\label{foursphereeom} 
\ee
which is satisfied by the fuzzy four sphere configurations $X_a \equiv Y_a$ for $\mu_1^2=-16$. Here $Y_a$ are $N \times N$ matrices carrying the $(0,n)$ UIR of $SO(5)$ with $N = \frac{1}{6}(n+1)(n+2)(n+3)$ and satisfying the defining properties given in (\ref{fuzzyS4}). Whereas, in the other extreme, one may set $X_a = 0 = X_9$, with the only remaining non-trivial equation of motion
\be
\ddot{X_i} + \lbrack X_j , \lbrack X_j , X_i \rbrack \rbrack  + \mu_2^2 X_i = 0 \,,
\label{twosphereeom}
\ee
which is solved by fuzzy two sphere configurations $X_i \equiv Z_i$ or their direct sum for $\mu_2^2 = - 2$. In this case, $X_i$ are $N \times N$ matrices carrying the spin $j = \frac{N-1}{2}$ UIR of $SO(3) \approx SU(2)$.  

In view of these observations we consider as ansätze configurations involving fuzzy two- and four- spheres with collective time dependence, which fulfill the Gauss law constraint given in (\ref{GaussLaw1}). Tracing over the fuzzy two- and four-sphere configurations, we aim to obtain reduced models with only four phase space degrees of freedom, whose dynamics can be investigated in considerable detail. 

\subsection{Ansatz {\it I} and the Effective Action}

A reasonably simple, yet non-trivial configuration is constructed by introducing two separate collective time-dependent functions multiplying the fuzzy four- and two-sphere matrices. Concretely, we have
\be  
X_a =  r(t) \, Y_a \,, \quad X_i =  y(t) \, Z_i \,, \quad X_9 = 0 \,,
\label{Anstz1}    
\ee 
where $r(t)$ and $y(t)$ are real functions of time. In this ansatz, we consider a single spin-$j = \frac{N-1}{2}$ IRR of $SU(2)$ as the fuzzy $S^2$ configuration, while taking direct sums of fuzzy two spheres with different IRRs of $SU(2)$, i.e. forming $Z_i$ as a block-diagonal matrix composed of a direct sum of different IRRs of $SU(2)$ with spin less than $j$ remains an open possibility. We will briefly discuss $Z_i$'s made up of spin $\frac{1}{2}$ fuzzy spheres in section $4$. $Z_i$ exist at every matrix level, while this is not so for $Y_a$. Fuzzy four spheres exist at the matrix levels $4,10,20 \cdots$ as given by the dimension $N = \frac{1}{6}(n+1)(n+2)(n+3)$ of the IRR $(0,n)$ of $SO(5)$. Accordingly the fuzzy two spheres are taken at the matrix levels matching these dimensions of the fuzzy four sphere. In what follows, we initially keep the mass parameters $\mu_1^2$ and $\mu_2^2$ as unspecified in some of the key equations, but will investigate the detailed dynamics for $\mu_1^2=-16$ and $\mu_2^2=-2$, which emerge as the limiting values for the static solutions of (\ref{foursphereeom}) and (\ref{twosphereeom}) and subsequently will also  briefly investigate the consequences of taking another set of values for the masses, namely $\mu_1^2=-8$ and $\mu_2^2=1$ on the chaotic dynamics of the reduced models.

Substituting the (\ref{Anstz1}) configuration in the action (\ref{YMDM}), we perform the trace over the fuzzy four- and two-sphere matrices at each of the matrix levels $N = \frac{1}{6}(n+1)(n+2)(n+3)$ for $n=1,2,\cdots\,, 6$. Using Matlab to evaluate the traces, we obtain the Lagrangian of the reduced models in the form  
\be
L_n = N^2( c_{1} \dot{r}^2 +  c_{2}  \dot{y}^2  - 8  c_{1}  r^4 - c_{2}  y^4 -  c_{1}  \mu_1^2 r^2 -  c_{2}  \mu_2^2 y^2 -  c_{3} r^2y^2) \,,
\label{Lalp}
\ee
where the coefficients $c_{\beta} = c_{\beta}(n)$ ($\beta=1,2,3$) depend on $n$ and their values (given up to one digit after the decimal point at most) for $n=1,2,\cdots\,, 7$ are listed in the table \ref{table:table1} given below. We suppress the label $n$ of the coefficients $c_{\beta}(n)$ in (\ref{Lalp}) in order not to clutter the notation. Coefficients given for $n=7$ in table \ref{table:table1} are evaluated by obtaining a polynomial function of $n$ approximating\footnote{These polynomial functions are given as
\beqa
c_1(n) &=& \frac{1}{2}n(n+4) \nn \,, \\ 
c_2(n) &=& \frac{1}{288}(n^6 + 12 n^5 + 58 n^4 + 144 n^3 + 193 n^2 + 132 n) \,, \nn \\
c_3(n) &=& 0.0093 n^7 + 0.20 n^6 + 1.35 n^5 + 4.22 n^4 + 6.99 n^3 + 5.44 n^2 + 2.43 n + 0.36 \,. \nn
\eeqa	
}$c_{\beta}(n)$ for $n=1,2,\cdots\,,6$ and interpolating this result to $n=7$.
\begin{table}
\centering
\begin{tabular}{ | c | c | c | c | c | c | c | c |}
\cline{2-8}
		\multicolumn{1}{c |}{} & $n=1$ & $n=2$ & $n=3$ & $n=4$ & $n=5$ & $n=6$ & $n=7$ \\ \hline 
		$c_{1}$ &$2.5$  & $6$ & $10.5$  & $16$ &$22.5$  &$30$ &$38.5$      \\ \hline 
		$c_{2}$ &$1.9$  & $12.4$  &$49.9$  &$153$ &$391.9$ &$881.9$ &$1800$    \\ \hline 
		$c_{3}$ &$21$ & $207.7$ &$1080$ &$3970$ &$11691$ &$29493$ &$66345$ \\ \hline
\end{tabular}
\caption{Numerical values of coefficients $c_\beta(n)$ for Ans\"{a}tz {\it I}.}
\label{table:table1}
\end{table}

The corresponding Hamiltonian is easily obtained from (\ref{Lalp}), and it is given by
\beqa
H_{n}(r,y,p_r,p_y) &=& \dfrac{{p_r}^2}{4c_{1}N^2} + \dfrac{{p_y}^2}{4c_{2}N^2} + N^2(8  c_{1} \, r^4 + c_{2} \, y^4 +  c_{1} \, \mu_1^2 r^2 +  c_{2} \, \mu_2^2 y^2 + c_{3} \, r^2y^2) \,, \nn \\
&= :& \dfrac{{p_r}^2}{4c_{1}N^2} + \dfrac{{p_y}^2}{4c_{2}N^2}  + N^2 V_n(r,y) \,,
\label{Hn1}
\eeqa
where $V_n(r,y)$ denotes the potential function.

To explore the dynamics of the models governed by $H_n$, we first evaluate the Hamilton's equations of motion. These take the form 
\begin{subequations}
	\begin{align}
	\dot{r} - \dfrac{p_r}{2 c_{1}N^2} = 0	\,, \quad \quad & \dot{y} - \dfrac{p_y}{2 c_{2}N^2} = 0 \,, \label{Heom1} \\
	\dot{p_r} + N^2( 32  c_{1} r^3 + 2  c_{1} \mu_1^2 r + 2  c_{3} r  y^2)  = 0 \,, \quad \quad  & \dot{p_y} + N^2(4 c_{2} y^3 + 2 c_{2} \mu_2^2 y + 2 c_{3} r^2 y)  = 0 	\,. \label{Heom2}
	\end{align}
	\label{Heom} 
\end{subequations}
Taking the mass parameter values as $\mu_1^{2}=-16$ \& $\mu_2^{2}=-2$, (\ref{Heom}) become
\begin{subequations}
	\begin{align}
		\dot{r} - \dfrac{p_r}{2 c_{1} N^2} = 0	\,, \quad \quad & \dot{y} - \dfrac{p_y}{2 c_{2}N^2} = 0 \,, \label{Heom3} \\
		\dot{p_r} + N^2(32  c_{1} r^3 - 32 c_{1} r + 2  c_{3} r  y^2)  = 0 \,, \quad \quad  & \dot{p_y} + N^2( 4 c_{2} y^3 - 4 c_{2}  y + 2 c_{3} r^2 y)  = 0 	\,. \label{Heom4}
	\end{align}
	\label{HeomV} 
\end{subequations}
\noindent In order to investigate the dynamics of these models governed by (\ref{HeomV}) in detail, it is quite useful to start the analysis by determining the fixed points corresponding to the equations of motion in (\ref{HeomV}) and addressing their stability at the linear order. Fixed points of a Hamiltonian system are defined as the stationary points of the phase space \cite{Percival,Ott,Hilborn,Campbell}, and can thefore be determined by the set of equations specified as
\be
(\dot{r}, \dot{y}, \dot{p_r}, \dot{p_y}) \equiv (0,0,0,0) \,.
\label{fpcond} 
\ee
Using (\ref{fpcond}) in (\ref{HeomV}) leads to four algebraic equations, two of which are trivially solved by $(p_r, p_y) \equiv (0,0)$, which means that all the fixed points are confined to the $(p_r, p_y) \equiv (0,0)$ plane in the phase space. The remaining two equations are
\beqa
- 32  c_{1} r^3 +32  c_{1}  r - 2  c_{3} r y^2 & = & 0 \,, \nn \\
- 4 c_{2} y^3 +4 c_{2} y - 2 c_{3} r^2 y & = & 0 \,,
\label{fpeqns1}
\eeqa
and have the general set of solutions given as
\be
(r,y) \equiv \lbrace (0, 0) \,, (\pm 1, 0) \,, (0, \pm 1) \,, (\pm h_1, \pm h_2) \,, (\pm h_1, \mp h_2) \rbrace \,,
\label{cpoints}
\ee
where $h_1$ and $h_2$ are given in terms of $c_\beta$ as
\be
h_1=- \sqrt{2} i \frac{\sqrt{-c_2c_3+ 16 c_1c_2}}{\sqrt{c_3^2-32c_1c_2}} \,, \quad
h_2= - 4 i \frac{\sqrt{2c_1c_2-c_1c_3}}{\sqrt{c_3^2-32c_1c_2}} \,.
\ee
Clearly, only real solutions of (\ref{fpeqns1}) are physically acceptable. From table \ref{table:table1}, it is straightforward to compute that both $h_1$ and $h_2$ are real except at $n=1$. For $n>1$ the set of fixed points are given as 
\begin{multline}
(r,y,p_r,p_y) \equiv \lbrace (0,0,0,0) , (  \pm1,0,0,0) \,, ( 0, \pm 1,0,0) ) \,,\\
(\pm h_1(n), \pm h_2(n),0,0) \,, (\pm h_1(n), \mp h_2(n),0,0)   \rbrace \,, 
\label{real2}
\end{multline}
where the values of $h_1$ and $h_2$ are presented in the table \ref{table:table2} below
\begin{table}
	\centering
	\begin{tabular}{ | c | c | c | c | c | c |c|}
		\cline{2-7}
		\multicolumn{1}{c |}{} & $n=2$ & $n=3$ & $n=4$ & $n=5$&$n=6$ &$n=7$ \\ \hline 
		$h_1$  & $0.26$ & $0.28$  & $0.27$ &$0.26$ &$0.24$ & $0.23$     \\ \hline 
		$h_2$ & $0.6$  &$0.38$  &$0.25$ &$0.17$ &$0.12$ & $0.093$  \\ \hline 
	\end{tabular}
	\caption{Numerical values of $h_1$ and $h_2$.}
	\label{table:table2}
\end{table}
\noindent while for the $n=1$ model we only have
\be
(r, y, p_r, p_y) \equiv \lbrace (0,0,0,0) \,, (\pm 1,0,0,0) \,, (0, \pm 1,0,0) \rbrace \,, 
\label{real1}
\ee
as the fixed points.

Let us note that (\ref{cpoints}) corresponds to the critical points of the potential $V_n$, since (\ref{fpeqns1}) are the equations determining the extrema of the latter. From the eigenvalues of the matrix $\frac{\partial^2 V_n}{\partial g_i \partial g_j}$, (with the help of the notation $(g_1, g_2) \equiv (r,y)$), we see that the points $(\pm 1,0)$ and $(0 \,,\pm 1)$ are local minima, $(0,0)$ is a local maximum, while $(\pm h_1(n), \pm h_2(n))$, $(\pm h_1(n), \mp h_2(n))$ are all saddle points of $V_n$. Evaluating $V_n$ at the local minima, we find $V_n(\pm 1,0) = -8 c_1$ and $V_n(0, \pm 1) = - c_2$. Using table \ref{table:table1}, we easily conclude that $(\pm 1,0)$ is the absolute minimum of $V_n$ for $n=1,2,3$, while $(0, \pm 1)$ is the absolute minima for the models with $n =4,5,6$. 

Fixed point energies are readily evaluated using (\ref{real2}), (\ref{real1}) and the mass squared values $\mu_1^{2} = -16$ and $\mu_2^{2} = -2$ and they are  
\be
E_F(0, 0, 0, 0)= 0 \,,\quad E_F(\pm 1,0,0,0) = -8 N^2 c_1 \,, \quad E_F(0,\pm 1,0,0) = - N^2 c_2 \,, 
\label{fpenergies1}
\ee
while the values of $\frac{E_F(\pm h_1(n), \pm (\mp) h_2(n), 0, 0)}{N^2}$ and minimum values of the potentials $V_n$ are listed in table 3 for quick reference.
\begin{table}[!htbp]
	\centering
	\begin{tabular}{|c| c | c | c | c | c | c |c|}
		\cline{2-8}
		\multicolumn{1}{c|}{} &$n=1$ & $n=2$ & $n=3$ & $n=4$ & $n=5$&$n=6$ &$n=7$ \\ \hline 
		$\frac{E_F(\pm h_1(n), \pm (\mp) h_2(n), 0, 0)}{N^2}$ & $-$ & $-8.6$ & $-13.8$ & $-18.4$  &$-23$ &$-27.7$  &$-32.3$ \\ \hline
		$\frac{V_{(n), min}}{N^2}$ & $-20$ & $-48$ & $-84$ & $-153$  &$-391.9$ &$-881.9$  &$-1800$ \\ \hline
	\end{tabular}
	\caption{Numerical values for $E_F/N^2$ at the critical points $(\pm h_1(n), \pm(\mp)h_2(n), 0, 0)$ and the minimum values of $V_n$.}
	\label{table:table3}
\end{table}
Note that minimum values of $V_n$ are negative in general. This is expected, due to the presence of the massive deformation terms in the matrix model. Indeed it is already known that, fuzzy sphere vacuum solutions of the BFSS model with qubic and/or quadratic deformation terms are of lower (negative) energy, compared to the zero-energy vacuum solutions of the pure BFSS model with diagonal matrices.

\subsection{Linear Stability Analysis in the Phase Space}

It must be clear that the properties of extrema of $V_n$ does not provide sufficient information to decide on the stability of the fixed points. We now perform a first order stability analysis around the fixed points of $H_n$ given in (\ref{real2}) and (\ref{real1}). Together with the Lyapunov spectrum and the Poincare sections that will be determined in the next section, this analysis will allow us to comment on the outset and variation of chaos, that is, the increase and decrease in the amount of chaotic orbits in the phase spaces of $H_n$, w.r.t. energy.

For the phase space coordinates it is useful to introduce the notation
\be
(g_1, g_2, g_3, g_4) \equiv (r, y, p_r, p_y) \,,
\label{psc}
\ee
From $g_\alpha$ and $\dot{g}_\alpha$, we may form the Jacobian matrix 
\be
J_{\alpha \beta} \equiv \frac{\partial \dot{g}_\alpha}{\partial g_\beta} \,,
\ee
whose eigenvalue structure allows us to decide on the stability character of the fixed points \cite{Percival,Ott,Hilborn,Campbell}. Written in explicit form, we have
\be
J(r,y) \equiv \begin{pmatrix}
	0 & 0 & \frac{1}{2 c_{1} N^2}   & 0  \\ 
	0 & 0 & 0 & \frac{1}{2 c_{2} N^2}   \\
	J_{31} & -2 N^2 c_{3} r y & 0 & 0 \\ 
	-2 N^2 c_{3} r y & J_{42} & 0 & 0
\end{pmatrix} \,,
\ee
where $J_{31} $ and $J_{42}$ are
\beqa
J_{31} &=& N^2(32c_{1} - 96 c_{1} r^2 -2c_{3}y^2) \,, \nn \\
J_{42} &=& N^2(4 c_{2}- 12c_{2} y^2 - 2c_{3} r^2) \,.  
\eeqa
Eigenvalues of $J(r,y)$ at the fixed points (\ref{real2}) are easily evaluated and listed in the table \ref{table:tablefp} below.
\begin{table}
	\centering	
	\begin{tabular}{| c | c |}
	\cline{1-2}
	Fixed Points & Eigenvalues of $J(r,y)$ \\ \hline 
		$(0,0,0,0)$ & ${\pm 4,\pm \sqrt{2}}$ \\ \hline
		$(\pm 1,0,0,0)$ & $(\pm4i \sqrt{2},\pm i \frac{\sqrt{c_3 c_2 c_1^2-2 c_2^2 c_1^2}}{c_1 c_2})$ \\ \hline
		$(0,\pm1,0,0)$ & $(\pm 2i,\pm i \frac{\sqrt{c_3 c_1 c_2^2 -16 c_1^2 c_2^2}}{c_1c_2})$ \\ \hline
		$(h_1(2),h_2(2),0,0)$ & $(\pm 2.5, \pm i 3.2)$ \\ \hline
		$(h_1(3),h_2(3),0,0)$ & $(\pm 2.9, \pm i 3.4)$ \\ \hline
		$(h_1(4),h_2(4),0,0)$ & $(\pm 3.1, \pm i 3.5)$ \\ \hline
		$(h_1(5),h_2(5),0,0)$ & $(\pm 3.1, \pm i 3.5)$ \\ \hline
		$(h_1(6),h_2(6),0,0)$ & $(\pm 3.2, \pm i 3.5)$ \\ \hline
		$(h_1(7),h_2(7),0,0)$ & $(\pm 3.2, \pm i 3.4)$ \\ \hline
		\end{tabular}
	\caption{Eigenvalues of the Jacobian $J(r, y)$ at thefixed points.}
	\label{table:tablefp}
\end{table}

General criterion of the linear stability analysis \cite{Percival,Ott,Hilborn,Campbell} states that a fixed point is stable if all the real eigenvalues of the Jacobian are negative, and unstable if the Jacobian has at least one real positive eigenvalue. It may be that all the eigenvalues of the Jacobian matrix are purely imaginary at a fixed point. This is called the borderline case and an analysis beyond first order is necessary to decide if the system is stable or unstable at such a point. Accordingly, we see that $(0,0,0,0)$ and $(h_1(n),h_2(n),0,0)$ are all unstable fixed points as the corresponding Jacobians have at least one real positive eigenvalue, as readily seen from the table \ref{table:tablefp}. From the same table and the values of $c_\alpha(n)$ in table \ref{table:table1}, we see that $(\pm1,0,0,0)$ and $(0,\pm 1, 0, 0)$ are borderline cases. We are not going to explore the structure of these borderline fixed points any further, as we expect that their impact on the chaotic dynamics should be rather small compared to those of the unstable fixed points, which we just identified at the linear level. Our numerical results on the Lyapunov spectrum indeed corroborates with this expectation as will be discussed shortly.

\section{Chaotic Dynamics}

\subsection{Lyapunov Spectrum and Poincar\'{e} Sections}

In order to probe the presence and analyze the structure of chaotic dynamics of the models described by the Hamiltonians $H_n$, we will first examine their Lyapunov spectrum. As it is well known, Lyapunov exponents measure the exponential growth in perturbations and therefore give a reliable and decisive means to establish the presence of chaos in a dynamical system \cite{Ott,Hilborn,Campbell}. Suppose that we denote the perturbations in phase space coordinates ${\bm g}(t) \equiv (g_1(t), g_2(t)\,,\cdots\,, g_{2n}(t))$ by $\delta {\bm g}(t)$. The system is chaotic if, at large $t$, $\delta {\bm g}(t)$ deviates exponentially from its initial value at $t= t_0$: $|\delta {\bm g}(t)| = e^{\lambda (t-t_0)} |\delta {\bm g}(t_0)|$, and $\lambda > 0$ are called the positive Lyapunov exponents. There are $2n$ of them for a phase space of dimension $2n$. This description is essentially equivalent to the statement that even slightly different initial conditions give trajectories in the phase space, which are exponentially diverging from each other and hence lead to chaos. Thus, in a dynamical system presence of at least one positive Lyapunov exponent is sufficient to conclude the presence of chaotic motion. In Hamiltonian systems, due to the symplectic structure of the phase space, Lyapunov exponents appear in $\lambda_i$ and $-\lambda_i$ pairs, a pair of the Lyapunov exponents vanishes as there is no exponential growth in perturbations along the direction of the trajectory specified by the initial condition and sum of all the Lyapunov exponents is zero as a consequence of Liouville's theorem. These facts are well-known and their details may be found in many of the excellent books on chaos \cite{Ott,Hilborn,Campbell}.  

\begin{figure}[!htb]
	\centering
	\begin{subfigure}[!htb]{.32\textwidth}
		\centering
		\includegraphics[width= 1\linewidth]{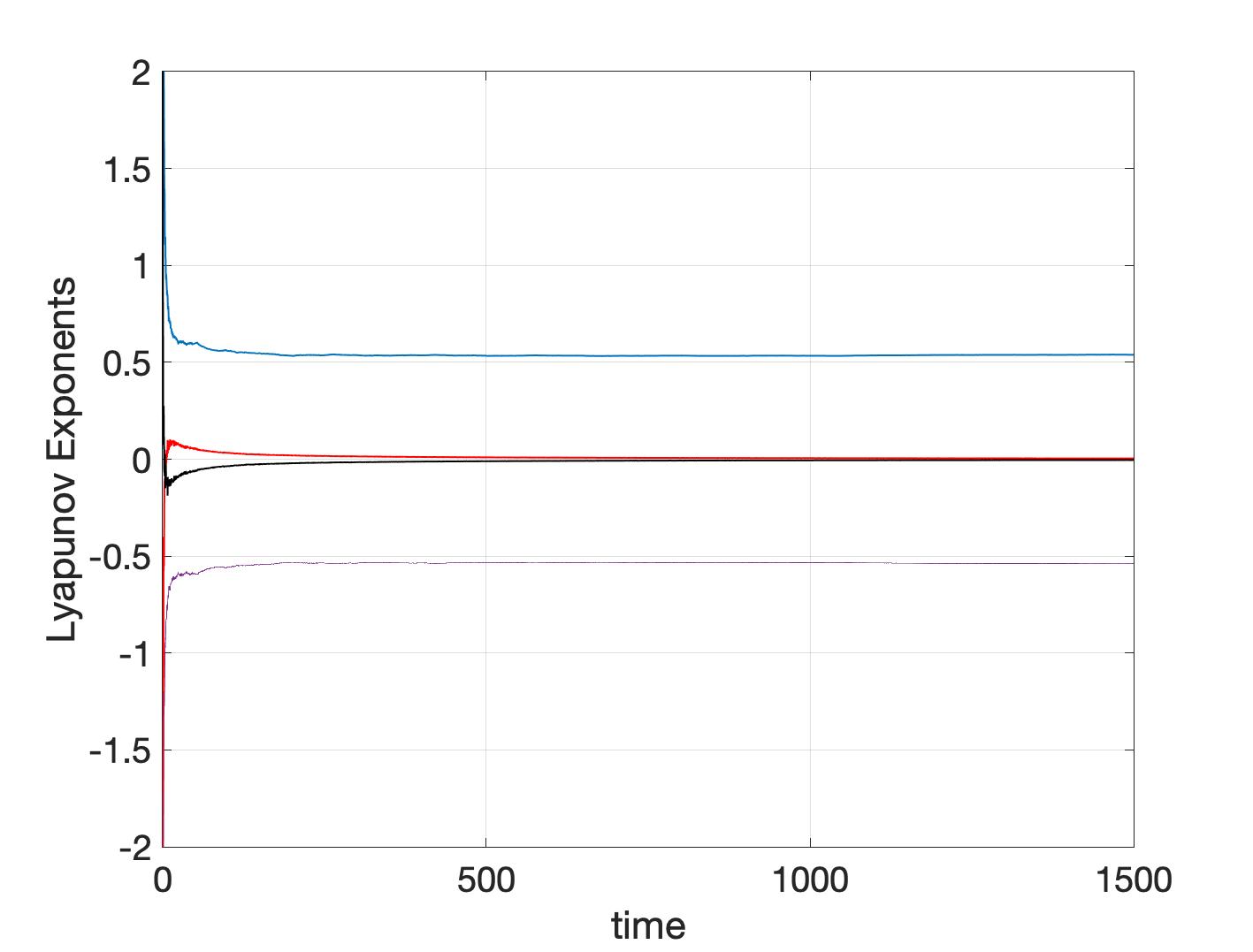}  
		\caption{$n=1$ and $E^\prime/N^2=30$}
		\label{fig:fig1a}
	\end{subfigure}	
	\begin{subfigure}[!htb]{.32\textwidth}
		\centering
		\includegraphics[width=1\linewidth]{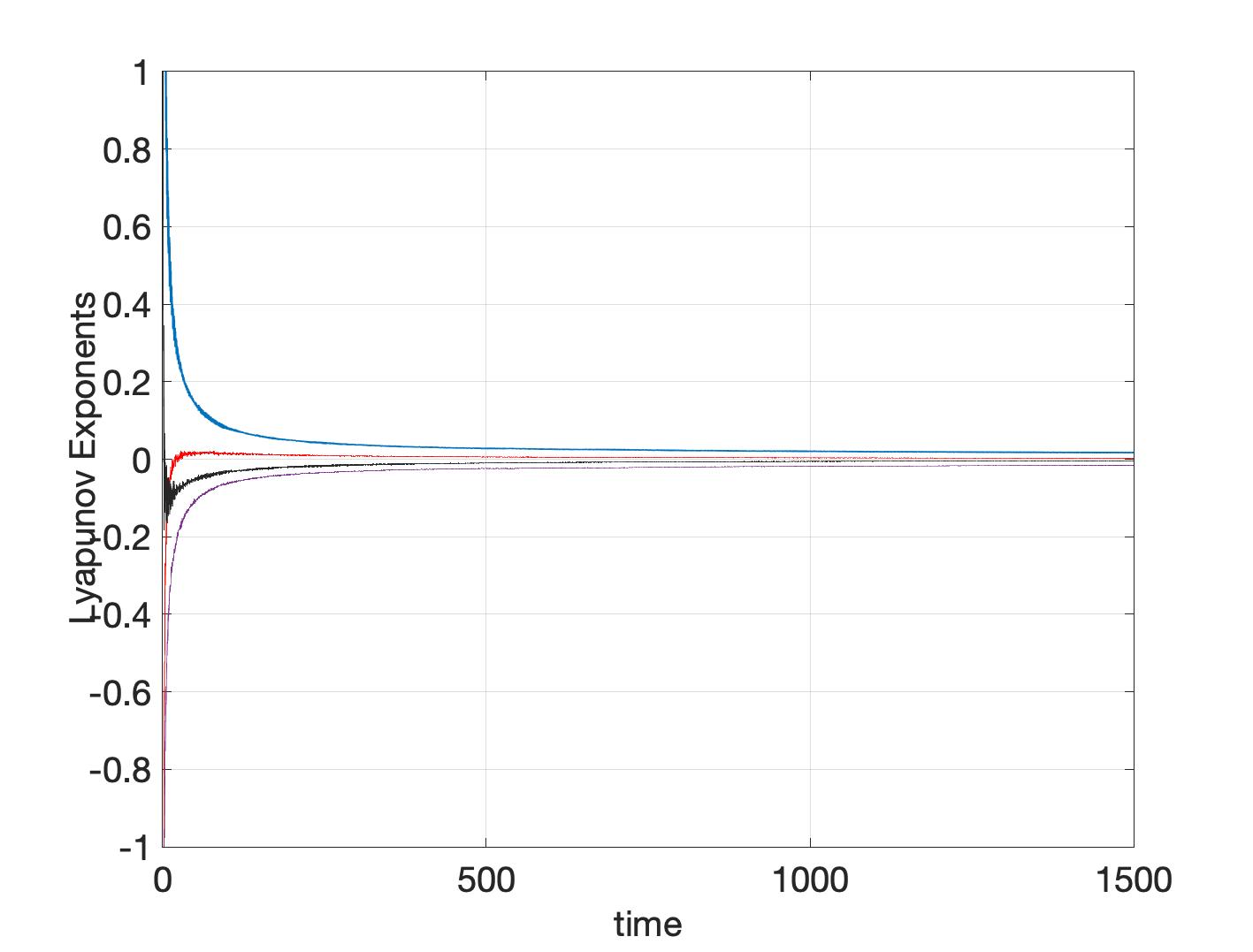}  
		\caption{$n=1$ and $E^\prime/N^2=500$}
		\label{fig:fig1b}
	\end{subfigure}	
	\begin{subfigure}{.32\textwidth}
		\centering
		\includegraphics[width= 1\linewidth]{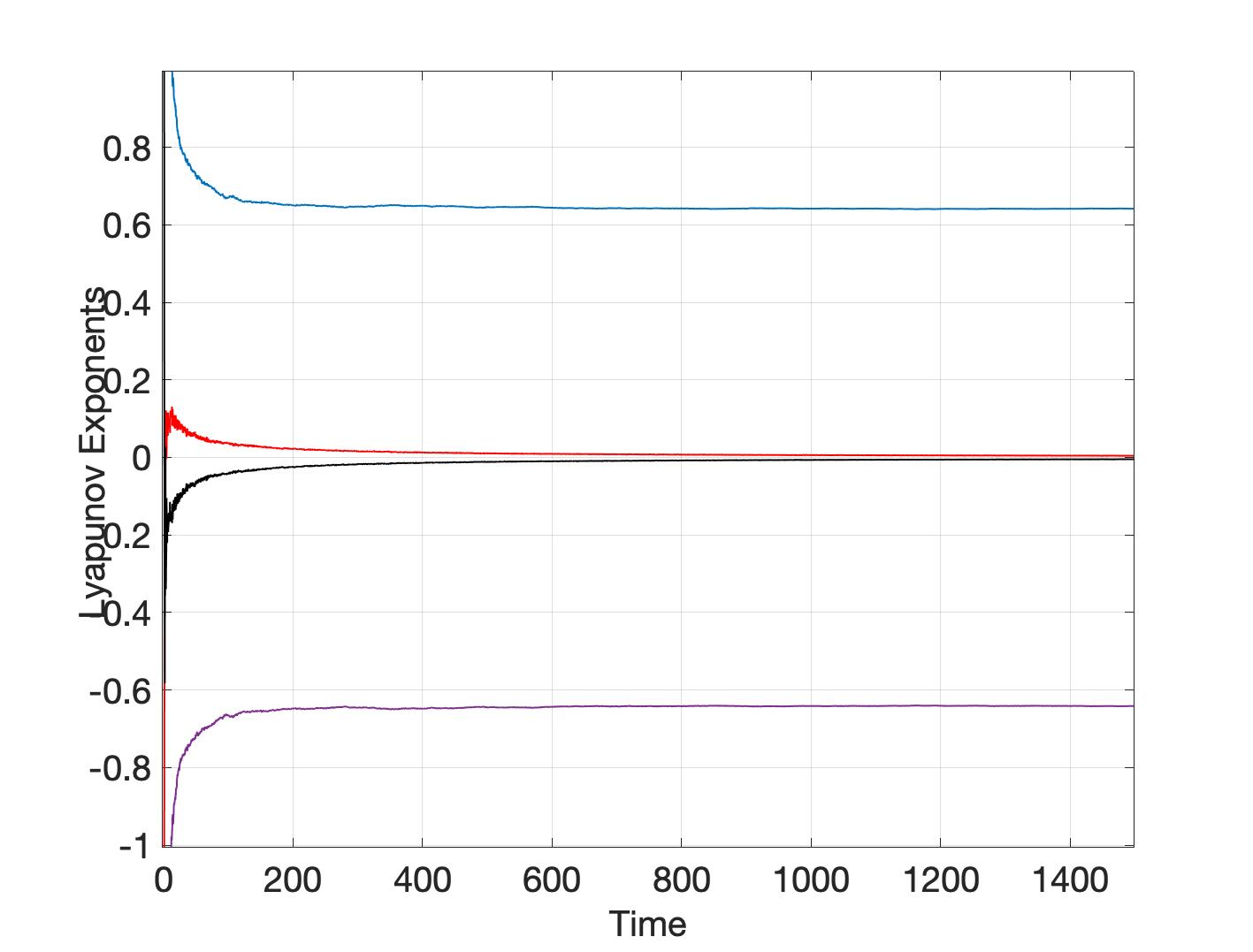}  
		\caption{$n=2$ and $E^\prime/N^2=50$}
		\label{fig:fig1c}
	\end{subfigure}	
	\begin{subfigure}{.32\textwidth}
		\centering
		\includegraphics[width=1\linewidth]{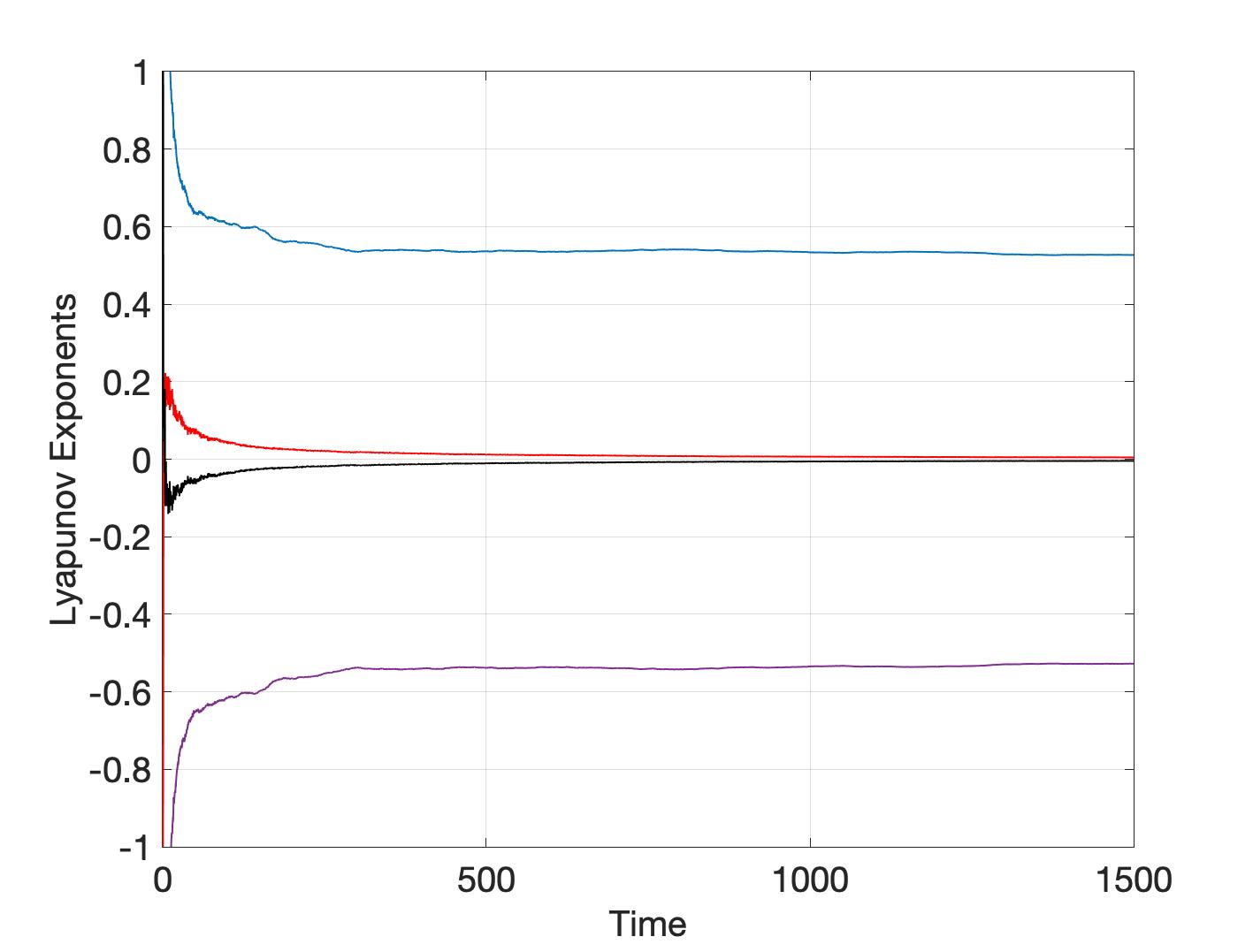}  
		\caption{$n=3$ and $E^\prime/N^2=100$}
		\label{fig:fig1d}
	\end{subfigure}	
	\begin{subfigure}{.32\textwidth}
		\centering
		\includegraphics[width= 1\linewidth]{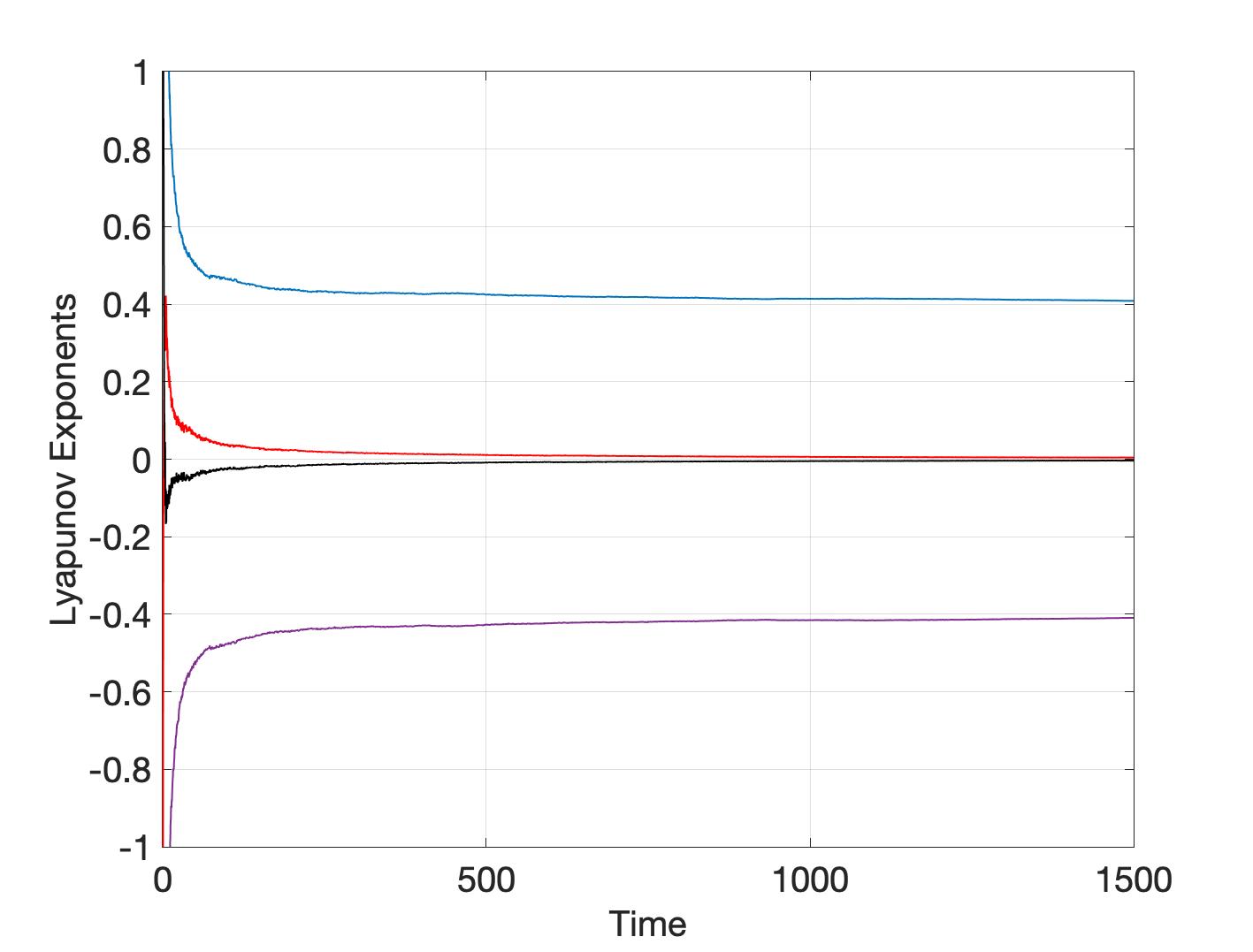}  
		\caption{$n=4$ and $E^\prime/N^2=150$}
		\label{fig:fig1e}
	\end{subfigure}	
	\begin{subfigure}{.32\textwidth}
		\centering
		\includegraphics[width=1\linewidth]{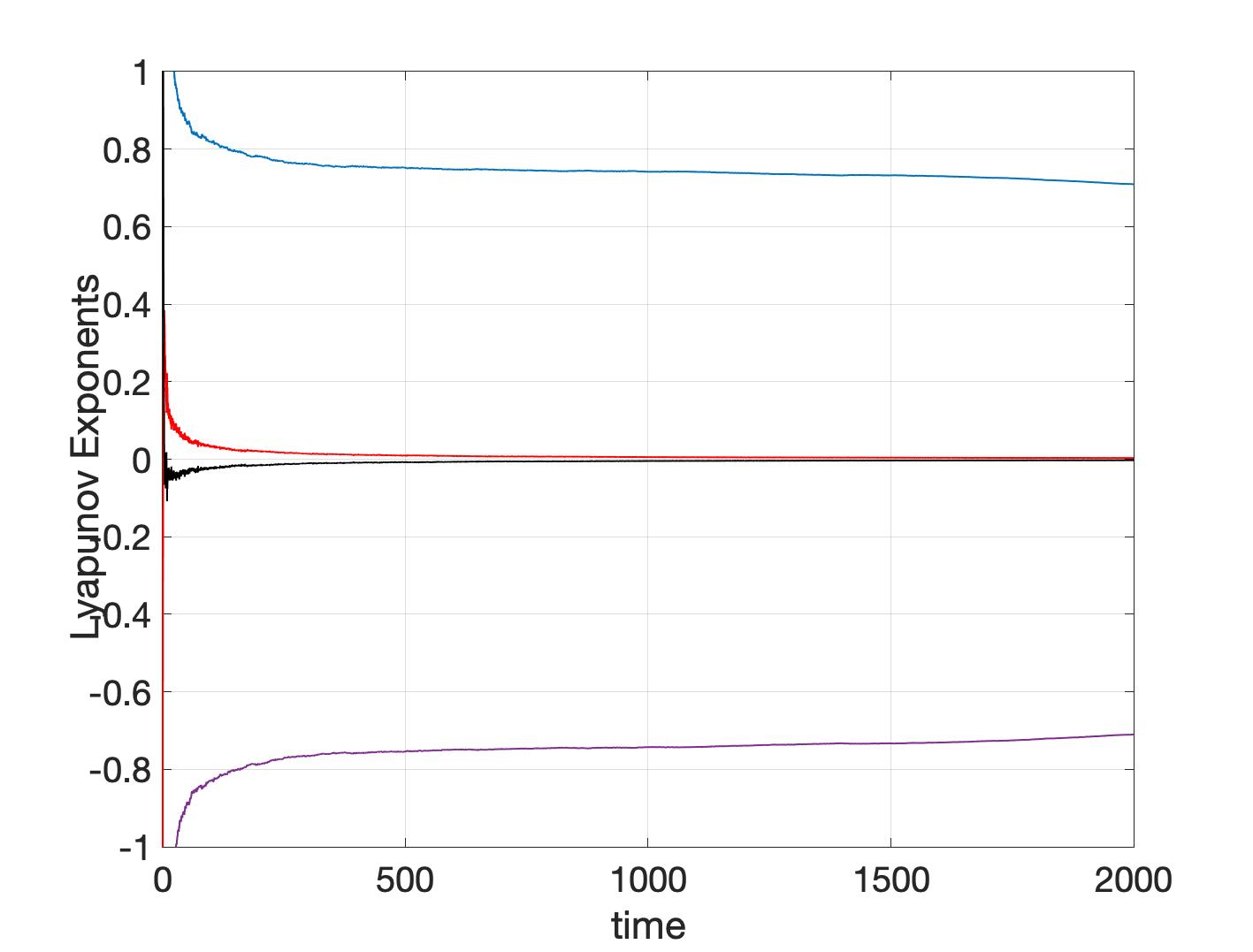}  
		\caption{$n=5$ and $E^\prime/N^2=500$}
		\label{fig:fig1f}
	\end{subfigure}		
	\begin{subfigure}{.32\textwidth}
		\centering
		\includegraphics[width= 1\linewidth]{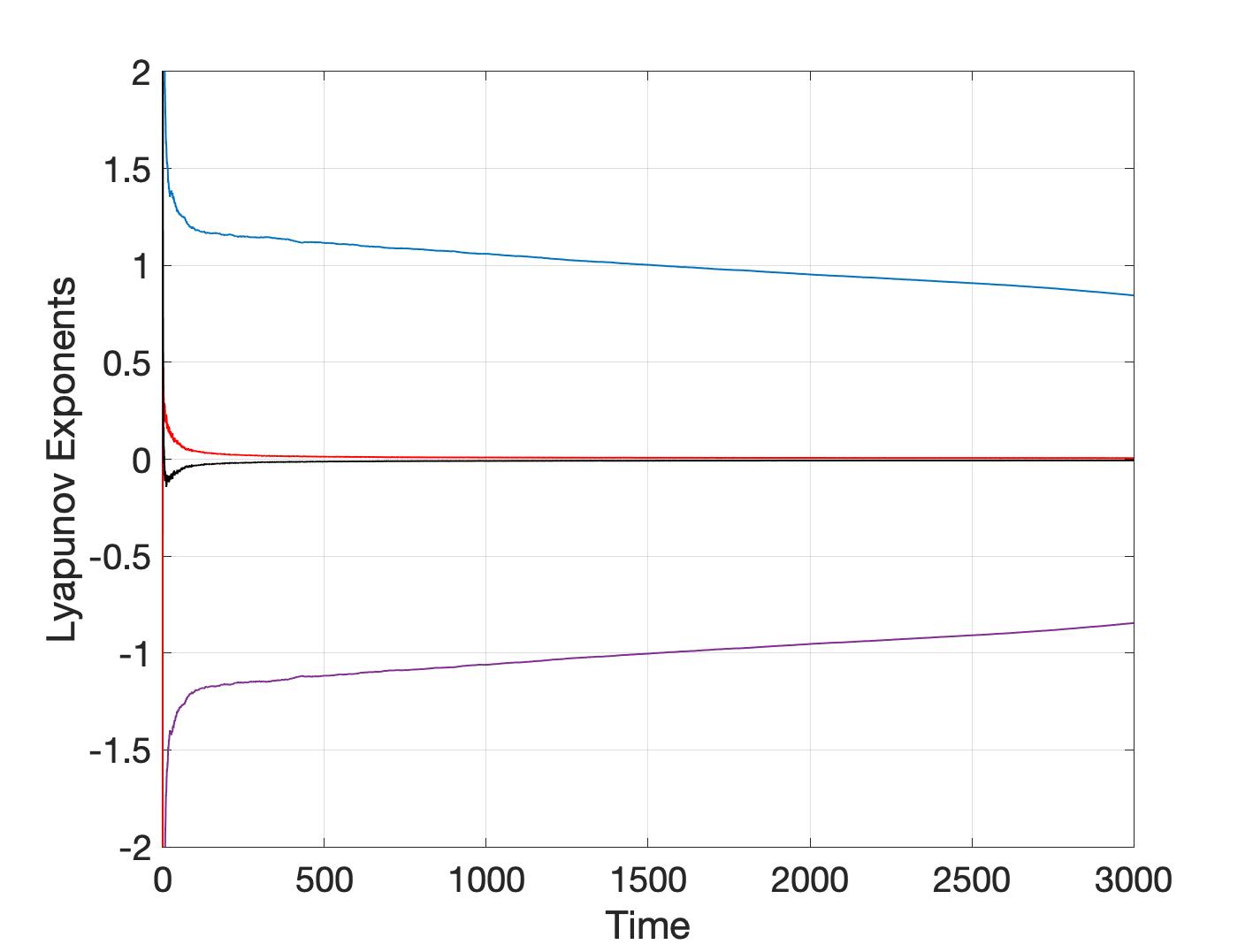}  
		\caption{$n=6$ and $E^\prime/N^2=1500$}
		\label{fig:fig1g}
	\end{subfigure}	
	\begin{subfigure}{.32\textwidth}
		\centering
		\includegraphics[width=1\linewidth]{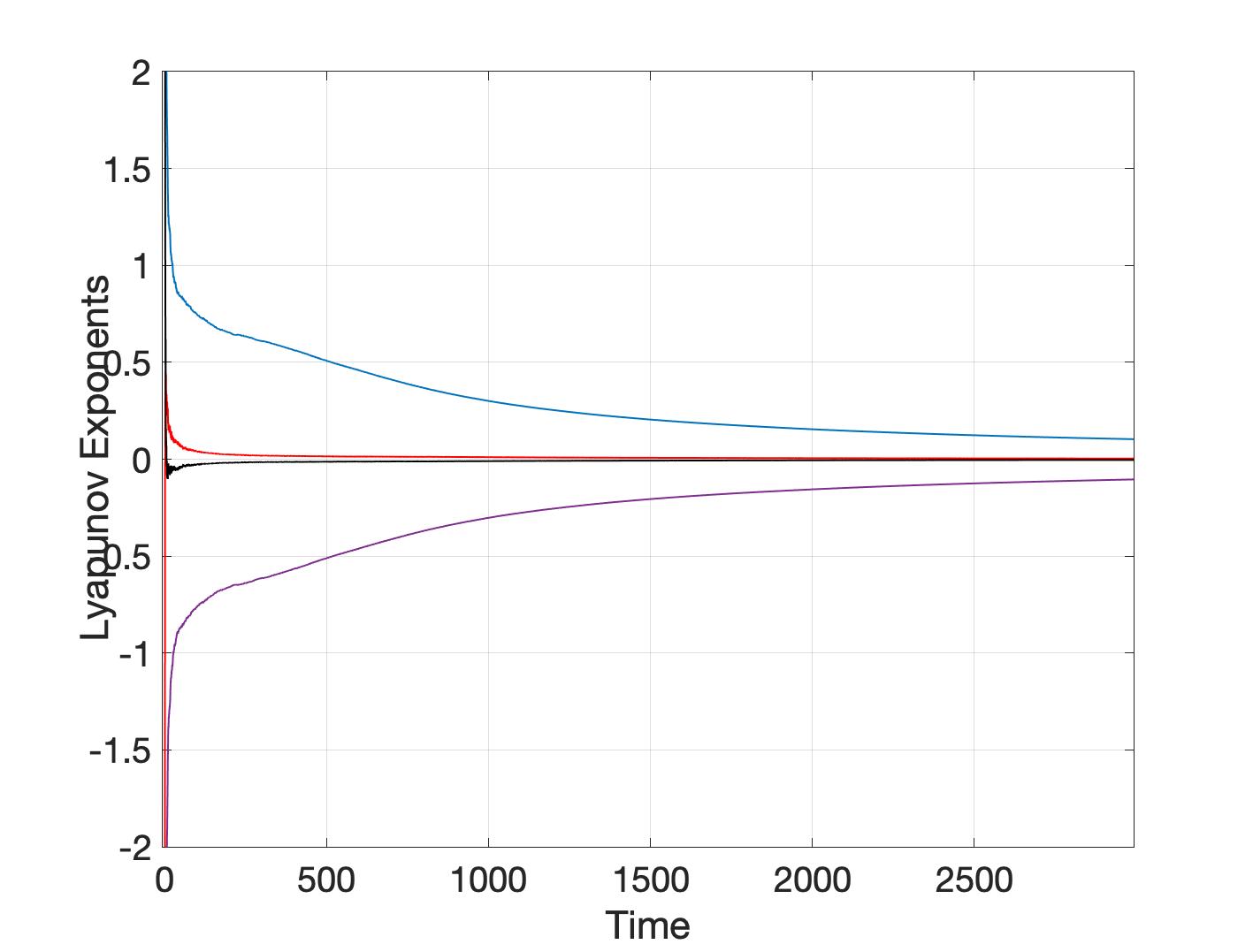}  
		\caption{$n=7$ and $E^\prime/N^2=2000$}
		\label{fig:fig1h}
	\end{subfigure}	
	\caption{Time Series for Lyapunov Exponents}
	\label{fig:fig1}
\end{figure}

In order to obtain the Lyapunov spectrum for our models we run a Matlab code, which numerically solves the Hamilton's equations of motion in (\ref{HeomV}) for all $H_n$ ($1 \leq n \leq 7$) at several different values of the energy. We run the code $40$ times with randomly selected initial conditions satisfying a given energy and calculate the mean of the time series from all runs for each of the Lyapunov exponents at each value of $n$. In order to give certain effectiveness to the random initial condition selection process we developed a simple approach which we briefly explain next. Let us denote a generic set of initial conditions at $t=0$ by $(r(0),y(0),p_r(0),p_y(0))$. For $H_{n < 4}$ we take $y(0)=0$ and for $H_{n \geq 4}$, $r(0)=0$ as part of the initial condition and subsequently generate three random numbers $\omega_i$ $(i=1,2,3)$ and define $\Omega_i = \frac{\omega_i}{\sqrt{\omega_i^2}}\sqrt{E^\prime}$ for a given energy $E^\prime = E +|V_{(n),min}|$ of the system\footnote{Let us note that, via introducing $E^\prime$, we have simply shifted the minimum of the potentials $V_{(n)}$ to zero value. This is convenient for describing and using the process of random selection of initial conditions for the reduced models, as well as, in presenting some of our results. For instance, the time series for Lyapunov exponents and the plots of the Poincar\'{e} sections. However, we will continue to use the energy variable $E$ while examining the variation of the largest Lyapunov exponents and discussing the relation of the latter to the temperature.}, so that $E^\prime = \Omega_i^2 = \Omega_1^2 + \Omega_2^2 +\Omega_3^2$. Subsequently, we take positive roots in the expressions\footnote{We have checked that randomly selecting positive and negative roots in (\ref{rain2}) does not cause any significant impact on our results.} 
\be
p_r(0) = N \sqrt{4c_1 \Omega_1^2} \,, \quad p_y(0) = N \sqrt{4c_2 \Omega_2^2} \,,
\label{rain2}
\ee
and the real roots of
\beqa
8 c_{1} r^4(0) - 16 c_{1} r^2(0) + 8c_1 - \Omega_3^2 &=& 0 \quad  \mbox{for} \quad H_{n < 4}  \nn \\
c_{2} y^4(0) -2 y^2(0) + c_2 - \Omega_3^2 &=& 0 \quad  \mbox{for} \quad H_{n \geq 4} \,,
\label{rain1}
\eeqa
where in the last step of the process our code randomly selects from the available real roots of the equations (\ref{rain1}). In the computations we use a time step of $0.25$ and run the code for a sufficient amount of computer time to clearly observe the values that the Lyapunov exponents converge to. Below, we present sample plots for the time series of Lyapunov exponents at each value of $n$. From the figures \ref{fig:fig1} chaotic dynamics of the models are clearly observed, as in each case (except in figure \ref{fig:fig1b}) a positive Lyapunov exponent is present. We also observe that the properties of Lyapunov spectrum for Hamiltonian systems summarized at the end of the first paragraph of this section are readily satisfied. Let us immediately note that for the model at $n=1$ have distinct features from the rest. This is already observed from the first two plots (figures \ref{fig:fig1a} and \ref{fig:fig1b}); for $E^\prime \approx 30 \cdot 4^2$ (i.e. $E \approx 10$) there is a positive Lyapunov exponent, while at $E^\prime \approx 500 \cdot 4^2$ all the Lyapunov exponents appear to be converging to zero indicating that very little chaos remains at this energy. These distinct and non-chaotic features of the $n=1$ model will also be noted in the ensuing discussions. 

\begin{figure}[!htb]
	\centering
	\begin{subfigure}[!htb]{.32\textwidth}
		\centering
		\includegraphics[width=1\linewidth]{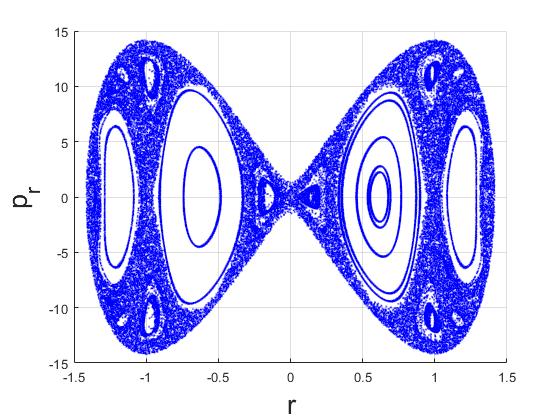}  
		\caption{$n=1$ and $E^\prime/N^2=20$}
		\label{fig:fig3a}
	\end{subfigure}	
	\begin{subfigure}[!htb]{.32\textwidth}
		\centering
		\includegraphics[width= 1\linewidth]{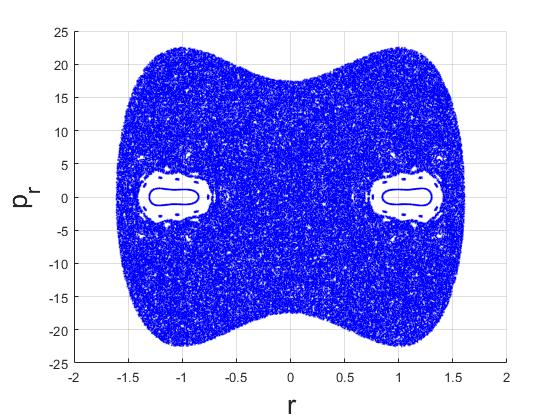}  
		\caption{$n=1$ and $E^\prime/N^2=50$}
		\label{fig:fig3b}
	\end{subfigure}	
	\begin{subfigure}[!htb]{.32\textwidth}
		\centering
		\includegraphics[width=1\linewidth]{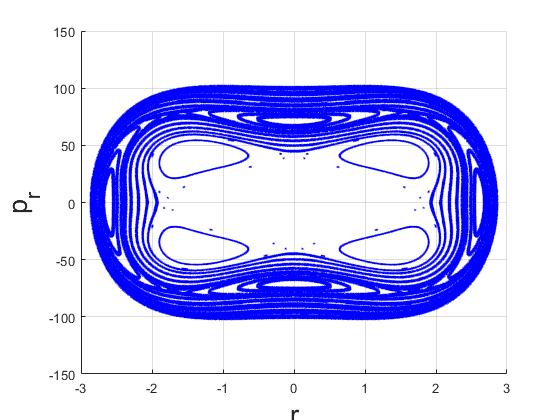}  
		\caption{$n=1$ and $E^\prime/N^2=1000$}
		\label{fig:fig3c}
	\end{subfigure}	
	\begin{subfigure}[H]{.32\textwidth}
		\centering
		\includegraphics[width= 1\linewidth]{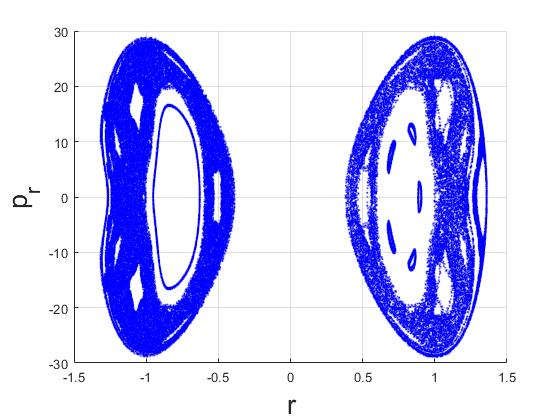}  
		\caption{$n=2$ and $E^\prime/N^2=34$}
		\label{fig:fig3d}
	\end{subfigure}	
	\begin{subfigure}[!htb]{.32\textwidth}
		\centering
		\includegraphics[width=1\linewidth]{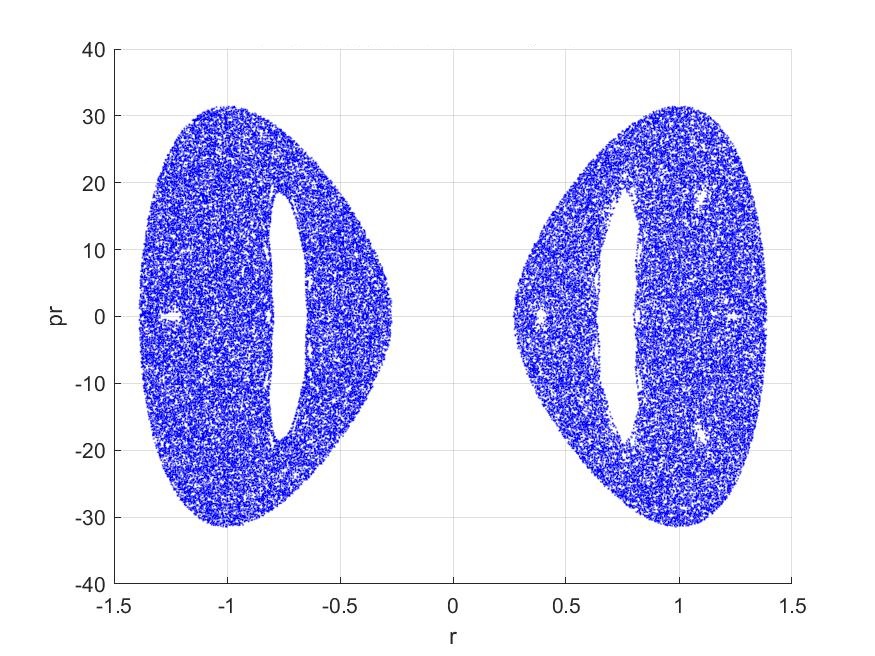}  
		\caption{$n=2$ and $E^\prime/N^2=41$}
		\label{fig:fig3e}
	\end{subfigure}	
	\begin{subfigure}[H]{.32\textwidth}
		\centering
		\includegraphics[width=1\linewidth]{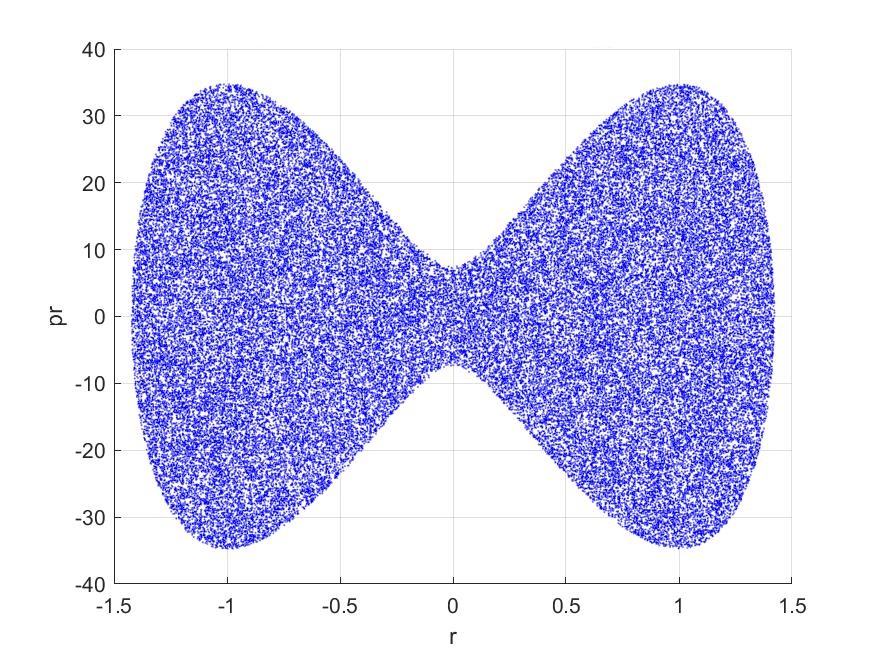}  
		\caption{$n=2$ and $E^\prime/N^2=50$}
		\label{fig:fig3f}
	\end{subfigure}
	\begin{subfigure}[!htb]{.32\textwidth}
		\centering
		\includegraphics[width= 1\linewidth]{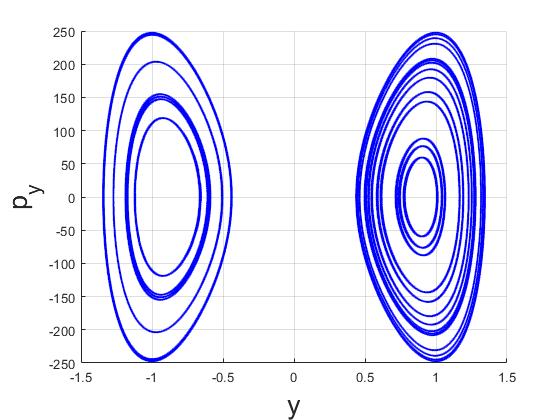}  
		\caption{$n=4$ and $E^\prime/N^2=100$}
		\label{fig:fig3g}
	\end{subfigure}	
	\begin{subfigure}[!htb]{.32\textwidth}
		\centering
		\includegraphics[width=1\linewidth]{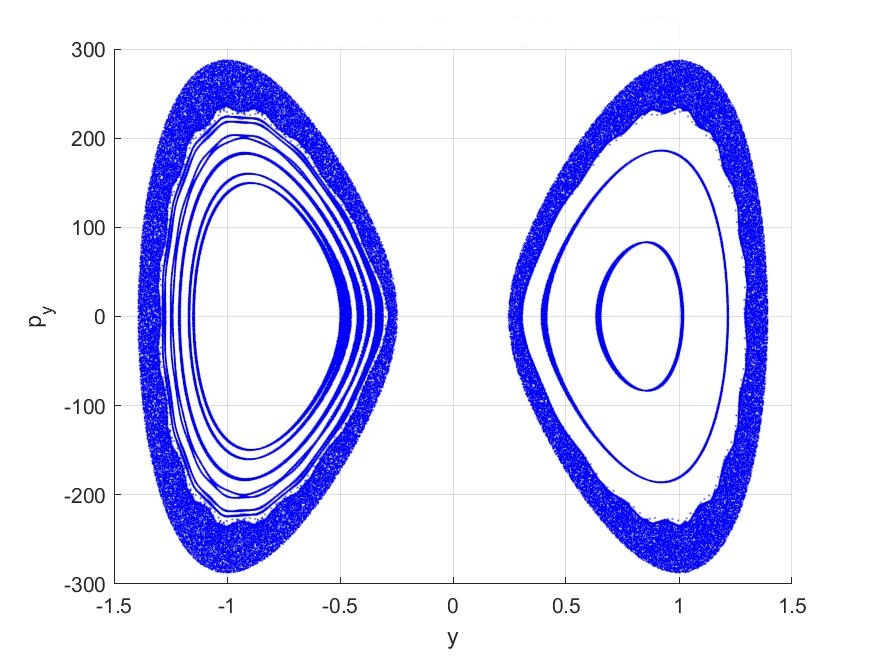}  
		\caption{$n=4$ and $E^\prime/N^2=135$}
		\label{fig:fig3h}
	\end{subfigure}	
	\begin{subfigure}[!htb]{.32\textwidth}
		\centering
		\includegraphics[width= 1\linewidth]{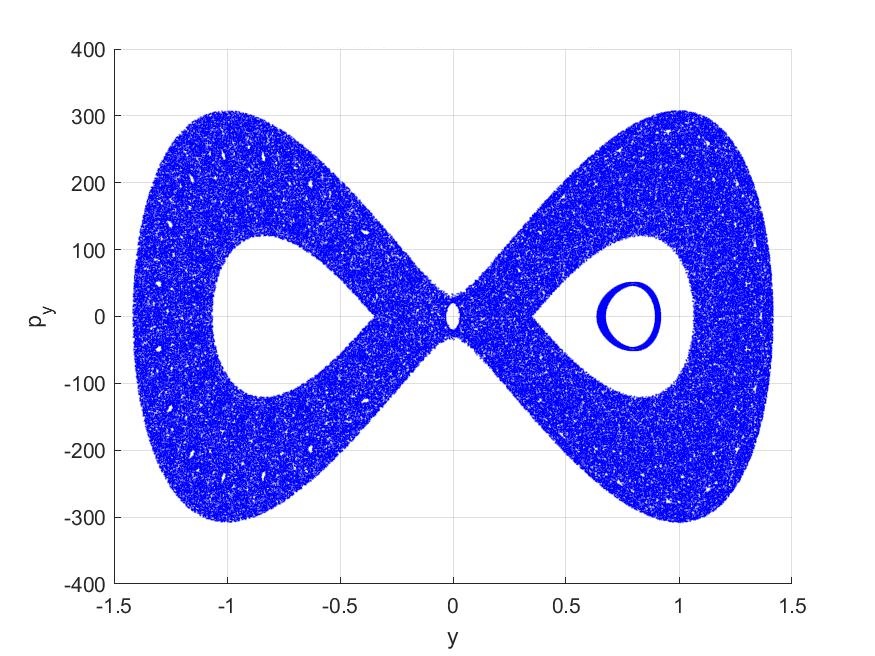}  
		\caption{$n=4$ and $E^\prime/N^2=155$}
		\label{fig:fig3i}
	\end{subfigure}	
	\begin{subfigure}[!htb]{.32\textwidth}
		\centering
		\includegraphics[width= 1\linewidth]{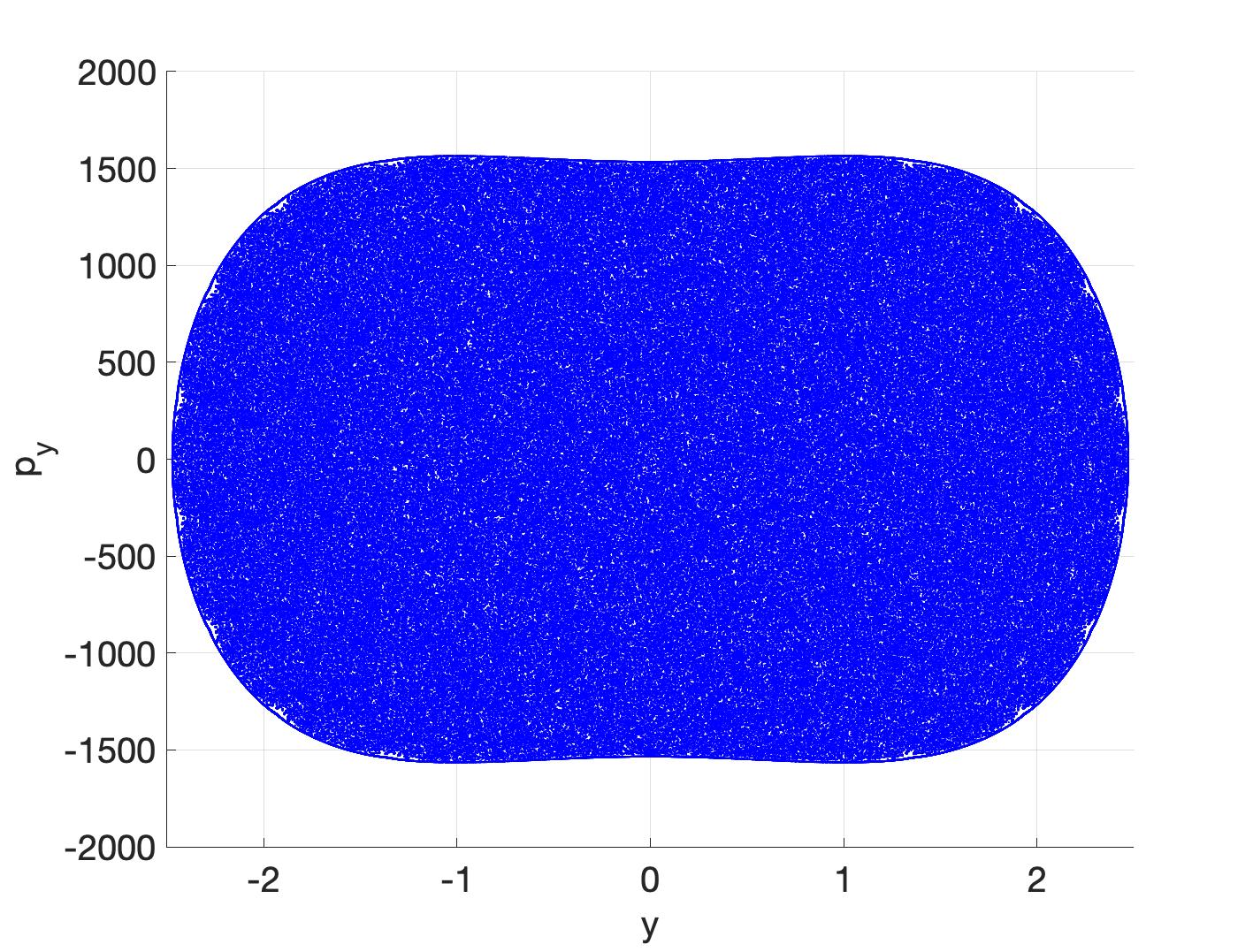}  
		\caption{$n=4$ and $E^\prime/N^2=4000$}
		\label{fig:fig3k}
	\end{subfigure}	
	\caption{Poincar\'{e} Sections (Plots are given w.r.t. the shifted energies $E^\prime = E +|V_{(n),min}|$)}
	\label{fig:fig3}
\end{figure}

To provide further evidence for the chaotic dynamics emerging from the reduced models, we have obtained the Poincar\'{e} sections at several different values of the energy. For each model we plot the Poincar\'{e} sections at energies below, around and above the energy of unstable fixed points $(\pm h_1(n), \pm (\mp) h_2(n), 0, 0)$ to visualize how the phase space trajectories develop and capture the onset of chaotic dynamics. As in the calculation of Lyapunov spectrum, we use $40$ randomly selected initial conditions for each model at a given energy, and the same procedure as in the analysis of the Lyapunov spectrum is followed. For $H_{n < 4}$, $y(0) = 0$ is a part of the initial conditions and for $H_{n \geq 4}$, $r(0)=0$ is so, therefore it is convenient to look at the Poincar\'{e} sections on the $r-p_r$-plane and the $y-p_y$-plane, respectively in these cases. Our plots for the models  $n=1,2,4$ are presented below in the figure \ref{fig:fig3}, while the results for the $n=3,5,6,7$ are carried over to the appendix and given in the figures \ref{fig:figA}. From the Poincaré sections we infer that the phase space is filled essentially with quasi-periodic orbits at energies below $E_F^\prime$, while at and around $E_F^\prime$ quasi periodic trajectories and chaotic ones coexist, while for energies larger than $E_F$ the phase space is either completely or strongly dominated by chaotic trajectories.

\subsection{Dependence of the Largest Lyapunov Exponent on Energy}

\begin{figure}[!htb]
	\centering
	\begin{subfigure}[!htb]{.32\textwidth}
		\centering
		\includegraphics[width= 1\linewidth]{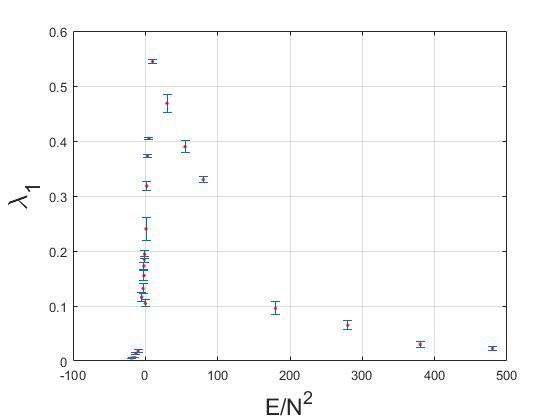}  
		\caption{$\lambda_1$ vs. $E/N^2$}
		\label{fig:fig2a}
	\end{subfigure}	
	\begin{subfigure}[!htb]{.32\textwidth}
		\centering
		\includegraphics[width=1\linewidth]{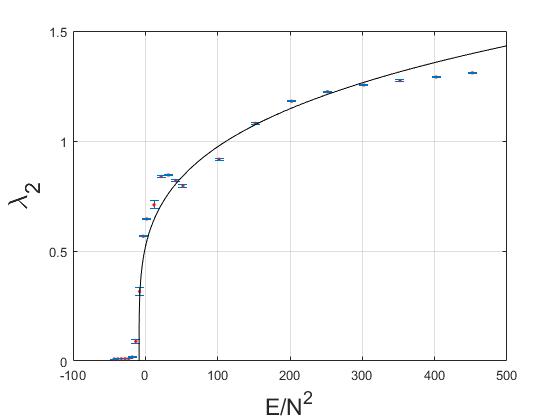}  
		\caption{$\lambda_2$ vs. $E/N^2$}
		\label{fig:fig2b}
	\end{subfigure}	
	\begin{subfigure}[!htb]{.32\textwidth}
		\centering
		\includegraphics[width=1\linewidth]{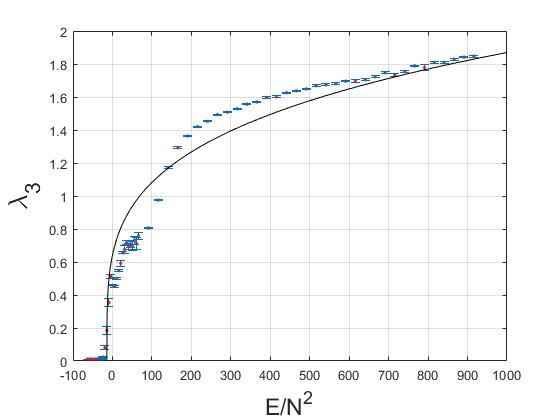}  
		\caption{$\lambda_3$ vs. $E/N^2$}
		\label{fig:fig2c}
	\end{subfigure}	
	\begin{subfigure}[!htb]{.32\textwidth}
		\centering
		\includegraphics[width=1\linewidth]{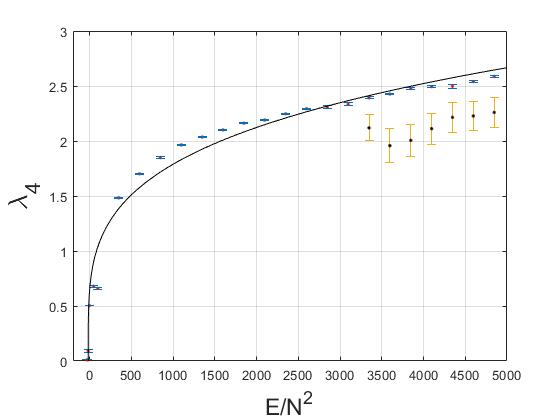}  
		\caption{$\lambda_4$ vs. $E/N^2$}
		\label{fig:fig2d}
	\end{subfigure}		
	\begin{subfigure}[!htb]{.32\textwidth}
		\centering
		\includegraphics[width= 1\linewidth]{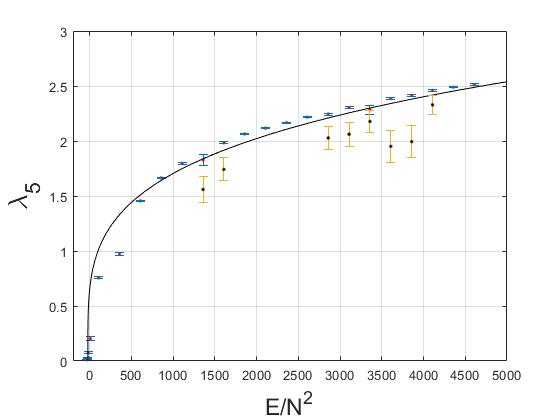}  
		\caption{$\lambda_5$ vs. $E/N^2$}
		\label{fig:fig2e}
	\end{subfigure}	
	\begin{subfigure}[!htb]{.32\textwidth}
		\centering
		\includegraphics[width=1\linewidth]{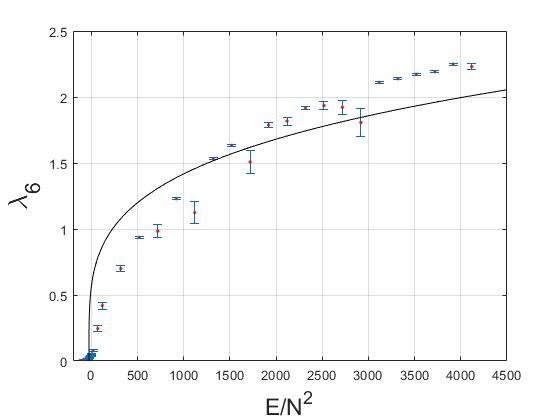}  
		\caption{$\lambda_6$ vs. $E/N^2$}
		\label{fig:fig2f}
	\end{subfigure}		
	\begin{subfigure}[!htb]{.32\textwidth}
		\centering
		\includegraphics[width=1\linewidth]{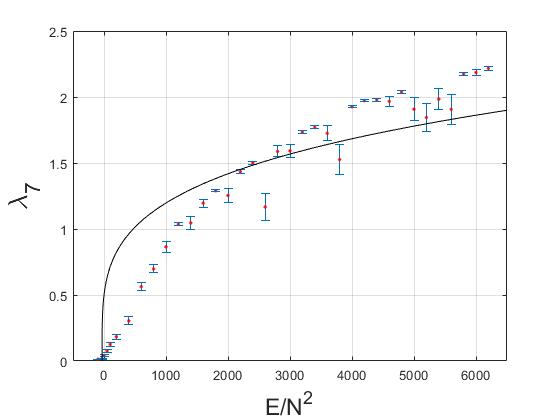}  
		\caption{$\lambda_7$ vs. $E/N^2$}
		\label{fig:fig2g}
	\end{subfigure}				
	\caption{MLLE vs. $E/N^2$ for $\mu_1^2=-16$ and $\mu_2^2 =-2$}
	\label{fig:fig2}
\end{figure}

In order to see the dependence of the mean largest Lyapunov exponent(MLLE), (denoted as $\lambda_n$ in the figures and the tables) to energy, we obtain the MLLE at a large span of energy values, which appears to be best suited to observe the onset and progression of chaotic dynamics in these models. As may be expected, the energies determined for the unstable fixed points in the previous section are of central importance here. From the figures \ref{fig:fig2}, we see that appreciable amount of chaotic dynamics starts to develop once the energy of systems exceeds $E_F$ of the models at the fixed points $(\pm h_1(n), \pm (\mp) h_2(n), 0, 0)$. The onset of the chaotic dynamics starting at and around the fixed point energies $E_F(\pm h_1(n), \pm (\mp) h_2(n), 0, 0)$ as observed from progression of the MLLE values with increasing energy is seen from the figures \ref{fig:fig2}. Error bars at each data point is found by evaluating mean square error using MLLE and the LLE values of each of the $40$ runs.

From figure \ref{fig:fig2d} - \ref{fig:fig2g}, we see that, at some data points with $E \gtrsim 1250$ for the models $n\geq 4$ there is an observable decrease in the value of MLLE and the mean square errors appear to be considerably larger than those computed for the rest of the data points. A closer analysis of the Lyapunov time series at these data points reveal that LLE values of less than a quarter of the $40$ initial conditions approach to zero, leading to the observed decrease in MLLE values and the increase in the mean square errors. From a physical point of view, approach of some of the LLE values to zero implies that the systems' development in time, starting from these initial conditions are of either periodic or quasi-periodic type and not chaotic. Nevertheless, the overall MLLE values are still quite large and the sample Poincar\'{e} section in figure \ref{fig:fig3k} taken at one of these energies are densely chaotic, showing no sign of Kolmogorov-Arnold-Moser(KAM) tori, which would have signaled the presence of quasi-periodic orbits. Therefore, we are inclined to think that such periodic or quasi-periodic orbits occur only at comparatively very small regions of the phase space. At $n=4$ and $n=5$ such data points are confined to a range of energy values and we  evaluated the MLLE values at these energies by excluding the initial conditions leading to vanishing LLE. The results of MLLE obtained this way are given in the plots in the figures \ref{fig:fig2d} and \ref{fig:fig2f} with blue error bars as with the rest of the data, while those points with error bars in yellow color represent the original data and they are kept in the same plots for comparison and facilitating this technical discussion. The main purpose of this analysis is that, it enables us to get improved best fit curves (the physical uses of which will be discussed in the next section), while for $n=6$ and $n=7$, we do not perform a similar analysis since, it appears from \ref{fig:fig2f} and \ref{fig:fig2g} that such data points appear to be distributed in a more scattered fashion and this yields no significant improvement for the type of best fit curves that we introduce and discuss in the next subsection.  We also see that in the $n=1$ model the MLLE acquires a peak value of about $\approx 0.55$ at an energy around $E \approx 12\cdot 4^2$($E^\prime \approx 32 \cdot 4^2$) and rapidly decreases toward zero with increasing energy. From the profile of MLLE w.r.t. $E/N^2$ in figure \ref{fig:fig2a}  as well as the time series plot (\ref{fig:fig1b}) of the model at $E = 500 \cdot 4^2$, we conclude that this model is not chaotic for energies $E \geq 500 \cdot 4^2$. These conclusions are also fully supported by the Poincar\'{e} sections given in the figures \ref{fig:fig3a} - \ref{fig:fig3c}. 

\subsection{Temperature Dependence of the Lyapunov Exponent for the Mass Deformed Matrix Model}

In \cite{Gur-Ari:2015rcq}, BFSS models is examined at the classical level and it was found that the largest Lyapunov exponent is given as $\lambda = 0.2924(3)(\lambda_{'t Hooft} T)^{1/4}$. The dependence of $\lambda$ on temperature can be readily understood, since in the 't Hooft limit the classical observables of the BFSS model can only depend on a power of the combination $\lambda_{'t Hooft} T$. The latter is determined by simply dimensional analysis, since $\lambda$ must have units of $\mbox{Length}^{-1}$, while $\lambda_{'t Hooft} T$ has units  $\mbox{Length}^{-4}$. The temperature is essentially introduced into to the model via the use of the virial and equipartition theorems, since the analysis is performed in real time\footnote{This is in contrast to several previous studies on the BFSS model and its deformations mainly targeting to explore their phase structure, which resort to the imaginary time formalism with period $\beta$, the inverse temperature \cite{Kawahara:2006hs, Kawahara:2007fn, DelgadilloBlando:2007vx, DelgadilloBlando:2008vi, Asano:2018nol, Berkowitz:2018qhn}.}. To be more precise, for a purely quartic potential, $\langle K \rangle =2 \langle V \rangle  = \frac{1}{2} n_{d.o.f} T$ (We work in the units, where the Boltzmann constant,$k_B$, is set to unity.) and $E = \langle K \rangle + \langle V \rangle = \frac{3}{4} n_{d.o.f} T$, with $ n_{d.o.f} = 8(N^2-1)-36$ being the independent degrees of freedom in the BFSS matrix model, after properly accounting for constraints and symmetries. The result $\lambda = 0.2924 T^{1/4}$ (recall that we are setting $\lambda_{'t Hooft}$ to unity) of \cite{Gur-Ari:2015rcq, Kawahara:2007fn}  is parametrically smaller than the MSS bound, $2 \pi T$, on quantum chaos for sufficiently large $T$ and clearly violates this bound only at small temperatures, at $(\frac{0.292}{2 \pi})^\frac{4}{3} = 0.015$ to be precise. This is, in fact, what is generally expected as the classical theory is good only in describing the dynamics in the high temperature regime.

In view of these developments, to elaborate on the dependence of our MLLE results on temperature we proceed with the following strategy. Due to the massive deformation terms in the matrix model, the potential is no longer a polynomial of homogeneous degree in the matrices and therefore the virial theorem is not sufficient to express the kinetic ,$\langle K \rangle$, and potential,$\langle V \rangle$, energies as a multiple of each other. Nevertheless, we can immediately see\footnote{From (\ref{YMDM}), we can express the Lagrangian as $L_{YMM}=K-V$. Applying the virial theorem, we find
\beqa
2 \langle K \rangle &=& 2 \langle V \rangle -  2 \, Tr \frac{1}{4} [X_I, X_J]^2 \,, \nn \\
&=&	 4 \langle V \rangle -  2 \, Tr \, \frac{1}{2} \mu_1^2 X_a^2 - 2 \, Tr \, \frac{1}{2} \mu_2^2 X_i^2 \,. \nn
\eeqa
First equality implies that $\langle V \rangle < \langle K \rangle $, since $Tr [X_I, X_J][X_I, X_J]^\dagger = - Tr [X_I, X_J]^2 \geq 0$ for Hermitian $X_I$, while the second implies $2\langle  V  \rangle \leq  \langle K \rangle$ for $\mu_1^2 <0 $, $\mu_2^2 < 0$.} that $\langle V \rangle < \langle K \rangle $ is always satisfied and for $\mu_1^2 <0 $, $\mu_2^2 < 0$ that $\langle V \rangle < \langle K \rangle $ is always satisfied and for $\mu_1^2 <0 $, $\mu_2^2 < 0$, we further have that $ 2\langle  V  \rangle \leq  \langle K \rangle $. Thus, we have $E \leq  \frac{3}{4}  n_{d.o.f} T $, where now $ n_{d.o.f} = 7(N^2-1)-28 $. The reason for the latter is that, $X_9$ is already set to zero in our model and this decreases the number of d.o.f accordingly. We may take $n_{d.o.f} \approx 7N^2$ for large $N$. Our ansatz configuration involving the fuzzy two- and four- sphere matrices with collective time dependence describes a configuration with particular orientation of coincident $D0$-branes and open string states stretching between them, and the action we obtain after tracing over this configuration effectively model their dynamics. This suggests that, we may consider the energy of the latter as due to the $\approx 7 N^2$ degrees of freedom of the matrix model and hence we may contemplate the inequality 
\be
\frac{E}{N^2} \leq \frac{21}{4} T \,.
\label{EvT}
\ee

Taking these considerations into account, we examine the best fit curves of the form 
\be
\lambda_n = \alpha_n \left (\frac{E}{N^2} - \frac{(E_n)_F}{N^2} \right)^{1/4} \,,
\label{lvE}
\ee
to the $\lambda_n$ versus $\frac{E}{N^2}$ data in figure \ref{fig:fig2}. The functional form (\ref{lvE}) is well-motivated, firstly because, it is readily observed from figure \ref{fig:fig2} that the reduced models do not develop any appreciable chaos up until the critical energies $(E_n)_F < 0$ are well exceeded. Next, we note that, at a given level $n$, $(E_n)_F$ is determined by the values of the masses $\mu_1^2$, $\mu_2^2$, and  (\ref{lvE}) embodies the implicit dependence of $\lambda_n$ on these additional dimensionful parameters of the model through $(E_n)_F$, in contrast to the pure BFSS model, whose only dimensionful parameter is $\lambda_{'t Hooft}$ and finally the power of $1/4$ is suggestive from the energy (and also temperature) dependence of the LLE in the pure BFSS model. The best fitting curves for the given functional form (\ref{lvE}) are also given in figure \ref{fig:fig2}, while the coefficients $\alpha_n$ are listed in table \ref{table:fitvalue1} below:
\begin{table}[!htb]
	\centering
	\begin{tabular}{|c|c|c|c|}
		\hline
	$n$ &  $\alpha_{n}$ &$T_c$ & $\beta_{n}$ \\ \hline 
	$2$ & {0.3017} & 0.0832 & 0.4567 \\ \hline  
	$3$	& {0.3313} & 0.1029 & 0.5015 \\ \hline
	$4$	& {0.3168} & 0.1046 & 0.4795 \\ \hline
	$5$	& {0.3016} & 0.1045 & 0.4565 \\ \hline
	$6$	& {0.2505} & 0.0911 & 0.3791 \\ \hline
	$7$	& {0.2113} & 0.0796 & 0.3198 \\ \hline	
	\end{tabular}
	\caption{$\alpha_n$ values for the best fit curves (\ref{lvE}), Upper bounds for $T_c$ and $\beta_n$ values for the inequality (\ref{lvT}).}
	\label{table:fitvalue1}
\end{table}

A number of comments regarding these best fit curves are in order. We immediately see that the $\alpha_n$ values are reasonably close to each other for $2\leq n \leq 4$ and the quality of the fitting curves are quite good, while they are smaller for $n = 6$ and $7$ and there is somewhat a decrease in the quality of the fitting curves. As we have already noted in the previous subsection, the improved quality of the fits for $n=4$ and $n=5$ are due to more detailed analysis of the data as we have already explained, which does not, however, benefit the cases of $n=6$ and $n=7$ due to the scattered distribution of relatively lower LLE values compared with those obtained at nearby energies. A more sophisticated  initial condition selection procedure, for instance by confining to small hyper-volumes of phase space, and averaging the LLE data over such hyper-volumes may help to decrease fluctuations on the data and subsequently enhance the quality of the fits too, while such more advanced numerical analysis is out of the scope of our present work. Let us also note that for $n=7$, another reason for a relatively poor fit could be the fact that the coefficients $c_{(1,2,3)}$ for this case are obtained through extrapolation of our exact results for $n \leq 6$ as noted after (\ref{Lalp}), and this may be influencing the MLLE spectrum w.r.t. the energy and in turn the quality of the fit too.

Let us now comment on the temperature dependence of the MLLEs. In view of the inequality (\ref{EvT}), we conclude that at zero temperature, $E \leq 0$, while the MLLE values are essentially non-vanishing until the energies drop below or around $(E_n)_F$ as clearly seen from the data presented in figure \ref{fig:fig2}. This means that already at zero temperature the MSS bound $ \lambda_L \leq 2 \pi T$ on quantum chaos is violated by this classical system. In fact, we can put an upper bound on the temperature below which the MSS bound is violated. Using (\ref{EvT}) and (\ref{lvE}) together, we see that MSS bound is violated for temperatures in the interval $0 \leq T \leq T_c$, where $T_c$ is the temperature saturating the inequality $\alpha_n \left (\frac{21}{4} T - \frac{(E_n)_F}{N^2} \right)^{1/4} \geq 2 \pi T$. For $n=2\,, \cdots\,,7$, numerical values of $T_c$ are given in  \ref{table:fitvalue1} and they are roughly an order of magnitude larger than the critical temperature $0.015$ determined for the BFSS model in \cite{Gur-Ari:2015rcq, Kawahara:2007fn}. Due to (\ref{EvT}) we can expect that exact values of $T_c$ should be less than these upper bounds and therefore closer to the value determined for the BFSS model.
  
At sufficiently high energies, we may estimate
\be
\lambda_n \approx \alpha_n \left( 1 - \frac{1}{4} \frac{(E_n)_F}{E}\right) \left (\frac{E}{N^2} \right)^{1/4}  \,, \quad \mbox{for}\, \quad  E \gg |(E_n)_F| > 0 \,,
\label{lvE2}
\ee 
and therefore 
\be
\lambda_n \leq \beta_n  \, T^{1/4} \,, \quad \beta_n := \left(\frac{21}{4} \right )^{1/4} \alpha_n  \,,
\label{lvT}
\ee
which is strictly valid in the high temperature regime for non-vanishing values of $\lambda_n$ and is parametrically smaller than the MSS bound $2 \pi T$. Numerical values of $\beta_n$ are provided in table \ref{table:fitvalue1} for easy access.

One may also wonder if and how the possible different values of the mass parameters could alter or modify the results obtained in this section. This question will be discussed in some detail in the next subsection.

Another important issue that needs to be addressed is how to incorporate and/or compute quantum corrections to the matrix model and to the reduced effective actions presented in this paper or more generally in the broader context for the BFSS type models, as well as their fully supersymmetric versions. We present our opinions on this issue in relation to recent developments in the conclusions and outlook section of our paper.   

\subsection{Another Set of Mass Values}

It is surely important to consider how assigning different values to the mass parameters could impact the dynamics of our models. Let us first note that $\mu_1^2 =-16$ and $\mu_2^2=-2$ is a suitable and immediate guiding choice for the mass values due to the reasons discussed around equations (\ref{foursphereeom}) and (\ref{twosphereeom}), but surely not canonical in the sense that they are not enforced on us by the full set of equations of motion (\ref{YMDMeom}). Nevertheless, noting the critical role played by the unstable fixed point $(h_1(n), h_2(n),0,0)$ with negative energies $E_F$ for the analysis given in the previous sections, we may wonder, how the physics could get altered if we work with values of $\mu_1^2$ and $\mu_2^2$ for which no fixed point of the type $(h_1(n), h_2(n),0,0)$, is present but only $(0,0,0,0)$ is the unstable fixed point with zero energy at each level $n$. It is not our aim in this section to provide a detailed survey of all possibilities for different mass-squared values, nor is this within the scope of our present work, but we simply confine ourselves to examining another choice for the mass values and to serve the aforementioned purpose we take $\mu_1^2= -8$ and $\mu_2^2 = 1$.   

With the Lagrangian and Hamiltonian given in the form (\ref{Lalp}) and (\ref{Hn1}) and the corresponding Hamilton's equations given as in (\ref{Heom}), we find that the local minimum of the potential is at $(r,y) = (\pm\frac{1}{\sqrt{2}},0)$ and the local maximum is at $(r,y) = (0,0)$ with the corresponding energies $V_{(n),min} = - 2c_1 N^2$ and $0$, respectively. For easy reference, the numerical values of the  $V_{(n), min}/N^2$ for $n=1\,, \cdots \,,7$, are $-5$, $-12$,  $-21$, $-32$, $-45$, $-60$ and $-77$, respectively.  

Fixed points of the phase space are
\be
(r,y,p_r,p_y) =  \lbrace (0,0,0,0)\,, (\pm\frac{1}{\sqrt{2}},0,0,0) \rbrace 
\ee    
with the corresponding energies 
\be
E_F(0,0,0,0) =0  \,, \quad E_F(\pm \frac{1}{\sqrt{2}}, 0,0,0) = - 2 c_1 N^2 = - n(n+4) N^2 \,.   
\ee
Linear stability analysis shows that $(0,0,0,0)$ is an unstable fixed point while $(\pm \frac{1}{\sqrt{2}},0,0,0)$ are of the borderline type that we encountered previously and play no significant role in the numerical analysis that follows next. 

\begin{figure}[!htb]
	\centering
	\begin{subfigure}[!htb]{.32\textwidth}
		\centering
		\includegraphics[width= 1\linewidth]{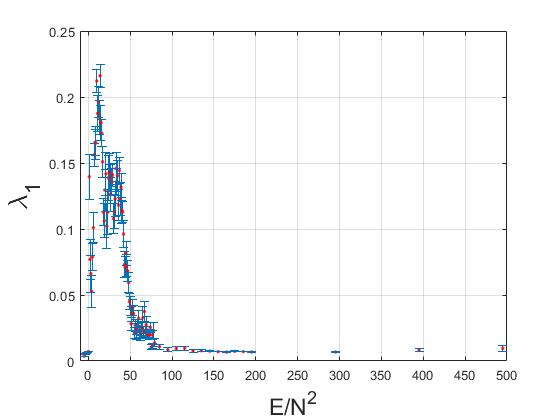}  
		\caption{$\lambda_1$ vs. $E/N^2$}
		\label{fig:fig4a}
	\end{subfigure}	
	\begin{subfigure}[!htb]{.32\textwidth}
		\centering
		\includegraphics[width=1\linewidth]{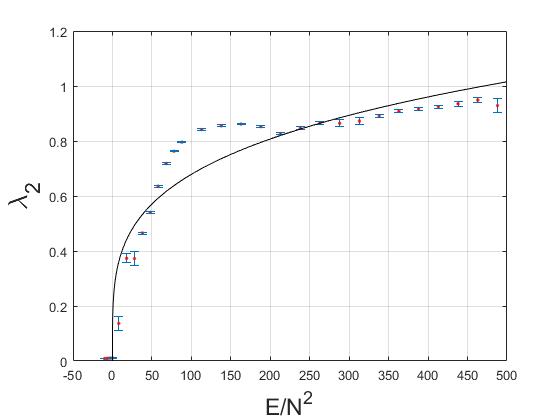}  
		\caption{$\lambda_2$ vs.$E/N^2$}
		\label{fig:fig4b}
	\end{subfigure}
	\begin{subfigure}[!htb]{.32\textwidth}
		\centering
		\includegraphics[width= 1\linewidth]{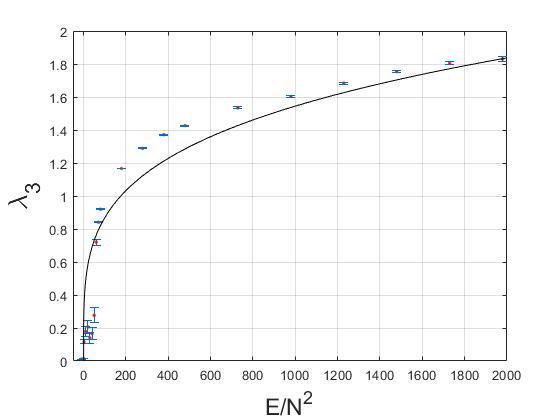}  
		\caption{$\lambda_3$ vs. $E/N^2$}
		\label{fig:fig4c}
	\end{subfigure}	
	\begin{subfigure}[!htb]{.32\textwidth}
		\centering
		\includegraphics[width=1\linewidth]{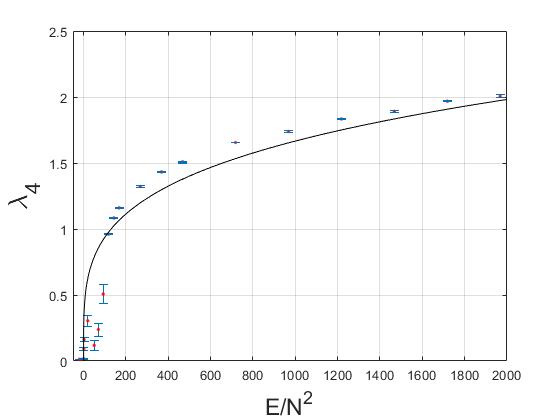}  
		\caption{$\lambda_4$ vs. $E/N^2$}
		\label{fig:fig4d}
	\end{subfigure}
	\begin{subfigure}[!htb]{.32\textwidth}
		\centering
		\includegraphics[width= 1\linewidth]{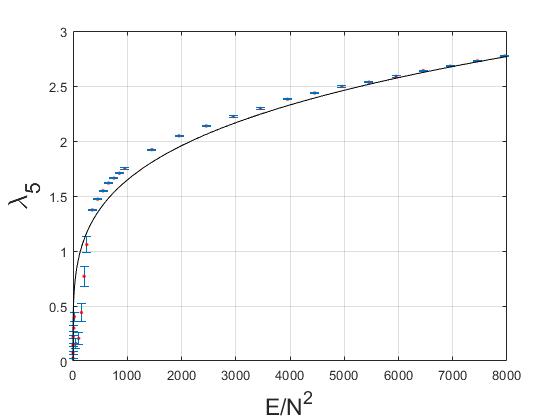}  
		\caption{$\lambda_5$ vs. $E/N^2$}
		\label{fig:fig4e}
	\end{subfigure}	
	\begin{subfigure}[!htb]{.32\textwidth}
		\centering
		\includegraphics[width=1\linewidth]{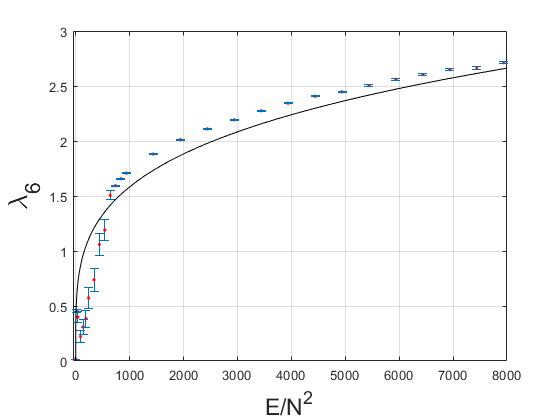}  
		\caption{$\lambda_6$ vs. $E/N^2$}
		\label{fig:fig4f}
	\end{subfigure}
	\begin{subfigure}[!htb]{.32\textwidth}
	\centering
	\includegraphics[width=1\linewidth]{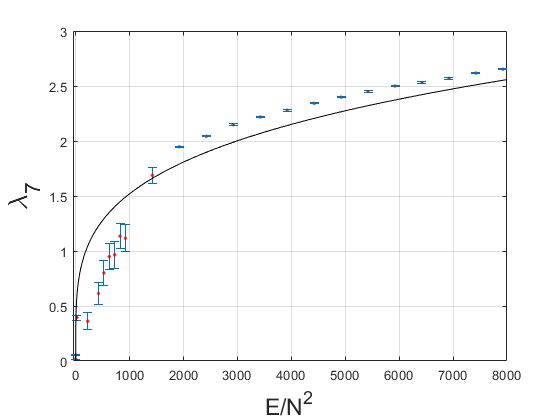}  
	\caption{$\lambda_7$ vs. $E/N^2$}
	\label{fig:fig4g}
\end{subfigure}				
	\caption{MLLE vs.$E/N^2$ for $\mu_1^2=-8$ and $\mu_2^2 = 1$}
	\label{fig:fig4}
\end{figure}

Following the same steps of the numerical analysis for the Lyapunov spectrum outlined previously in section 3.1, we find that, in this case too, the models exhibit chaotic dynamics for $n=2,3,4,5,6,7$, while the model at  $n=1$ is essentially not chaotic for energies $E \gtrapprox 100$, but retains some chaos only in a narrow band of energy from around $E \approx 0$ (i.e. the fixed point energy) to $E \gtrapprox 100 \cdot 4^2$ (See figure \ref{fig:fig4}). The transition to chaos for $n=2,3,4,5,6,7$ appears to happen around the fixed point energies as can be clearly seen from these figures. This fact is also captured by inspecting the Poincar\'{e} sections, which are provided in figure \ref{fig:figB} in appendix B.

Proceeding in a similar manner as in the previous subsection, we consider fitting curves of the form 
\be
\lambda_n = {\tilde \alpha}_n \left ( \frac{E}{N^2} \right)^\frac{1}{4} \,,
\label{lvE2}
\ee
since now the unstable fixed point energy is zero at each value of $n$. We see that fits given in figure \ref{fig:fig4} are reasonably good, while they predict somewhat lower values of $\lambda_n$ at larger energies, except at the level $n=2$. The latter case displays a sharper increase the MLLE at intermediate energies, which is not captured well by the fit. ${\tilde \alpha}_n$ are provided in the table \ref{table:fitvalues2} and happen to be quite close to each other except for $\alpha_2$.  
\begin{table}[!htb]
	\centering
	\begin{tabular}{ | c | c | c |c|}
		\cline{1-4}
		$n$ &  ${\tilde \alpha}_n$ & $T_c$ & ${\tilde \beta}_n$ \\ \hline
		$2$ &$0.2147$ &$0.021$ & 0.349 \\ \hline
		$3$ &$0.2747$ &$0.029$ & 0.447 \\ \hline
		$4$ &$0.2963$ &$0.033$ & 0.481 \\ \hline
		$5$ &$0.2927$ &$0.032$ & 0.476 \\ \hline 
		$6$ &$0.2815$ &$0.030$ & 0.458 \\ \hline
		$7$ &$0.2705$ &$0.029$ & 0.440 \\ \hline  
	\end{tabular}
	\caption{${\tilde \alpha}_n$ values for the best fit curves (\ref{lvE2}), upper bounds for ${\tilde T}_c$ and ${\tilde \beta}_n$ values for the inequality (\ref{lvT2}).}
	\label{table:fitvalues2}
\end{table}

Since $\mu_2^2 = 1 >0$, virial theorem no longer implies $2 \langle V \rangle \leq \langle K \rangle $ in general, but $ \langle V \rangle \leq \langle K \rangle $ is still valid. Thus, we have $E \leq  n_{d.o.f} T $ and therefore the bound
\be
\frac{E}{N^2} \leq 7 T \,,
\label{EvT2}
\ee
instead of (\ref{EvT}).

Data in figure \ref{fig:fig4} indicates that all $\lambda_n \rightarrow 0$ as $E \rightarrow 0$ and are vanishingly small for $V_{(n), min} \leq E \leq 0$ too. These facts clearly show that, as opposed to the previous case, there is no violation of the MSS bound at zero temperature. 

As in the previous case, we can put an upper bound on the temperature below which MSS bound will eventually be violated. Using (\ref{lvE2}) and (\ref{EvT2}),  we see that MSS bound is exceeded for temperatures in the interval $0 \leq T \leq {\tilde T}_c$, where ${\tilde T}_c$ is the temperature saturating the inequality ${\tilde \alpha}_n (7 T)^{1/4} \geq 2 \pi T$. Numerical values of ${\tilde T}_c$ are given in table \ref{table:fitvalues2} and they are significantly lower than the $T_c$ values listed in table \ref{table:fitvalue1} and closer to $0.015$ of the BFSS model. 

Finally, we see that 
\be
\lambda_n \leq {\tilde \beta}_n \, T^{1/4} \,, \quad{\tilde \beta}_n := 7 ^{1/4} {\tilde \alpha}_n  \,,
\label{lvT2}
\ee
with ${\tilde \beta}_n$ values provided in table \ref{table:fitvalues2}, and they are all parametrically less than $2 \pi T$ at temperatures above $T_c$. 

\section{Ans\"{a}tz {\it II}}

We would like to briefly discuss the consequences of changing the fuzzy two sphere part of ans\"{a}tz {\it I}. Namely, we consider the configurations 
\be  
X_a =  r(t) \, Y_a \,, \quad X_i =  y(t) \, Z_i \,, \quad X_9 = 0 \,,
\label{Anstz2}    
\ee 
at the matrix levels $N = \frac{1}{6}(n+1)(n+2)(n+3)$ for $n=2,3,5$. In (\ref{Anstz2}), $Y_a$ are the same as in ans\"{a}tz {\it I}, while we take $Z_i = \oplus_{k=1}^{K_n} \Sigma_i(k)$, with $\Sigma_i(k)$ spanning the spin $\frac{1}{2}$ UIR of $SU(2)$, in the $k^{th}$ block of the direct sum and $K_n$ is the number of $2 \times 2$ blocks, which are $5,10,28$ for $n=2,3,5$, respectively. Thus, we have a direct sum of $2 \times 2$ fuzzy two spheres.\footnote{For odd values of $N$, it is not possible to form $Z_i$'s by $2 \times 2$ blocks only. In this case we can fill the last block of the matrices simply with the $0$-matrix, i.e. the trivial representation of $SU(2)$.} We call this the ans\"{a}tz {\it II}.  

The Lagrangian and Hamiltonian are given in the same form as in (\ref{Lalp}) and (\ref{Hn1}), where now the coefficients $c_\beta$ are given in the table \ref{table:table7}.
\begin{table}
	\centering
	\begin{tabular}{ | c | c | c | c |}
		\cline{2-4}
		\multicolumn{1}{c |}{} & $n=2$ & $n=3$ & $n=5$ \\ \hline 
		$c_{1}$  & $6$  & $21/2$  & $45/2$      \\ \hline 
		$c_{2}$ & $3/8$ & $3/8$ & $3/8$     \\ \hline 
		$c_{3}$ & $7$& $11.45$ &$3.44$  \\ \hline
	\end{tabular}
	\caption{Numerical values ofcoefficients $c_\beta(n)$ for Ans\"{a}tz {\it II}.}
	\label{table:table7}
\end{table}
Corresponding fixed points and their energies are
\be
(r,y,p_r,p_y) \equiv \lbrace (0,0,0,0)\,, (\pm 1,0,0,0)\,, (0,\pm 1,0,0) \rbrace \,,
\ee
\be
E_F(0,0,0,0) = 0 \,, \quad E_F(\pm 1,0,0,0) = -c_2 N^2 \,, \quad E_F(0,\pm 1,0,0) =  - 8 c_1 N^2 \,.
\ee
Linear stability analysis reveals that $(0,0,0,0)$ is the only unstable fixed point in these models. The minimum value of the potential is $V_{(n),min} = - 8 c_1 N^2$,and from table \ref{table:table7}, numerical values of $V_{(n),min}/N^2$ are, $-48,-84,-180$, respectively. 

\begin{figure}[!htb]
	\centering
	\begin{subfigure}[!htb]{.32\textwidth}
		\centering
		\includegraphics[width= 1\linewidth]{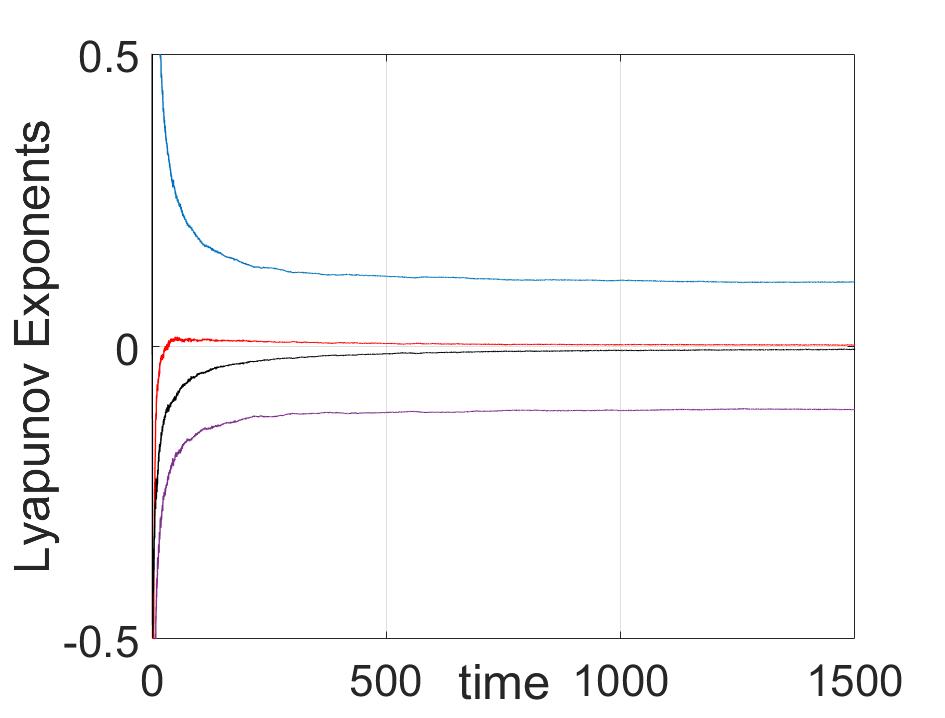}  
		\caption{$n=2$ and $E^\prime/N^2=50$}
		\label{fig:fig5a}
	\end{subfigure}	
	\begin{subfigure}[!htb]{.32\textwidth}
		\centering
		\includegraphics[width=1\linewidth]{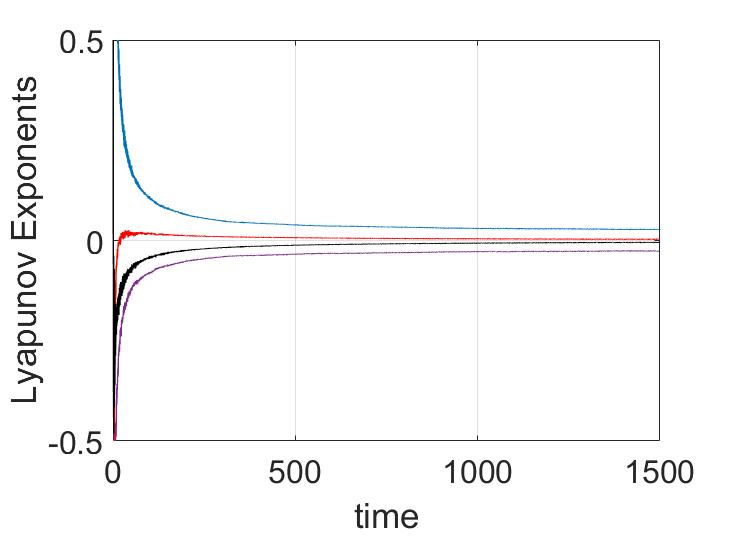}  
		\caption{$n=2$ and $E^\prime/N^2=500$}
		\label{fig:fig5b}
	\end{subfigure}
	\begin{subfigure}[!htb]{.32\textwidth}
		\centering
		\includegraphics[width= 1\linewidth]{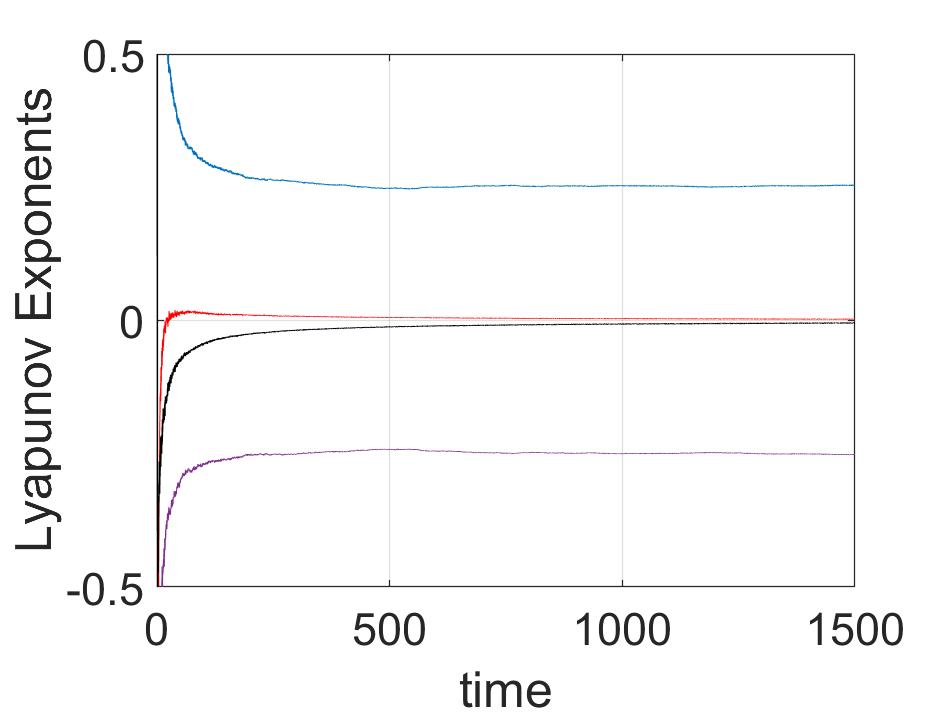}  
		\caption{$n=3$ and $E^\prime/N^2=100$}
		\label{fig:fig5c}
	\end{subfigure}	
	\begin{subfigure}[!htb]{.32\textwidth}
		\centering
		\includegraphics[width= 1\linewidth]{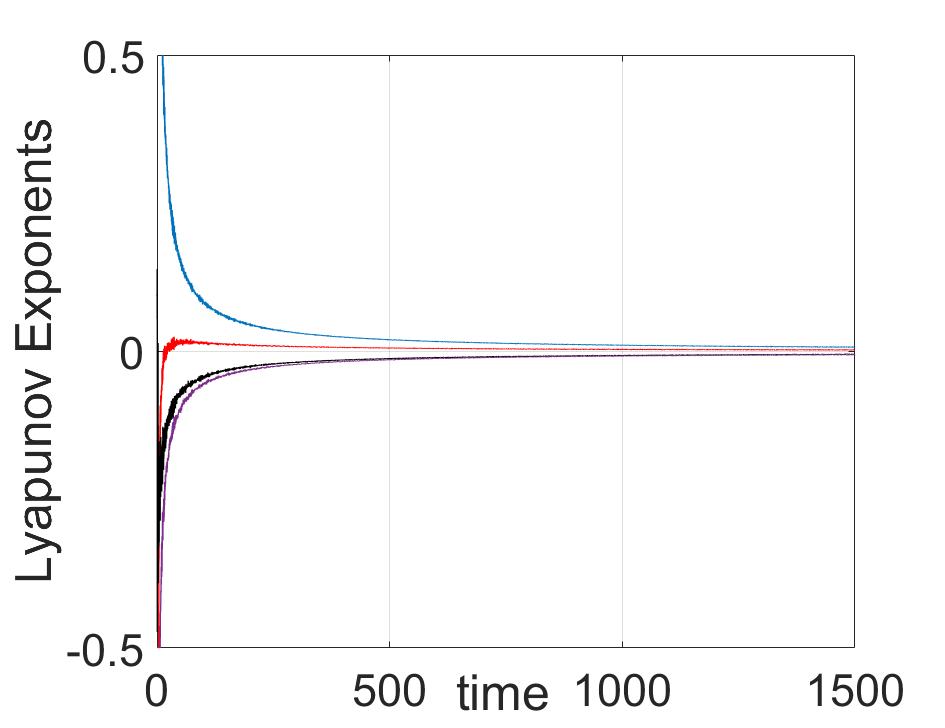}  
		\caption{$n=3$ and $E^\prime/N^2=500$}
		\label{fig:fig5d}
	\end{subfigure}	
	\begin{subfigure}[!htb]{.32\textwidth}
		\centering
		\includegraphics[width= 1\linewidth]{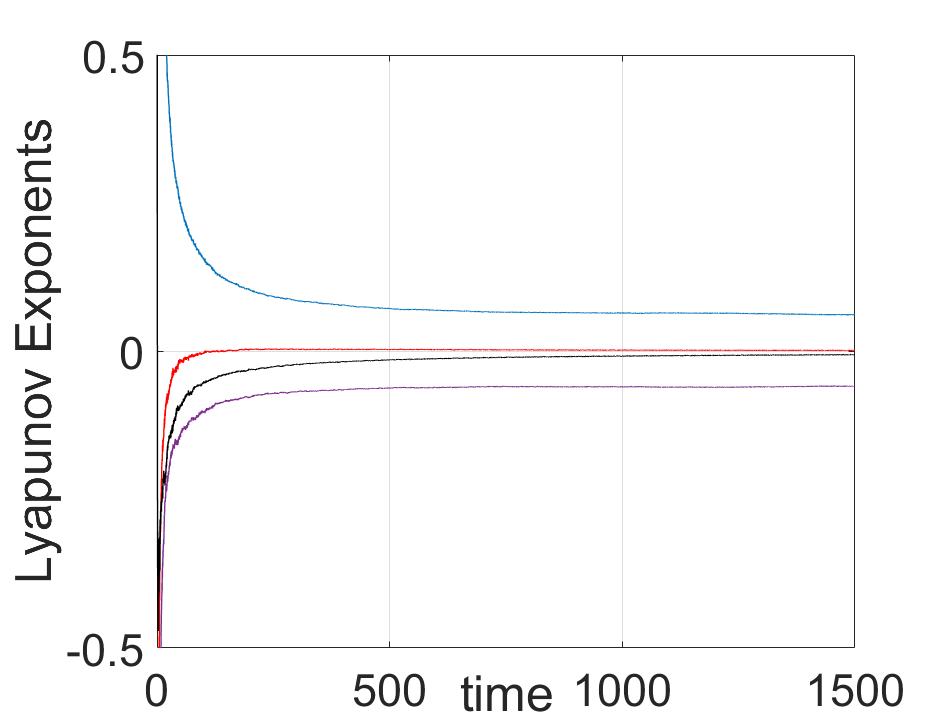}  
		\caption{$n=5$ and $E^\prime/N^2=180$}
		\label{fig:fig5e}
	\end{subfigure}	
	\begin{subfigure}[!htb]{.32\textwidth}
		\centering
		\includegraphics[width= 1\linewidth]{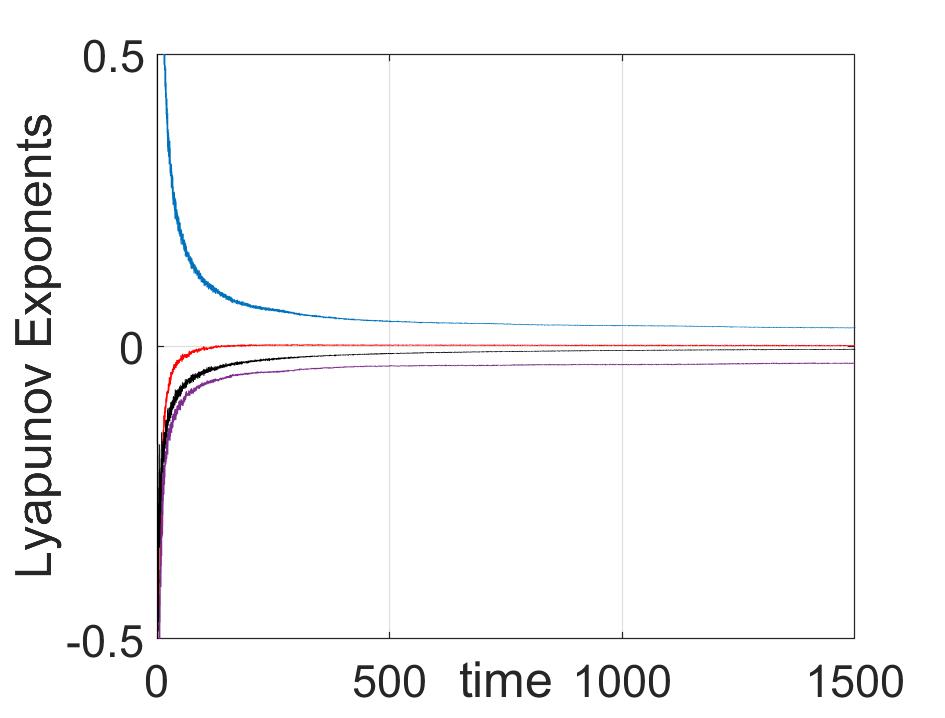}  
		\caption{$n=5$ and $E^\prime/N^2=1000$}
		\label{fig:fig5f}
	\end{subfigure}	
	\caption{Times Series for Lyapunov Exponents}
	\label{fig:fig5}
\end{figure}

\begin{figure}[!htb]
	\centering
	\begin{subfigure}[!htb]{.32\textwidth}
		\centering
		\includegraphics[width= 1\linewidth]{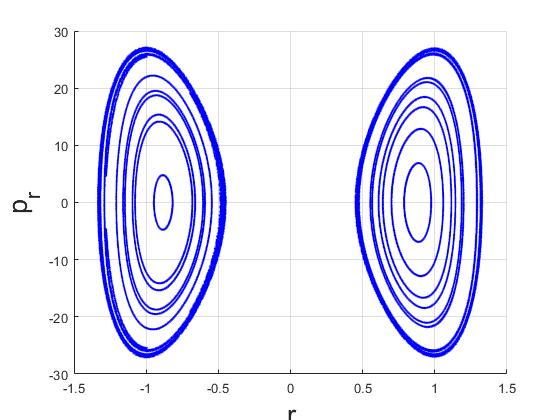}  
		\caption{$n=2$ and $E^\prime/N^2=30$}
		\label{fig:fig6a}
	\end{subfigure}	
	\begin{subfigure}[!htb]{.32\textwidth}
		\centering
		\includegraphics[width=1\linewidth]{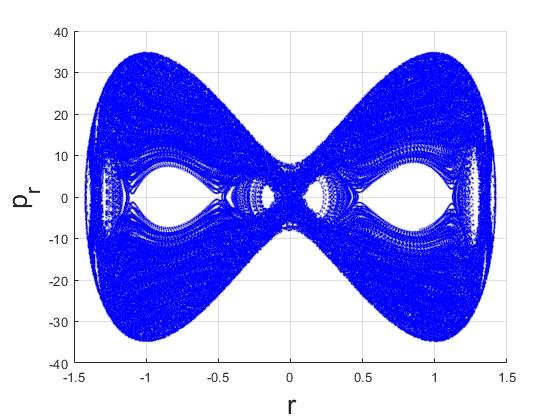}  
		\caption{$n=2$ and $E^\prime/N^2=50$}
		\label{fig:fig6b}
	\end{subfigure}	
	\begin{subfigure}[!htb]{.32\textwidth}
		\centering
		\includegraphics[width=1\linewidth]{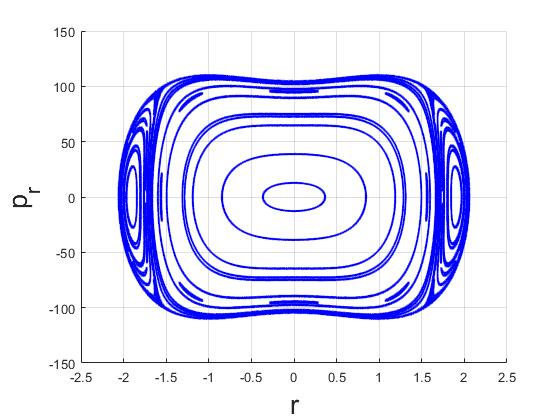}  
		\caption{$n=2$ and $E^\prime/N^2=500$}
		\label{fig:fig6c}
	\end{subfigure}	
	\begin{subfigure}[!htb]{.32\textwidth}
		\centering
		\includegraphics[width= 1\linewidth]{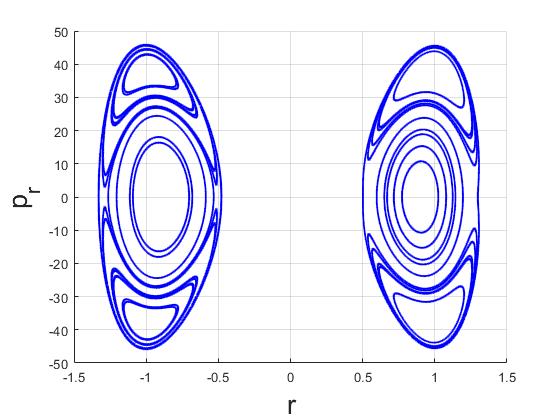}  
		\caption{$n=3$ and $E^\prime/N^2=50$}
		\label{fig:fig6d}
	\end{subfigure}	
	\begin{subfigure}[!htb]{.32\textwidth}
		\centering
		\includegraphics[width=1\linewidth]{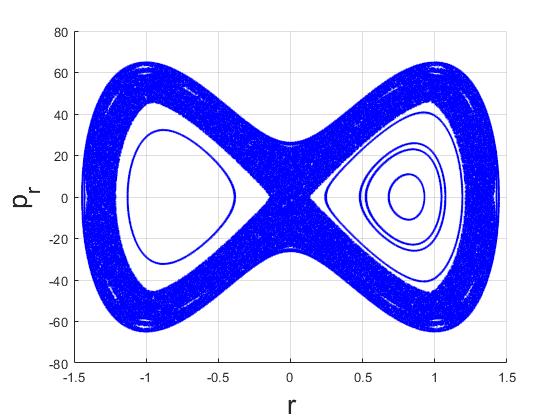}  
		\caption{$n=3$ and $E^\prime/N^2=100$}
		\label{fig:fig6e}
	\end{subfigure}	
	\begin{subfigure}[!htb]{.32\textwidth}
		\centering
		\includegraphics[width=1\linewidth]{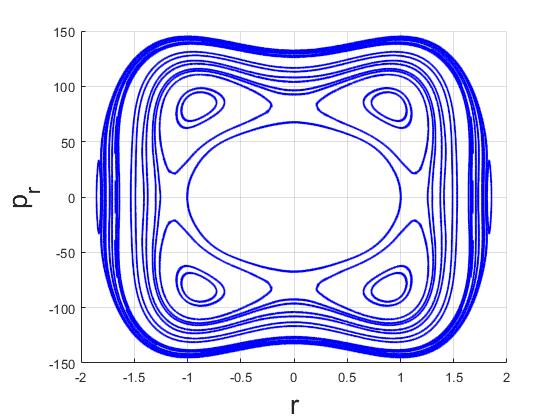}  
		\caption{$n=3$ and $E^\prime/N^2=500$}
		\label{fig:fig6f}
	\end{subfigure}	
	\begin{subfigure}[!htb]{.32\textwidth}
		\centering
		\includegraphics[width= 1\linewidth]{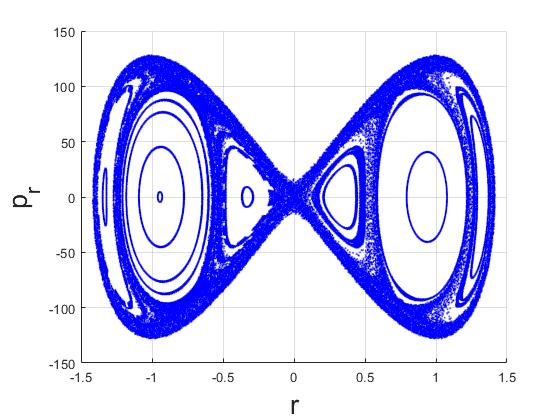}  
		\caption{$n=5$ and $E^\prime/N^2=180$}
		\label{fig:fig6g}
	\end{subfigure}	
	\begin{subfigure}[!htb]{.32\textwidth}
		\centering
		\includegraphics[width=1\linewidth]{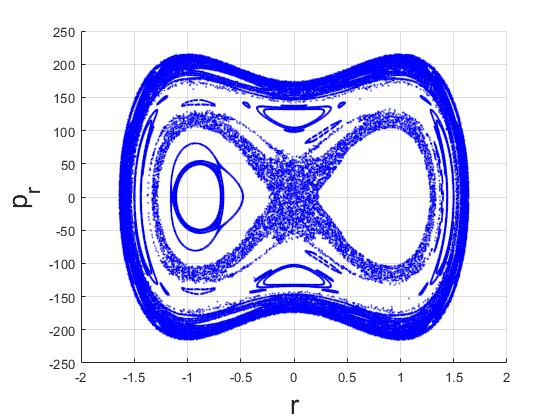}  
		\caption{$n=5$ and $E^\prime/N^2=500$}
		\label{fig:fig6h}
	\end{subfigure}	
	\begin{subfigure}[!htb]{.32\textwidth}
		\centering
		\includegraphics[width=1\linewidth]{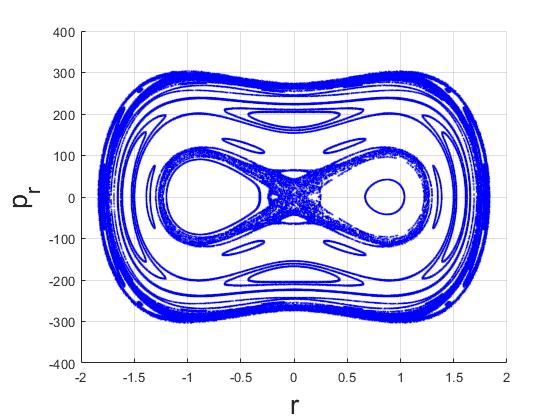}  
		\caption{$n=5$ and $E^\prime/N^2=1000$}
		\label{fig:fig6i}
	\end{subfigure}	
	\caption{Poincar\'{e} Sections}
	\label{fig:fig6}
\end{figure}

We have numerically studied the Lyapunov spectrum at several different values of $E^\prime/N^2$($E^\prime = E+ V_{(n),min}$) in these systems and determined that for an interval starting around the fixed point energies $E_F^\prime/N^2 = 48,84,180$, respectively and going up to $\approx 500$ for the first two cases and $\approx 1000$  for $n=5$, there is a positive Lyapunov exponent indicating the presence of chaotic motion. From the times series plots in figure \ref{fig:fig5} and the Poincar\'{e} sections in figure \ref{fig:fig6}, we see that chaos is not dense, coexists together with quasi periodic motion and remains local in the phase space. At higher energies chaos ceases to exist and the phase space becomes dominated by periodic and/or quasi periodic orbits. 

\section{Conclusions and Outlook}

In this paper, we have studied the emergence of chaos in a massive deformation of a YM matrix theory, which has the same matrix content as that of the bosonic part of the BFSS model. Using an ansatz configuration involving fuzzy two- and four- spheres as backgrounds and assuming collective time dependence of the matrices, we were able to obtain a family of effective models, which descend from tracing over the fuzzy spheres at matrix levels $N = \frac{1}{6}(n+1)(n+2)(n+3)$, for $n=1\,,\cdots \,,7$. We have performed a detailed numerical analysis and demonstrated the chaotic dynamics in these reduced models by obtaining their Lyapunov spectrum and Poincar\'{e} sections. Exploiting the data obtained for the development of MLLE w.r.t energy, and deriving inequalities that relate energy and temperature using the virial and equipartition theorems, we have shown that for the family of reduced models with mass parameters $\mu_1^2=-16$ and $\mu_2^2 = -2$, the MSS bound is violated at low temperatures as well as at zero temperature, the latter being due to the presence of an unstable fixed point at each $n$ in the phase space, whose energy $E_F$, although being well above the minimum of the potential, is still negative, while the MLLE already assumes non-vanishing values around $E_F$ and hence at zero temperature too due to (\ref{EvT}). Making use of best fit curves of the functional form (\ref{lvE}) for $\lambda_n$ versus $E/N^2$ data, and the inequality (\ref{EvT}), we have obtained upper bounds on the temperature  below which the MSS bound is violated. In the high temperature regime, we have determined an upper bound on $\lambda_n$ as a function of temperature, which is parametrically smaller than the MSS bound as can be inspected from (\ref{lvT}). Similar results are also reached for the models with mass-squared values $\mu_1^2=-8$ and $\mu_2^2 = 1$, the main difference being that the only unstable fixed point of the phase space is $(0,0,0,0)$ with zero energy at each level $n$, leading to $\lambda_n \rightarrow 0$ for $E \rightarrow 0$ and hence there is no violation of the MSS bound at zero temperature. We found that critical temperatures ${\tilde T}_c$ in this case are quite smaller than $T_c$ and closer to the value obtained in \cite{Gur-Ari:2015rcq, Kawahara:2007fn} for the BFSS model. 

Let us now discuss some recently developed approaches to go beyond the classical treatment to obtain reasonably good quantum mechanical descriptions of the BFSS and related matrix models. To the best of our knowledge, this important question was not directly addressed up until very recently due to the lack of viable methods to incorporate quantum effects in real time calculations in such matrix models. In contrast, early (as well as some relatively recent) literature on the BFSS model and its deformations at finite temperature resort to the imaginary time formalism, and aim at obtaining a description of their phase diagram via both analytical perturbative and numerical (Monte Carlo) calculations \cite{Kawahara:2006hs, Kawahara:2007fn, DelgadilloBlando:2007vx, DelgadilloBlando:2008vi, Asano:2018nol, Berkowitz:2018qhn}. However, very recently, adaptation of real-time techniques used in the quantum chemistry literature for solving many-body problems \cite{qch1,qch2} proved to be applicable to both BFSS type matrix models and also on the general type of effective Hamiltonians studied in this paper with very interesting consequences \cite{Buividovich:2018scl,Buividovich:2017kfk}. The essence of this new approach and its relevance for our present work can be briefly described as follows. Expectation values of the Heisenberg equations of motion in a quantum state for the generalized coordinates and momentum of the matrix model Hamiltonian, for instance the one that can be obtained from our $L_{YMM}$ or Hamiltonians of the form $H_n$ in (\ref{Hn1}) involve, in turn, expectation values of higher degree polynomial operators in the generalized coordinates and momenta, whose equations of motion generate expectation values of even higher order operators, and in this way an infinite number of Heisenberg equations are generated. The idea proposed in \cite{Buividovich:2018scl,Buividovich:2017kfk}, is to truncate these infinite number of equations by choosing the density matrix of the quantum state, in which the expectation values are evaluated, as the most general time-dependent Gaussian function, or more precisely a Wigner function, characterized by the wave packet centers and wave packet dispersions, which are evaluated using the expectation values of the generalized coordinates and momentum. Use of Wick's theorem truncates the equations to be solved to a set of coupled first order differential equations for the expectation values and the dispersions. This is named as the Gaussian state approximation of the full quantum theory \cite{Buividovich:2018scl,Buividovich:2017kfk}. Application of this method to the BFSS model in \cite{Buividovich:2018scl,Buividovich:2017kfk} showed, among other things, that the Lyapunov exponents are in general smaller than the those obtained in the classical regime, and they appear to vanish below some temperature, ensuring that the MSS bound is not at all exceeded, in contrast to the violation of this bound at small temperatures in the classical treatment.

From the preceding brief description, we conclude that, for our models too, this new method is also directly applicable to go beyond the classical description. In fact, for the Hamiltonian $H_n$ involving the phase space variables $(g_1, g_2, g_3, g_4) \equiv (r, y, p_r, p_y)$ as given in (\ref{psc}), Gaussian state approximation will yield equation of motion for $14$ variables, which are, namely, for the expectation values $\langle g_i \rangle$ and the dispersions $\langle \langle g_i g_j \rangle \rangle \equiv \langle \frac{g_i g_j +g_j g_i}{2} \rangle - \langle g_i \rangle \langle g_j \rangle$, compared to classical equations for only $4$ variables. This will increase the number of directions the trajectories can diverge, which makes it very reasonable to expect a decrease in the value of the LLE, in accord with the results of \cite{Buividovich:2018scl, Buividovich:2017kfk} for the pure BFSS model. Therefore, we think that the Gaussian state approximation can be used to incorporate the effect of quantum dynamics to our analysis and based on the reasoning given above and the present results for the BFSS model, we expect that the resulting LLE spectrum will comply with the MSS bound. Our expectation is further strengthened by the fact that the Gaussian state approximation appears to give very accurate results for the BFSS model, when compared with those of Monte-Carlo simulations performed in the Euclidean time formalism \cite{Berkowitz:2018qhn}.

Let us also note that in \cite{Han:2020bkb}, development of bootstrapping methods for matrix quantum mechanics for small (one- and two-) matrix models have been initiated and it aims at either computing or constraining the expectation values of energy as well as other observables such as $\langle tr X_i^2 \rangle$ in the ground and excited states. We think that this method too, can be applied to our Hamiltonians $H_n$, since in \cite{Han:2020bkb} it has been already applied to the quantum anharmonic oscillator with single degree of freedom with $H = p^2 + x^2 + \alpha x^4$. The results of such an analysis may give further clues on the quantum mechanical properties of the mass deformed BFSS models.  These developments point to exciting new directions, in which the future work can be focused. 

There is already a vast literature on the quantum mechanical treatment of the BFSS, BMN and other related matrix models via both analytical perturbative and non-perturbative Monte Carlo methods in the Euclidean time formalism via imposing periodic boundary conditions on the bosonic matrices with period $\beta$ \cite{Kawahara:2006hs, Kawahara:2007fn, DelgadilloBlando:2007vx, DelgadilloBlando:2008vi, Asano:2018nol, Berkowitz:2018qhn}. The extent of the Euclidean time direction has the meaning of the inverse temperature $\beta = T^{-1}$. In order to make contact with these studies, we must lift the classical treatment appliciable in the real-time setting to quantum mechanical level by suitable means. As clearly demonstrated in \cite{Buividovich:2018scl}  for the bosonic as well as the fully supersymmetric BFSS model (in their ungauged \cite{Maldacena:2018vsr} form), this can be achieved via the Gaussian state approximation, whose essential features are already sketched in the previous few paragraphs. In particular, the Gaussian state approximation allows for expressing observables like $E/N^2$ and the coordinate dispersion $\frac{1}{N} \langle Tr \, X_i^2 \rangle$ in terms of temperature and allow for a direct comparison with the observables computed in the Euclidean time formalism, namely the expectation values of the operators $F^2 = - \frac{1}{N \beta} \int_0^{\beta} dt \, \left (\lbrack X_i \,, X_j \rbrack^2 \right)$ and the "extent of space" $R^2 = \frac{1}{N \beta} \int_0^{\beta} dt \, Tr \, X_i^2$. While the connection of the latter to the coordinate dispersion is apparent, the relation of the former to energy follows, since $\frac{N^2}{4 \lambda_{' t Hooft}} F^2 $ is the potential energy and $\frac{E}{N^2} = \frac{3}{4 \lambda_{' t Hooft}} \langle F^2 \rangle$ due to the virial theorem $\langle K \rangle = 2 \langle V \rangle$ for the quartic potential. In \cite{Buividovich:2018scl}, it is shown that there is indeed very good agreement between these observables, obtained in the Gaussian state approximation and the Monte Carlo studies performed in \cite{Berkowitz:2018qhn}. In a related article \cite{Buividovich:2017kfk}, it is also speculated that the critical temperature $T_c \approx 0.6$ at which the LLE tends to zero for the bosonic part of the BFSS model within the Gaussian state treatment, resembles the confinement-deconfinement  phase transition, which is known to occur in this model at a very close temperature $T^\prime_c \approx 0.9$. Reversing the logic, it is probable that the confinement-deconfinement phase transition prevents the model from violating the MSS bound.


These considerations suggest that, prospective work using Gaussian state approximation on massive (quadratic) as well as cubic (Myers type term) deformations of the BFSS models or the related effective models, such as those discussed in this paper, may be compared with already existing results in the Euclidean time approach; in particular with the recent non-perturbative study on the BMN model \cite{Asano:2018nol} as well as several earlier works \cite{Kawahara:2006hs, Kawahara:2007fn, DelgadilloBlando:2007vx, DelgadilloBlando:2008vi}, by examining how well the temperature dependence of observables like $E/N^2$ and $R^2$ agree or otherwise, and what the temperature dependence of the Lyapunov exponents on the real time side could imply for the transitions between different phases, such as the pure matrix phase with no background geometry, and the fuzzy sphere phase indicating emergent spherical geometry identified for these models in the Euclidean time formalism. To do so, a starting point could be the bosonic part of the BMN model and forming an ansatz configuration involving collective dependence among fuzzy two-sphere matrices only, and obtaining results on the dynamics of the reduced effective action in the Gaussian state approximation.

As a final remark, we want to look at the results from ansatze {\it I} and {\it II} from a broader perspective. In the former case, fuzzy two sphere part is taken as the maximal one for the given matrix size, while in the latter, it is made up of blocks of lowest level spin $1/2$ fuzzy spheres. Thus, we may conclude that fragmentation of the maximal fuzzy sphere into smallest fuzzy spheres results in the plummeting of the chaotic motion. Put it in another way, chaotic motion is favored in configurations with a big blob of a fuzzy two sphere than those in which small bubbles of fuzzy spheres proliferate. Although in both cases, $Z_i$'s describing  block diagonal and the maximal fuzzy sphere are both sparse, i.e. mostly filled with zeros and become even more so with increasing $n$, in the former case, interactions among $D0$-branes in some directions remain confined to distinct pairs only, while for the maximal fuzzy sphere, they interact on a chain of nearest neighbors and this makes a difference. Viewing these facts in the reverse order suggests that as small bubbles of fuzzy spheres coalesce to form fuzzy spheres of higher spin, it also triggers the entire system to evolve from phase space filled with quasi periodic motion to that dominated by chaos. A possible mechanism for splitting or fragmentation of fuzzy spheres into fuzzy spheres of smaller spin was suggested some time ago in \cite{Balachandran:2003wv} and it involves the introduction of an appropriate $SU(2)$ equivariant co-product and an associated Hopf algebra structure. We suspect that it may be possible to make use of this co-product in the present context to get a handle on the topology change of the fuzzy background and its impact on the dynamics of $D0$-branes and therefore on chaos. We hope to report on any developments in these directions elsewhere.

\acknowledgments{Authors thank G.C.Toga for collaboration and participation at the early stages of this project. Part of S.K.'s work was carried out during his sabbatical stay at the physics department of CCNY of CUNY and he thanks V.P. Nair and D. Karabali for the warm hospitality at CCNY and the metropolitan area. S.K. thanks A.P. Balachandran for reading the manuscript and critical comments. K.B.,S.K., C.T., acknowledge the support of TUBITAK under the project number 118F100 and the METU research project GAP-105-2018-2809.}

\appendix

\section{Review of Fuzzy Spheres}\label{rA1}

In this appendix, we briefly collect the definition and main features of the fuzzy $S^2$ and the fuzzy $S^4$. For detailed discussions the references \cite{Balachandran:2005ew} and \cite{Ydri:2016dmy} may be consulted.

\subsection{Fuzzy $S^2$}

Let us denote the $SU(2)$ representations in the spin-$j=\frac{N-1}{2}$ irreducible representation by $L_i$ $(i =1,2,3)$. They satisfy the commutation relations  
\be
\lbrack L_i\,, L_j \rbrack = i \epsilon_{ijk} L_k \,.
\ee
The fuzzy two-sphere, $S_F^2$ at the matrix level $N$ is defined in terms of the rescaled generators of $SU(2)$ 
\be
\widehat{ L_i} = \frac{1}{\sqrt{j(j+1)}} L_i
\ee
and their commutation relations are simply given as 
\be
\label{NnCmm}
\lbrack \widehat{L_i} ,\widehat{L_j} \rbrack = i \frac{1}{\sqrt{j(j+1)}} \epsilon_{ijk} \widehat{ L_k}.  
\ee
They satisfy
\be
{\widehat{ L_i}}^2  = \mathds{1}_{N} \,.
\ee
As $N$ goes to infinity, the standard commutative $S^2$ is recovered.

\subsection{Fuzzy $S^4$}
   
Fuzzy four-sphere, $S_F^4$ can be constructed as follows. We introduce the $\gamma$-matrices in five dimensions that are associated to the isometry group $SO(5)$ of $S^4$. They are $4\times4$ matrices satisfying the anticommutation relations
\be
\left \{ \gamma_a,\gamma_b \right \} = 2 \delta_{ab} {\mathds{1}}_4 \,,
\ee 
where $a,b=1,..,5$. 

Generators of the spinor representation $(0,1)$ of $SO(5)$ are given as
\be
\label{spin5}
G_{ab} = -\frac{i}{4} \lbrack \gamma_a\,, \gamma_b \rbrack \,,
\ee
and fulfill the $SO(5)$ commutation relations
\be
\lbrack G_{ab}\,, G_{cd} \rbrack = i(\delta_{a c} G_{b d} + \delta_{b d} G_{a c} - \delta_{a d} G_{b c} - \delta_{b c} G_{a d})\,.
\ee 
$\gamma_{a}$ act on the four-dimensional spinor space $\mathbb{C}^4$ which is the carrier space of the $(0,1)$ IRR of $SO(5)$. Let us define the Hilbert space $\mathcal{H}_n$, as the carrier space of the $(0,n)$ IRR of $SO(5)$, which may be formed as the $n$-fold symmetric tensor product of $\mathbb{C}^4$ as
\be
\mathcal{H}_n = \big(\mathbb{C}^4 \otimes \cdots \otimes \mathbb{C}^4\big)_{Sym} \,.
\ee
and has the dimension
\be
\label{dimFS4}
N = dim(0,n) = \frac{1}{6}(n+1)(n+2)(n+3) \,.
\ee  
We may as well form the $n$-fold symmetric tensor product of the $\gamma$-matrices 
\be
\label{fSph} 
X_a = \frac{1}{2}(\gamma_a \otimes {\mathds{1}}_4 \otimes ... \otimes {\mathds{1}}_4 + ... + {\mathds{1}}_4 \otimes ... \otimes {\mathds{1}}_4 \otimes \gamma_a )
\ee
which carry the $(0,n)$ IRR of $SO(5)$ and naturally act on $\mathcal{H}_n$. $X_a$ are $N \times N$ Hermitian matrices, which satisfy the defining relations for the fuzzy four-sphere, which are given as
\beqa
X_a X_a  &=&  \frac{1}{4}n(n+4) {\mathds{1}}_N \,, \nn \\
\epsilon^{a b c d e} X_a X_b X_c X_d &=& (n+2) X_e \,, 
\label{fuzzyS4}     
\eeqa
where $X_a X_a$ is an $SO(5)$ invariant operator.Commutators of $X_a$ yields 
\be
\lbrack X_a, X_b \rbrack = i M_{ab} \,, 
\label{XXcomm}
\ee 
where $M_{ab}$ are the generators of $SO(5)$ in the $(0,n)$ IRR and satisfy the same commutation relations as $G_{ab}$. We also have that $X_a$ transform as vectors under the $SO(5)$ rotations generated by $M_{ab}$, i.e.
\be
\lbrack M_{ab}, X_c \rbrack = i(\delta_{a c} X_b - \delta_{b c} X_a) \,. 
\ee   
From (\ref{XXcomm}) we see that, as opposed to the construction of $S_F^2$, the algebra of $X_a$'s defining the $S^4_F$ do not close. A more detailed analysis yields that $S_F^4$ may be understood as a squashed $CP^3_F$ with $S_F^2$ fibers  In other words, we may state that $S_F^4$ comes attached with a fuzzy two sphere at its every point. Further details of these features may be found in \cite{Ydri:2017ncg}.    

\section{Poincar\'{e} Sections}

Here we give we give the Poincar\'{e} sections taken for the models at the levels $n=3,5,6,7$ at various values of $E^\prime/N^2$.

\begin{figure}[!htb]
		\begin{subfigure}{.32\textwidth}
			\centering
			\includegraphics[width=1\linewidth]{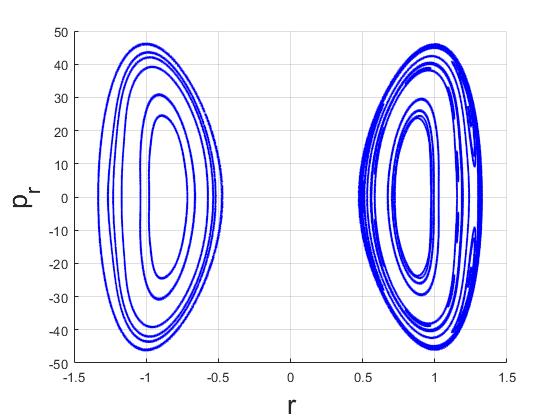}  
			\caption{$n=3$ and $E^\prime/N^2=50$}
			\label{fig:figAa}
		\end{subfigure}	
		\begin{subfigure}{.32\textwidth}
			\centering
			\includegraphics[width= 1\linewidth]{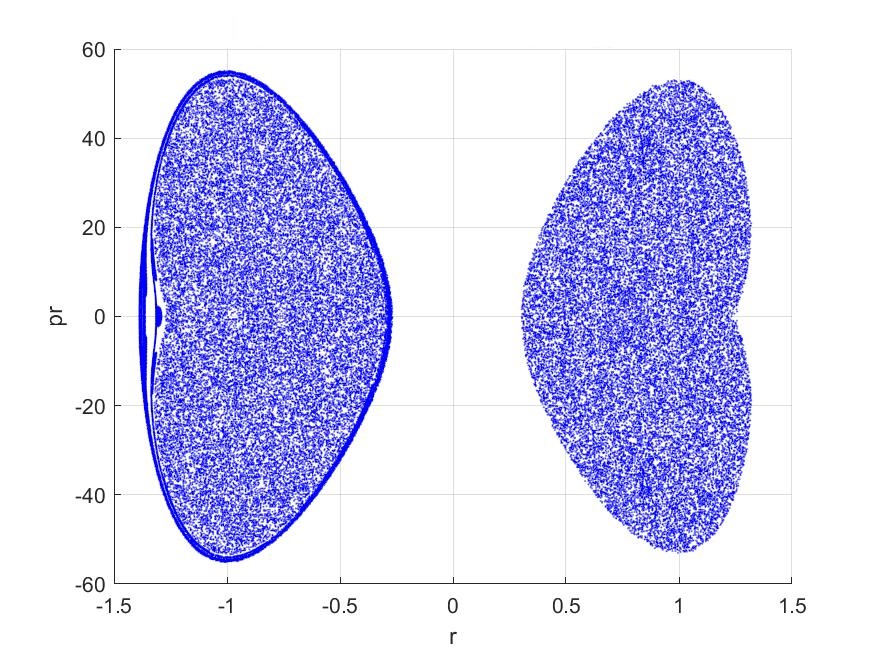}  
			\caption{$n=3$ and $E^\prime/N^2=71$}
			\label{fig:figAb}
		\end{subfigure}	
		\begin{subfigure}{.32\textwidth}
			\centering
			\includegraphics[width=1\linewidth]{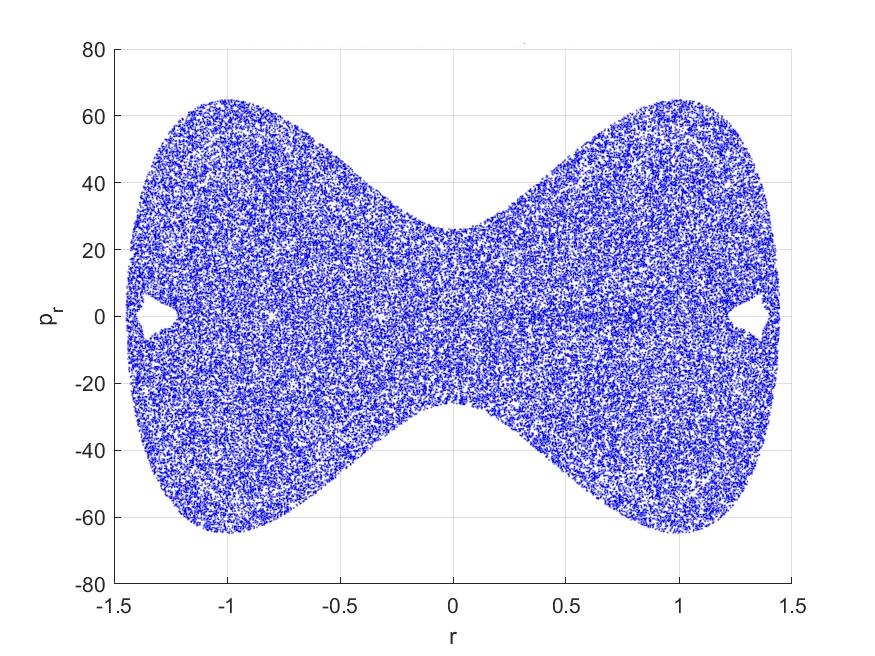}  
			\caption{$n=3$ and $E^\prime/N^2=100$}
			\label{fig:figAc}
		\end{subfigure}	
	\begin{subfigure}[!htb]{.32\textwidth}
		\centering	
			\includegraphics[width= 1\linewidth]{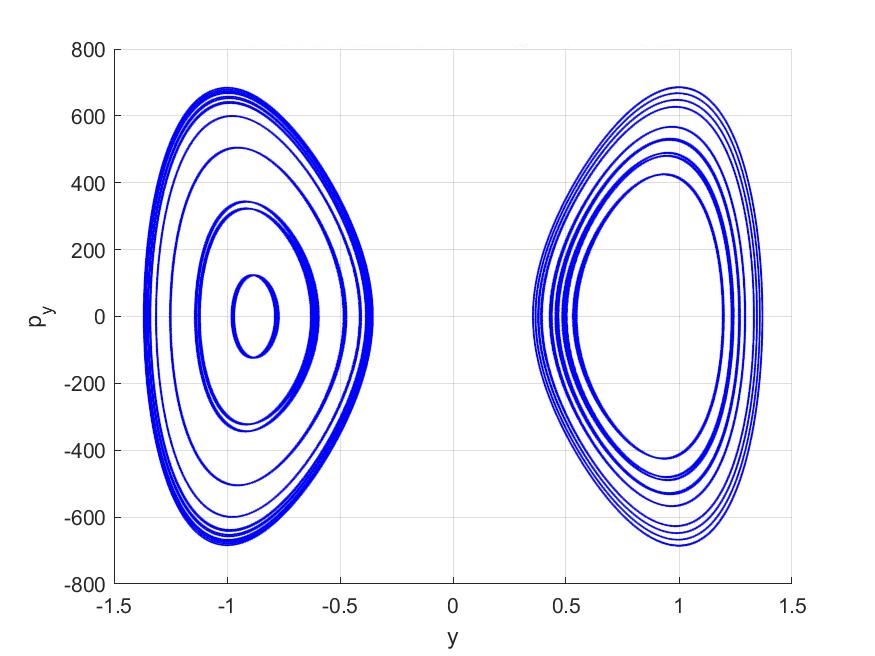}  
			\caption{$n=5$ and $E^\prime/N^2=300$}
			\label{fig:figAd}
		\end{subfigure}	
		\begin{subfigure}[!htb]{.32\textwidth}
			\centering
			\includegraphics[width=1\linewidth]{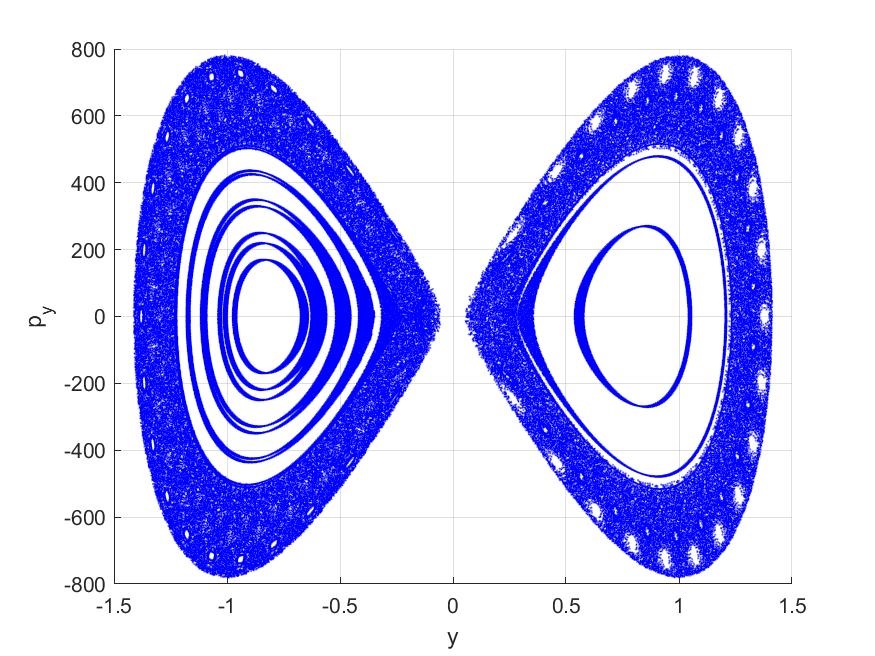}  
			\caption{$n=3$ and $E^\prime/N^2=390$}
			\label{fig:figAe}
		\end{subfigure}	
		\begin{subfigure}[H]{.32\textwidth}
			\centering
			\includegraphics[width=1\linewidth]{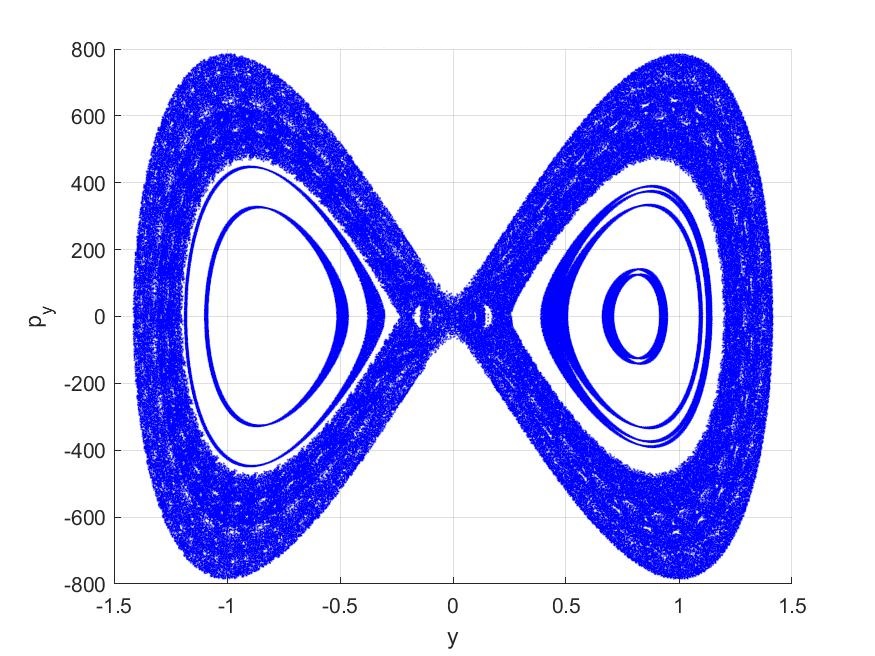}  
			\caption{$n=3$ and $E^\prime/N^2=395$}
			\label{fig:figAf}
		\end{subfigure}
		\begin{subfigure}[!htb]{.32\textwidth}
			\centering
			\includegraphics[width= 1\linewidth]{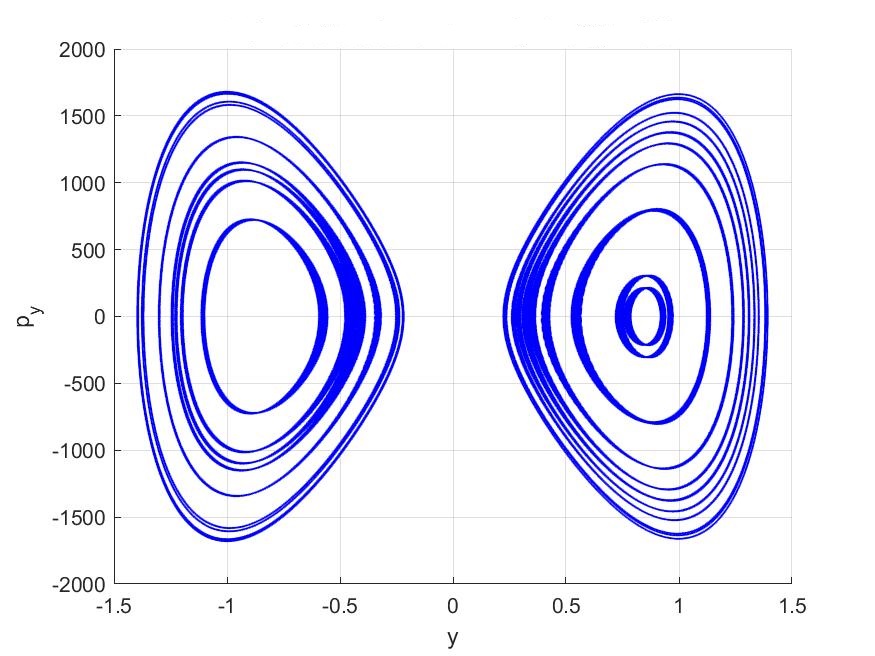}  
			\caption{$n=6$ and $E^\prime/N^2=800$}
			\label{fig:figAg}
		\end{subfigure}	
		\begin{subfigure}[!htb]{.32\textwidth}
			\centering
			\includegraphics[width=1\linewidth]{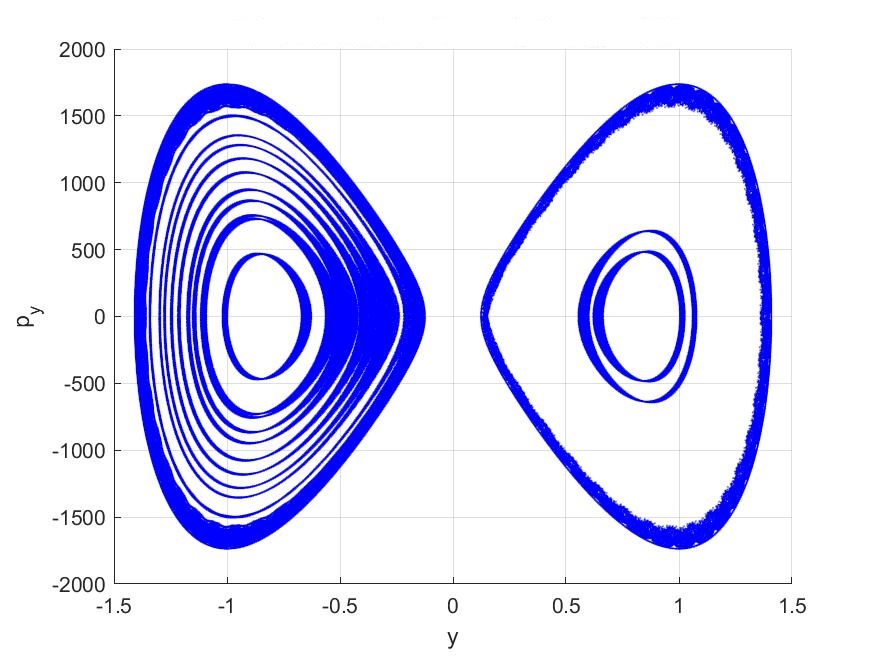}  
			\caption{$n=6$ and $E^\prime/N^2=855$}
			\label{fig:figAh}
		\end{subfigure}	
		\begin{subfigure}[!htb]{.32\textwidth}
			\centering
			\includegraphics[width= 1\linewidth]{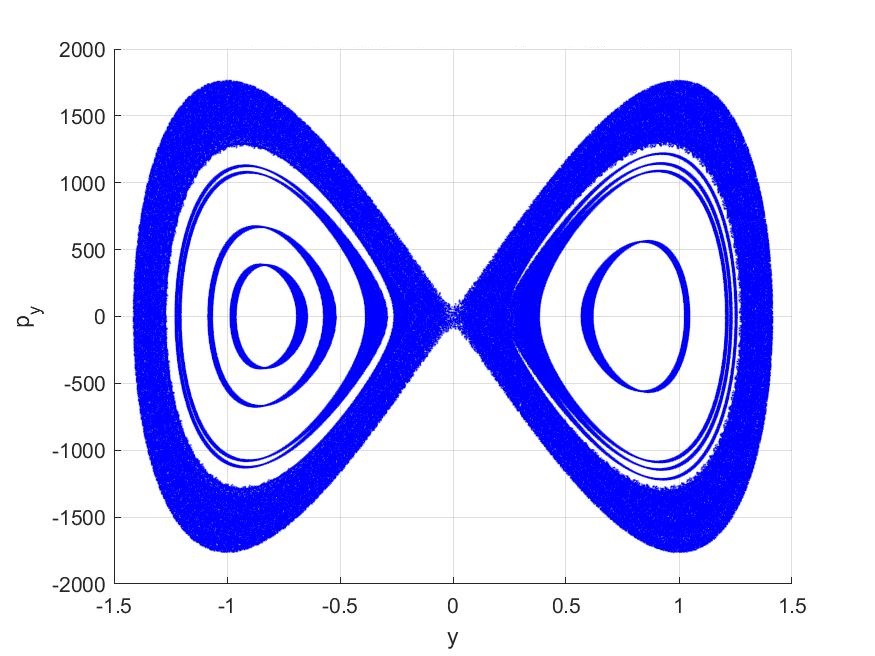}  
			\caption{$n=6$ and $E^\prime/N^2=885$}
			\label{fig:figAi}
		\end{subfigure}	
		\begin{subfigure}[!htb]{.32\textwidth}
			\centering
			\includegraphics[width= 1\linewidth]{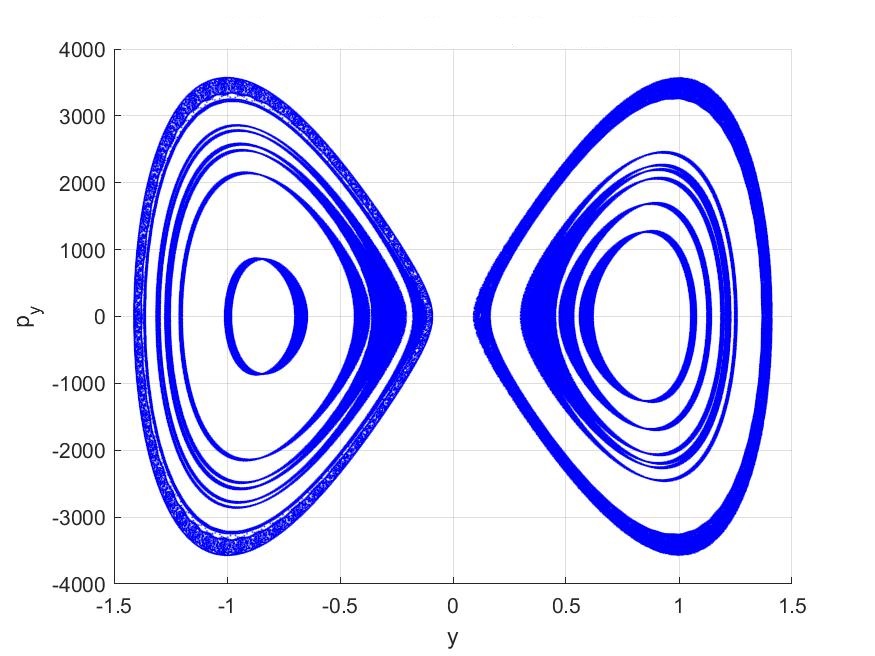}  
			\caption{$n=7$ and $E^\prime/N^2=800$}
			\label{fig:figAj}
		\end{subfigure}	
		\begin{subfigure}[!htb]{.32\textwidth}
			\centering
			\includegraphics[width=1\linewidth]{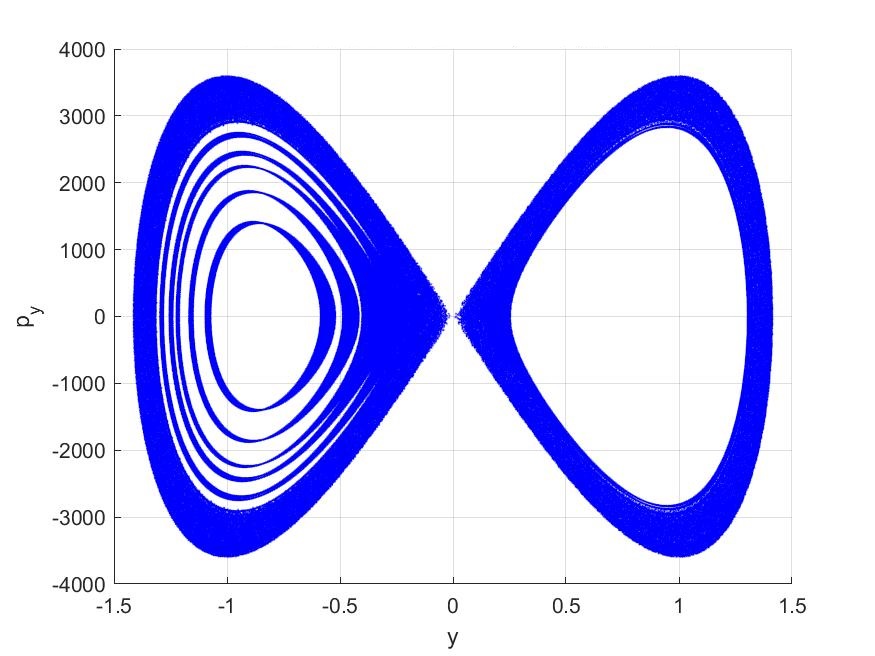}  
			\caption{$n=7$ and $E^\prime/N^2=855$}
			\label{fig:figAk}
		\end{subfigure}	
		\begin{subfigure}[!htb]{.32\textwidth}
			\centering
			\includegraphics[width= 1\linewidth]{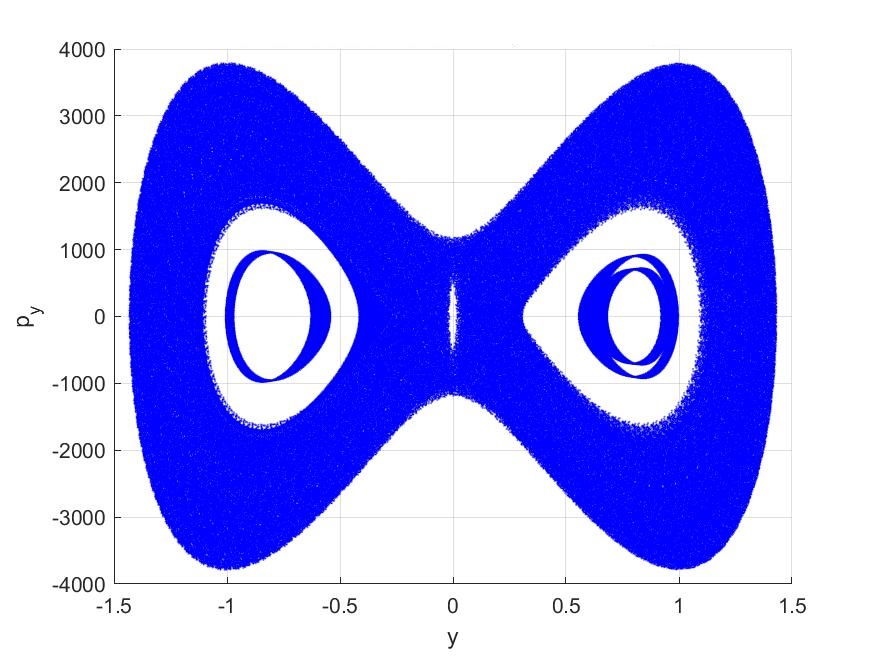}  
			\caption{$n=7$ and $E^\prime/N^2=885$}
			\label{fig:figAl}
		\end{subfigure}	
		\caption{Poincar\'{e} Sections at $\mu_1^2=-16$ and $\mu_2^2 = -2$}
		\label{fig:figA}
	\end{figure}

	\begin{figure}[!htb]
		\begin{subfigure}[!htb]{.32\textwidth}
			\centering
			\includegraphics[width=1\linewidth]{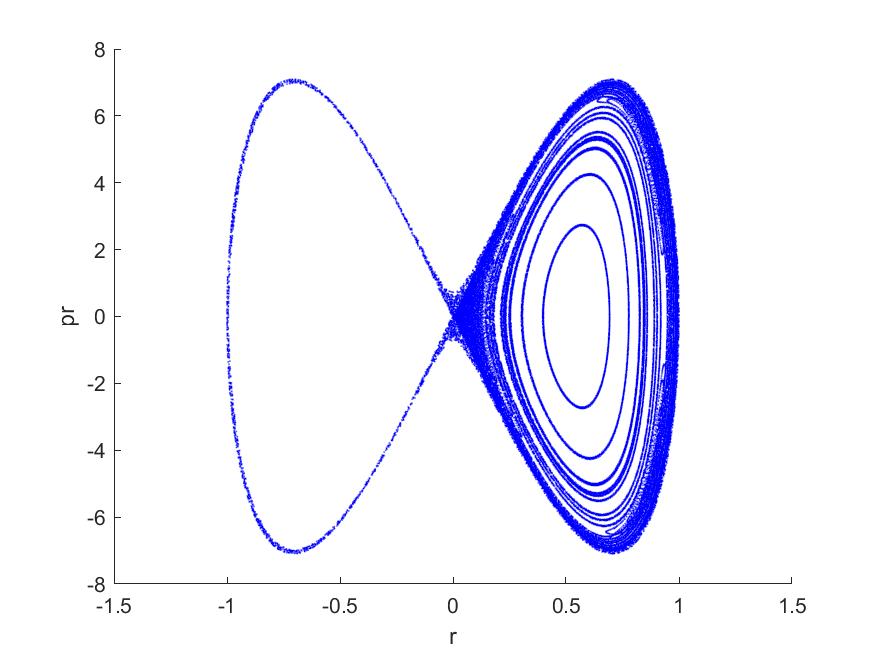}  
			\caption{$n=1$ and $E^\prime/N^2=5$}
			\label{fig:figBa}
		\end{subfigure}	
		\begin{subfigure}[!htb]{.32\textwidth}
			\centering
			\includegraphics[width= 1\linewidth]{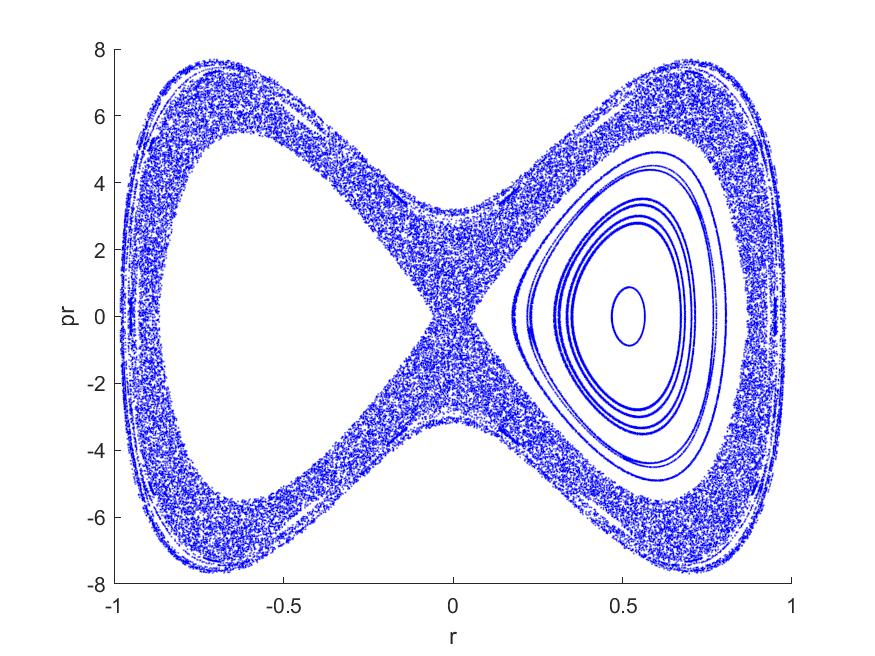}  
			\caption{$n=1$ and $E^\prime/N^2=6$}
			\label{fig:figBb}
		\end{subfigure}	
		\begin{subfigure}[!htb]{.32\textwidth}
			\centering
			\includegraphics[width=1\linewidth]{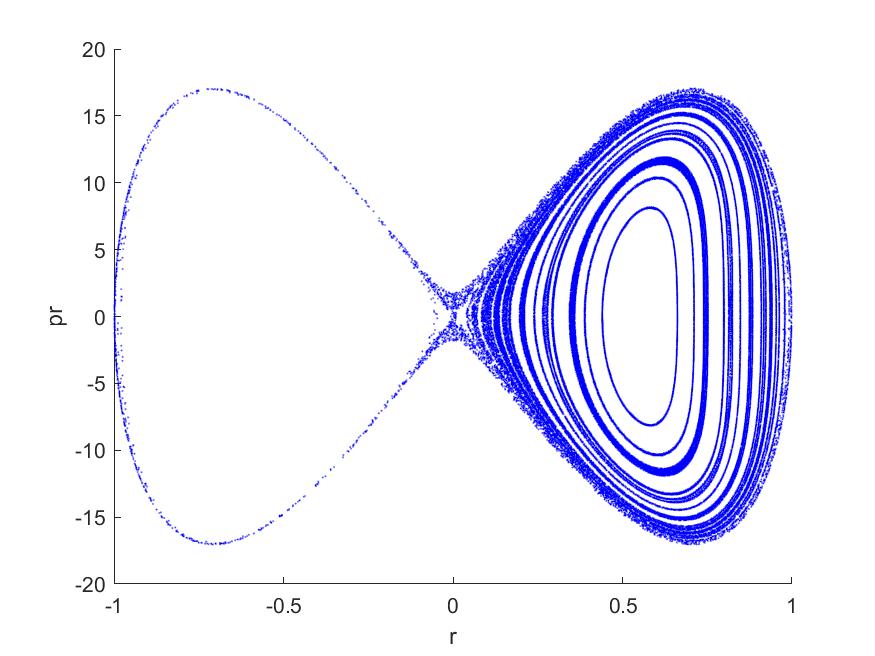}  
			\caption{$n=2$ and $E^\prime/N^2=12$}
			\label{fig:figBc}
		\end{subfigure}	
		\begin{subfigure}[H]{.32\textwidth}
			\centering
			\includegraphics[width= 1\linewidth]{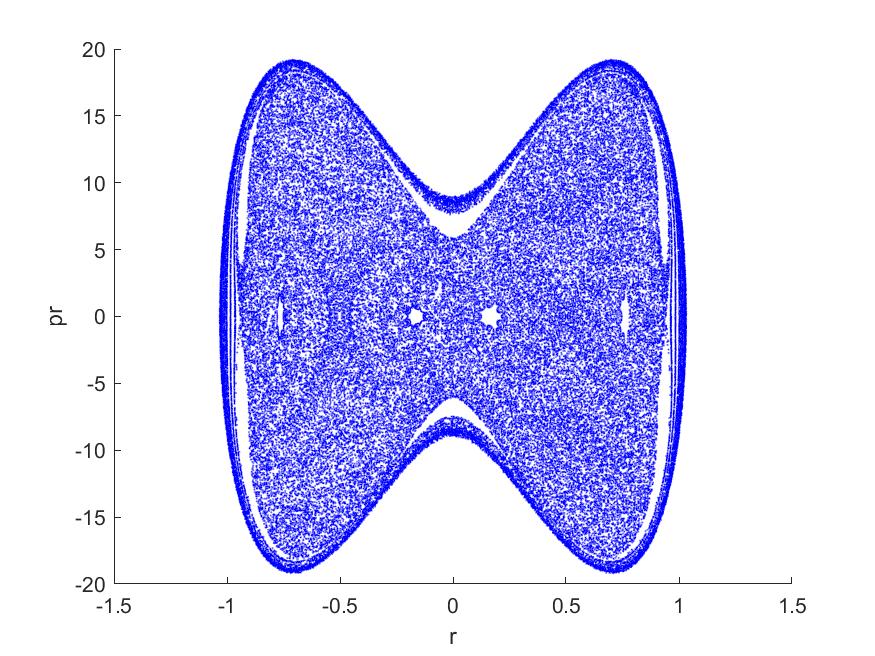}  
			\caption{$n=2$ and $E^\prime/N^2=15$}
			\label{fig:figBd}
		\end{subfigure}	
		\begin{subfigure}[!htb]{.32\textwidth}
			\centering
			\includegraphics[width=1\linewidth]{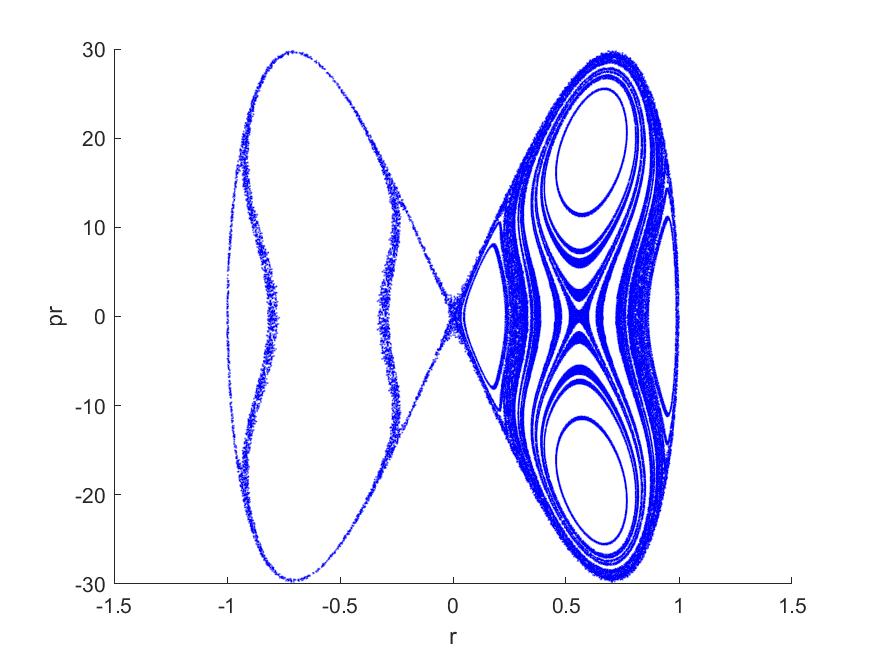}  
			\caption{$n=3$ and $E^\prime/N^2=21$}
			\label{fig:figBe}
		\end{subfigure}	
		\begin{subfigure}[H]{.32\textwidth}
			\centering
			\includegraphics[width=1\linewidth]{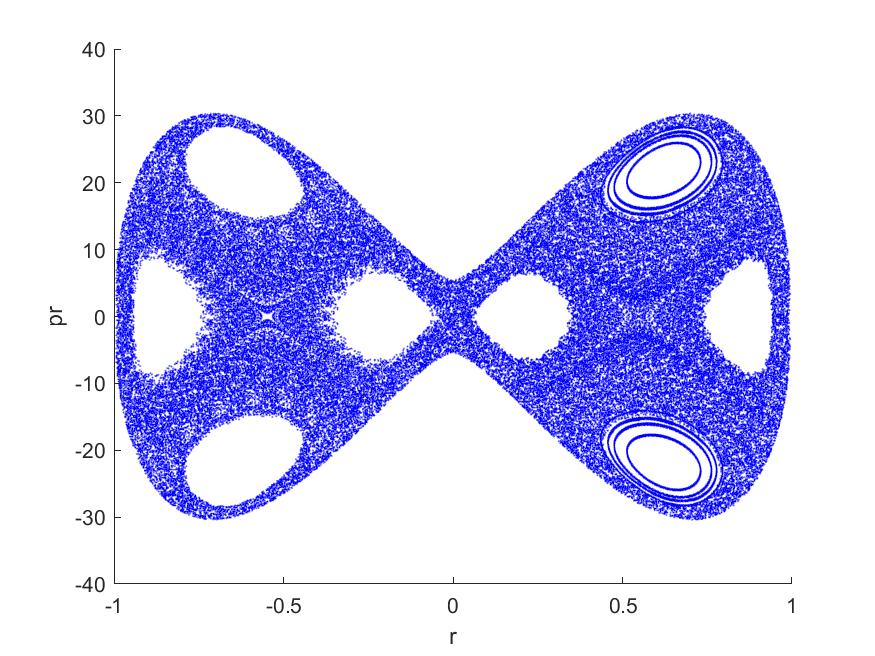}  
			\caption{$n=3$ and $E^\prime/N^2=22$}
			\label{fig:figBf}
		\end{subfigure}
		\begin{subfigure}[!htb]{.32\textwidth}
			\centering
			\includegraphics[width=1\linewidth]{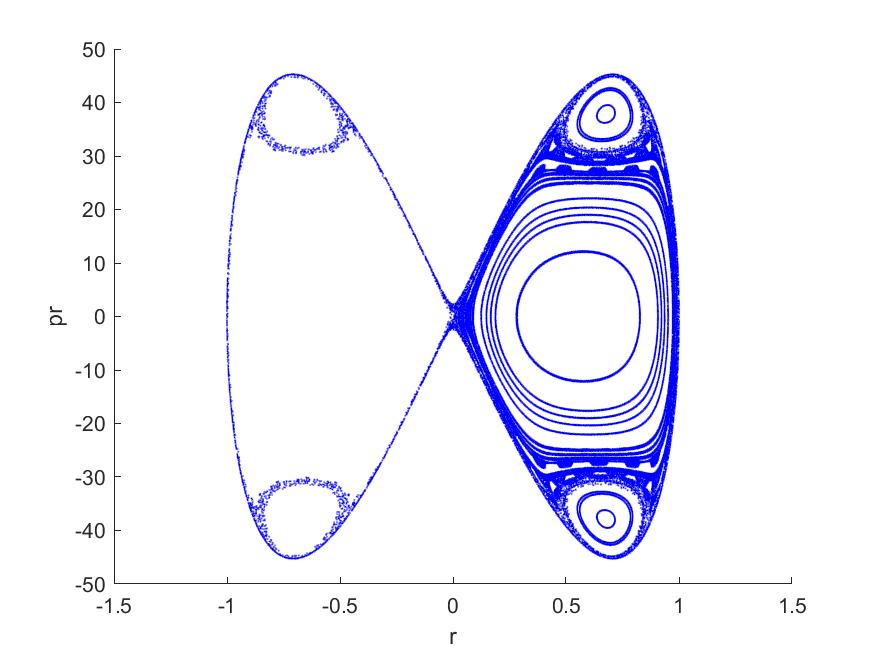}  
			\caption{$n=4$ and $E^\prime/N^2=32$}
			\label{fig:figBg}
		\end{subfigure}	
		\begin{subfigure}[!htb]{.32\textwidth}
			\centering
			\includegraphics[width= 1\linewidth]{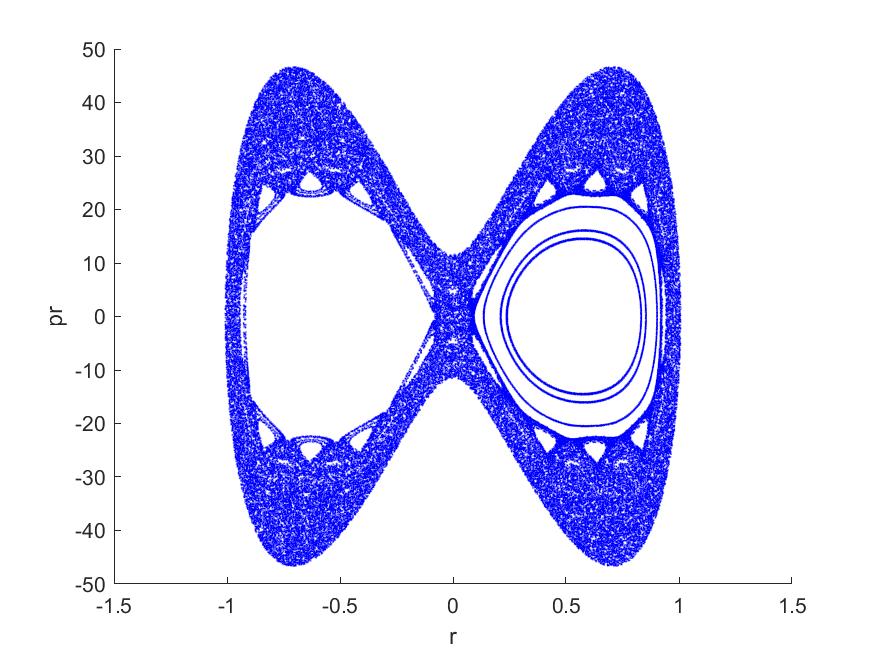}  
			\caption{$n=4$ and $E^\prime/N^2=34$}
			\label{fig:figBh}
		\end{subfigure}	
		\begin{subfigure}[!htb]{.32\textwidth}
			\centering
			\includegraphics[width=1\linewidth]{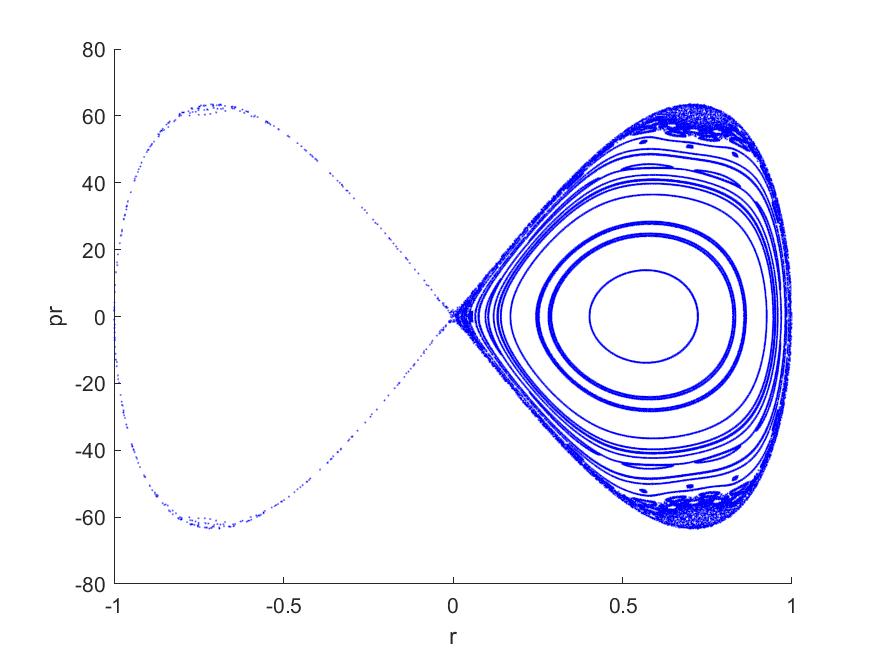}  
			\caption{$n=5$ and $E^\prime/N^2=45$}
			\label{fig:figBi}
		\end{subfigure}	
		\begin{subfigure}[H]{.32\textwidth}
			\centering
			\includegraphics[width= 1\linewidth]{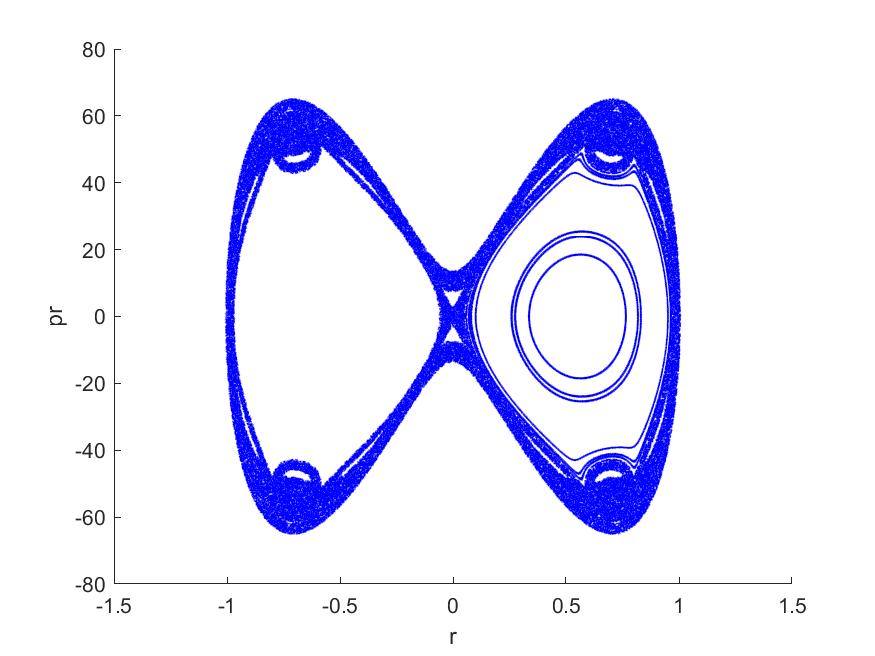}  
			\caption{$n=5$ and $E^\prime/N^2=47$}
			\label{fig:figBj}
		\end{subfigure}	
		\begin{subfigure}[!htb]{.32\textwidth}
			\centering
			\includegraphics[width=1\linewidth]{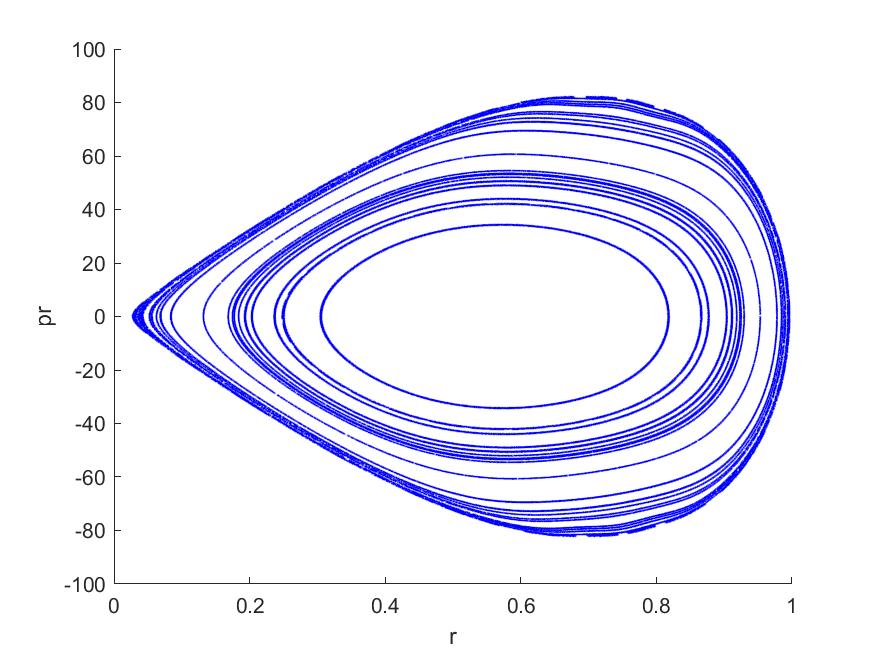}  
			\caption{$n=6$ and $E^\prime/N^2=60$}
			\label{fig:figBk}
		\end{subfigure}	
		\begin{subfigure}[H]{.32\textwidth}
			\centering
			\includegraphics[width=1\linewidth]{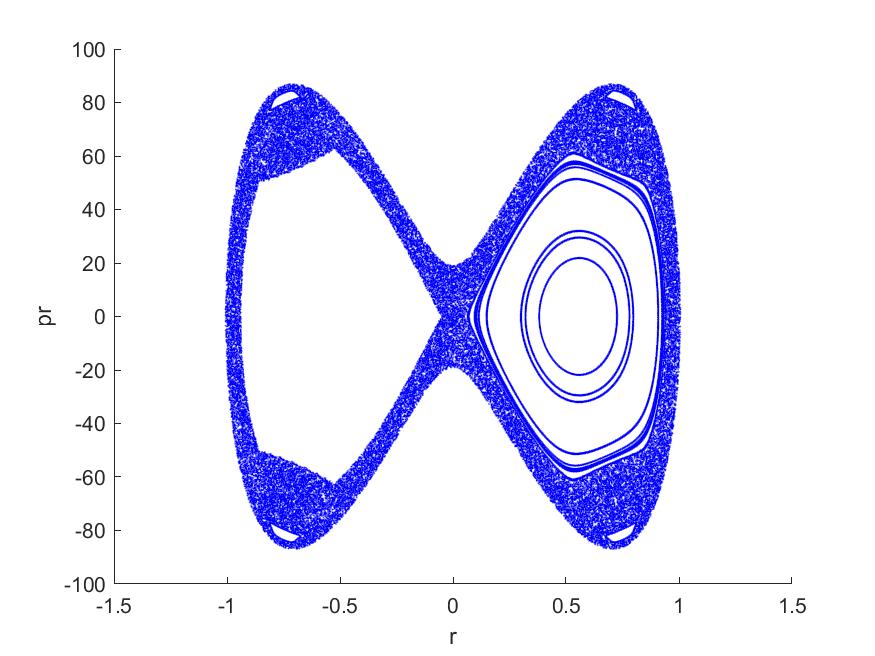}  
			\caption{$n=6$ and $E^\prime/N^2=63$}
			\label{fig:figBl}
		\end{subfigure}
		\caption{Poincar\'{e} Sections at $\mu_1^2=-8$ and $\mu_2^2 =1$}
		\label{fig:figB}
	\end{figure}
	
\end{document}